\documentclass[10pt,aps,pra,amsfonts,amssymb,amsmath,%showpacs,showkeys,
floatfix,nofootinbib, groupedaddress,superscriptaddress,citesort,twocolumn]{revtex4}%

\usepackage{amsthm}
\usepackage{mathrsfs}
\usepackage{amsfonts}
\usepackage{amstext}
\usepackage{amsmath}
\usepackage{amssymb}
\usepackage{color,xcolor}
\usepackage{bm}
\usepackage{multirow}
\usepackage{adjustbox}
\usepackage{hyperref}
\usepackage{float}
\usepackage{bbm}
\usepackage{graphicx}
\usepackage{booktabs}
\usepackage{algorithm}
\usepackage{algorithmic}

\def\qed{\leavevmode\unskip\penalty9999 \hbox{}\nobreak\hfill
     \quad\hbox{\leavevmode  \hbox to.77778em{%
              \hfil\vrule   \vbox to.675em%
               {\hrule width.6em\vfil\hrule}\vrule\hfil}}
     \par\vskip3pt}
\numberwithin{equation}{section}

\newtheorem{theorem}{Theorem}[section]

\newtheorem{lemma}{Lemma}[subsection]

\newtheorem{remark}[theorem]{Remark}

\newtheorem{definition}[theorem]{Definition}

\makeatletter
\newenvironment{breakablealgorithm}
  {
   \begin{center}
     \refstepcounter{algorithm}% New algorithm
     \hrule height.8pt depth0pt \kern2pt% Top rule
     \renewcommand{\caption}[2][\relax]{% Custom caption
       {\raggedright\textbf{\ALG@name~\thealgorithm} ##2\par}% Number includes subsection
       \ifx\relax##1\relax % #1 is \relax
         \addcontentsline{loa}{algorithm}{\protect\numberline{\thesubsection.\thealgorithm}##2}%
       \else % #1 is not \relax
         \addcontentsline{loa}{algorithm}{\protect\numberline{\thesubsection.\thealgorithm}##1}%
       \fi
       \kern2pt\hrule\kern2pt
     }
  }
  {
     \kern2pt\hrule\relax% Bottom rule
   \end{center}
  }
  \makeatother
\begin{document}

\title[$k$-entanglement measure]
{A $k$-entanglement measure complying with the many-body resource theory
 and its evaluation}

\author{Jie Guo}%\email{jieguo11@163.com}
\address{College of Mathematics, Taiyuan University of Technology, Taiyuan, 030024, P. R.
China}

\author{Shuyuan Yang}
\address{School of Mathematics, North University of China,
Taiyuan 030051,  P. R. China}

\author{Jinchuan Hou} \email{jinchuanhou@aliyun.com}

\address{College of Mathematics, Taiyuan University of Technology, Taiyuan, 030024, P. R.
China}
\author{Xiaofei Qi}\email{xiaofeiqisxu@aliyun.com}
\address{School of Mathematical Mathematics and Statistics, Shanxi University, Taiyuan 030006,
P. R. China; Key Laboratory of Complex Systems and Data Science of Ministry of Education,
Shanxi University, Taiyuan  030006,  P. R. China}
\author{ Kan He}\email{hekan@tyut.edu.cn}
\address{College of Mathematics, Taiyuan University of Technology, Taiyuan, 030024, P. R.
China}

\begin{abstract}
{$k$-entanglement provides a hierarchical characterization of multipartite correlations according to their entanglement structure, ranging from  multipartite entanglement ($k=n$) to genuine multipartite entanglement ($k=2$). Its quantitative treatment for mixed states faces three related obstacles:  the standard axioms for such  quantifications does not incorporate the requirements from many-body resource theory,
as it
 does not regulate how to quantify the entanglement assigned to part systems of the same multipartite state and how to describe the rule that the entanglement possessed by a part system never exceeds the entanglement possed by whole system;
 no explicit  measures are known that satisfy the above requirements from  many-body resource theory; and the extreme  difficulty in numerical evaluation at mixed states of known quantifications. In this work,  using the notion of sub-repartitions, we first formulate an axiomatic framework for \emph{true} $k$-entanglement measures that obeys the many-body resource theory, by supplementing usual faithfulness, LOCC monotonicity and permutation symmetry with two new ones: unification condition and hierarchy condition. Then, for every $2\leq k\leq n$, we construct a witness-based $k$-entanglement measure $E_w^{(k,n)}$  without the convex-roof extension, which is valid  for both finite- and infinite-dimensional $n$-partite systems. We prove that it satisfies all  five axioms above, together with convexity and subadditivity. Finally, we derive a method of constructing finite-dimensional $k$-entanglement witnesses from positive operator and $k$-separable-state optimizations, design  an evaluation scheme of $E_w^{(k,n)}$ based on pre-established database of  $k$-entanglement witnesses sampled by positive operators. To check the feasibility of this computing scheme, we develop a software tool, which implements the computation of the approximation of $E_w^{(k,n)}$  for systems  up to four qubits, with finite databases  containing $3\times 10^4$ samples of positive operators. Numerical tests show that the software
 is sufficiently accurate and reliable, and thus can be utilized to detect $k$-entanglement and describe the strength of $k$-entanglement efficiently. Four-qubit state evaluations typically take less than 200 seconds on the workstation used in this study, following database construction.}

\end{abstract}

\thanks{{\bf Keywords}: {multipartite systems, entanglement states, entanglement witnesses, $k$-nonseparability, $k$-entanglement measures}}
%Uncomment for PACS numbers title message
\thanks{{\bf PACS numbers}: {03.65.Ud, 03.65.Db, 03.67.Mn}}

\maketitle

\section{Introduction}\label{Sec1}

{ $k$-entanglement is an important quantum resource, which} provides a hierarchical description of how quantum correlations are distributed { across} the subsystems of a multipartite system. Experimental progress in trapped ions, ultracold atoms, superconducting circuits, quantum dots, and nitrogen-vacancy centers now enables increasingly precise control of multipartite systems \cite{GA}. These platforms support communication and teleportation protocols \cite{AP1,AP2,AP3}, quantum algorithms and error correction \cite{DB1,GCS,TB,RB,MS,CW,AR}, and quantum-enhanced metrology \cite{TG1,RL,KH1}; they also provide access to entanglement in quantum phase transitions and spin systems \cite{TR06,AM10,DB05}. The exponential growth of the state space and the presence of noise, however, make the rigorous characterization and quantification of $k$-entanglement a fundamental challenge \cite{LB,ZY,EM}.

{\bf $k$-separability and $k$-entanglement}. Multipartite entanglement has a richer structure than bipartite entanglement. Let ${\mathcal S}(H_1\otimes\cdots\otimes H_n)$ denote the set of states on $H=H_1\otimes\cdots\otimes H_n$, and let ${\mathcal Pur}(H_j)$ denote the pure states on $H_j$. A state $\rho$ is fully separable if
\begin{equation}
\rho = \sum_{r=1}^h p_r\, \rho_{1,r}\otimes\cdots\otimes\rho_{n,r},
\label{eq1.1}
\end{equation}
where $\rho_{j,r}\in{\mathcal Pur}(H_j)$, $p_r\geq0$, and {$\sum_r p_r=1$}, or if it is the trace-norm limit of such convex combinations in the infinite-dimensional case. For $n=2$, separability and full separability coincide; for $n>2$, { {multipartite systems can exhibit} different {separability or entanglement} {structures} {associated} {with} {different} {subsystem-groupings}. This leads to the concepts of  $k$-separability and $k$-entanglement. Recall that, for} $2\leq k\leq n$, a pure state $|\psi\rangle\in H$ is $k$-separable if it factorizes with respect to a $k$-partition $P=P_1|P_2|\cdots|P_k$ of $1|2|\cdots|n$:
\begin{equation*}
|\psi\rangle=|\psi_1\rangle|\psi_2\rangle\cdots|\psi_k\rangle,
\qquad |\psi_j\rangle\in H_{P_j},
\end{equation*}
where $H_{P_j}=\bigotimes_{s\in P_j}H_s$. A mixed state is $k$-separable if it is a convex combination of $k$-separable pure states, whose associated $k$-partitions need not be identical{, or if it is the trace-norm limit of such combinations in the infinite-dimensional case}. Otherwise, it is $k$-nonseparable, or equivalently $k$-entangled. {Within this $k$-entanglement framework}, multipartite entanglement (ME)  corresponds to $n$-entanglement, whereas genuine multipartite entanglement (GME) corresponds to $2$-entanglement.
The set ${\mathcal S}_k$ of $k$-separable states is closed and convex, and satisfies  ${\mathcal S}_k\subset{\mathcal S}_l$ whenever $2\leq l<k\leq n$. This nested structure makes multipartite entanglement intrinsically hierarchical. The index $k$ therefore provides a natural parameter for characterizing the hierarchy of $k$-entanglement beyond the binary distinction between separable and entangled states or the certification of GME alone.

 The detection and quantification of GME have been studied extensively because of its relevance to quantum metrology, measurement-based quantum computation, quantum phase transitions, spin chains, and biological processes \cite{VG04,HJB09,TR06,AM10,DB05,MS10,FC10}. Numerous GME criteria are available \cite{CS05,PK09,DL09,MH10,MH11}. GME measures and computable lower bounds have been proposed and refined \cite{ZHM11,YL22,ZXL23,XS23,MC23,YL23,YG24,JMG}. {Geometric} and concurrence-based approaches have {also} been developed {for} {tripartite} {entanglement} \cite{Ge,Dong,Wang}. More generally, a necessary-and-sufficient criterion for $k$-entanglement in arbitrary finite-dimensional $n$-partite systems was established in Ref.~\cite{GHQH}. Existing constructions include the $k$-entanglement concurrence \cite{ME2}, parameterized concurrence-based measures \cite{Li}, the $k$-geometric mean concurrence \cite{LiG}, and $q$-concurrence-based measures \cite{Ma}.

{\bf Three challenges.} Despite great progress in the study of $k$-entanglement, three basic theoretical bottlenecks remain in the quantitative study of $k$-entanglement: {

(i) The absence of an axiomatic framework for quantifying the $k$-entanglement that reflects the basic requirements from many-body resource theory. The current axiomatic framework for $k$-entanglement measures is the same as that for bipartite systems with three basic rules: nonnegativity and faithfulness, nonincreasing under LOCCs, invariance under permutation of subsystems. However, such an axiomatic framework does not work well for $n$-partite systems if $n\geq 3$, for instance, it
 does not regulate how to quantify the entanglement content assigned to part systems of the same multipartite state and how to describe the rule that the entanglement content possed by part system never exceed the entanglement content  possed by whole system;

(ii) the lack of a $k$-entanglement measure satisfying the above requirements from  many-body resource theory;
and

(iii) the extreme  difficulty in numerical evaluation at mixed states of existing $k$-entanglement measures, especially those quantifications constructed using convex-roof extension techniques
 which makes  evaluation at mixed and high-dimensional states computationally impossible.

{\bf A solution.} The present work makes three efforts to address the corresponding issues.

First, based on the notion of sub-repartitions, we define a partial order relation on pairs $(k,P)$ consisting of an entanglement index $k$ and a part system $P$. This allows us to establish an axiomatic framework for \emph{true} $k$-entanglement measures with five requirements. In addition to usual faithfulness, LOCC monotonicity and permutation symmetry, two new rules, called  unification condition and  hierarchy condition are introduced in the new axiomatic framework. The unification regulates how to measure  the entanglement content in part  systems, and the hierarchy condition
regulates the meaning of ``the entanglement content in part  systems can never exceed that in the whole system".

Second, for every $2\leq k\leq n$, we construct an explicit witness-based quantification  $E_w^{(k,n)}$ of $k$-entanglement for any finite- and infinite-dimensional $n$-partite systems without convex-roof extension. We prove that $E_w^{(k,n)}$ satisfies all five rules of the new axiomatic framework, as well as is convex and subadditive. Thus $E_w^{(k,n)}$ is a {\it true} $k$-entanglement measure, distinguishing from current $k$-entanglement measures, which confirms that
 the $k$-entanglement, including GME and ME, is a many-body quantum resource within the proposed framework.

 Finally, for finite dimensional systems,  we derive a method of constructing  $k$-entanglement witnesses from positive operator and $k$-separable-state optimizations, which are performed once to build a reusable witness database, and then
 design  a evaluation scheme of $E_w^{(k,n)}$ based on pre-established database of  $k$-entanglement witnesses sampled by positive operators. To check the feasibility of this computing scheme, we develop a software tool, which implements the computation of the approximation of $E_w^{(k,n)}$  for systems  up to four qubits, with finite-databases  containing $3\times 10^4$ samples of positive operators \cite{Software}. Numerical tests show that the software
 is sufficiently accurate and reliable as it recovers many known $k$-entanglement criteria,
 and thus can be utilized to improve known results, find new $k$-entanglement states, and detect $k$-entanglement as well as the strength of $k$-entanglement in any $n$-qubit states with $n\leq 4$ efficiently. Four-qubit state evaluations typically take less than 200 seconds on the workstation used in this study, following database construction.}
 \if false
 we reduce the evaluation of $E_w^{(k,n)}$ to product-state optimizations over finitely many inequivalent partitions. These optimizations are performed once to build a reusable witness database; subsequent state evaluations require only trace calculations and maximizations. The implementation covers two-, three-, and four-qubit systems and admissible coarser groupings \cite{Software}. It reproduces selected analytical thresholds, compares the additional positive-output intervals for noisy four-qubit W, Dicke, and singlet states with published sufficient GME criteria, and evaluates a known PPT-entangled state. Section~\ref{SecRepresentativeExamples} distinguishes these numerical onsets from established separability boundaries.\fi

{ The remainder of the paper follows these three clues. Section~\ref{Sec2}, as a warm-up,  presents a discussion of the case ME, that is, the case $k=n$, constructs  $E_w^{(n)}=E_w^{(n,n)}$ and shows how it meets the many-body resource theory.  Section~\ref{Sec3} treats the general case of $2\leq k\leq n$, proposes an axiomatic framework, constructs $E_w^{(k,n)}$ and proves that it is a true $k$-entanglement measure. Section~\ref{Sec4} derives an evaluation scheme for finite-dimensional systems {and provides a method of accuracy control}. Section~\ref{Sec5} provides an   implementation of the evaluation scheme for systems of up to four qubits. Section~\ref{Sec6} reports the numerical tests. Technical derivations, explicit qubit formulas, supplementary figures, proofs, and algorithms are kept in the Appendix; the software is available at Ref.~\cite{Software}.}

\section{A multipartite entanglement measure $E_w^{(n)}$}\label{Sec2}

{We first consider the case $k=n$, for which $k$-entanglement reduces to ME in an $n$-partite system. In this special case, an axiomatic framework that obeys many-body resource theory was already proposed in Refs.~\cite{GY20,HLQ22} by adding unification and hierarchy condition to the standard axioms.  We call a quantification $E^{(n)}$  a {\it true} ME measure if it satisfies all mentioned axioms. The purpose of this section is a warm-up. We construct a witness-based quantification $E_w^{(n)}$ of ME
for any $n$-partite systems and show that it is a true ME measure. This provides a clue for how to deal with quantifications of general $k$-entanglement.}\if false {The definition and relevant properties are presented below, with auxiliary characterizations, examples, and proofs in the Appendix.} Besides the standard axioms, $E_w^{(n)}$ satisfies the unification and hierarchy requirements of many-body resource theory.\fi

Denote by $\mathcal B(H)$ the set of all bounded linear operators acting on Hilbert space
$H=H_1\otimes H_2\otimes\cdots\otimes H_n.$
Recall that a self-adjoint operator $W\in{\mathcal B}(H)$ is a multipartite entanglement witness (MEW) if $W$ is not positive and
\begin{equation*}
{\rm Tr}\bigl[W(\eta_1\otimes\eta_2\otimes\cdots\otimes\eta_n)\bigr]\geq0
\end{equation*}
for all $\eta_j\in{\mathcal Pur}(H_j)$, $j=1,2,\ldots,n$. Then, a state $\rho\in{\mathcal S}(H)$ is multipartite entangled  if and only if there exists an MEW of the form
\begin{equation*}
W=\lambda I-L,\qquad L\in {\mathcal B}^+_1(H),\qquad {\rm Tr}(W\rho)<0,
\end{equation*}
where ${\mathcal B}^+_1(H)=\{L\in\mathcal B(H):0\leq L\leq I\}$ \cite{GHQH}.
Let
\begin{equation}
\mathcal{EW}_1 = \mathcal{EW}_1(H_1 \otimes H_2 \otimes \cdots \otimes H_n)
\label{eq2.1}
\end{equation}
be the set of all operators of the form $\lambda I-L$ with $L\in {\mathcal B}_1^+(H)$ and {$0\leq\lambda\leq \|L\|$} such that either $\lambda I-L$ is an MEW or $\lambda I-L\geq 0$. For $L\in{\mathcal B}^+(H)$, let
\begin{equation}
\begin{aligned}
g(L) = \sup\{ & {\rm Tr}((\eta_1\otimes \eta_2\otimes\cdots\otimes \eta_n)L) : \eta_j\in{\mathcal Pur}(H_j), \\
& j=1,2,\ldots, n\}.
\end{aligned}
\label{eq2.2}
\end{equation}
One readily verifies that
\begin{equation}
\begin{aligned}
&g(I)= 1, \quad g(\alpha L) = \alpha g(L),\\
&g(D)\leq g(L) \text{ if } D \leq L, \\
&g(L_1+L_2)\leq g(L_1) + g(L_2) \quad \text{and}\\
&g(L\otimes M)= g(L) g(M)
\end{aligned}
\label{eq2.3}
\end{equation}
whenever $\alpha>0$ and $D,L,L_1,L_2, M$ are positive operators. \if false {, with the tensor-product identity understood on the combined list of parties of the two systems}.\fi Further details concerning the normalization of the admissible witnesses and the equivalent separable-state characterization of $g(L)$ are given in Appendix~\ref{SecWarmupDetails}.

We define a witness-based quantification $E^{(n)}_w$ by
\begin{equation}
\begin{aligned}
E^{(n)}_w(\rho)  = &  \sup_{W\in\mathcal{EW}_1} |\min\{{\rm Tr}(W\rho),0\}| \\  = & \sup_{W\in\mathcal{EW}_1} \max\{-{\rm Tr}(W\rho),0\}.
\end{aligned}
\label{eq2.4}
\end{equation}
The witness characterization above immediately yields faithfulness and permutation symmetry.

\begin{theorem}\label{T2.1} {\rm (Faithfulness and symmetry)} {\it  For every state $\rho$, we have $0\leq E^{(n)}_w(\rho)\leq 1$, and $E^{(n)}_w(\rho)=0$ if and only if $\rho$ is fully separable. Moreover, for every permutation $\pi$ of the subsystems, $E^{(n)}_w(\rho^\pi)=E^{(n)}_w(\rho)$, where $\rho^\pi$ is obtained from $\rho$ by changing the order of subsystems according to a permutation $\pi$ of $(1,2,\ldots,n)$.}
\end{theorem}

\begin{theorem}\label{T2.2}{\rm (Monotonicity under LOCC)} {\it For any LOCC operation $\Phi$ and any $n$-partite state $\rho$, we have
\begin{equation*}
E^{(n)}_w(\Phi(\rho))\leq E^{(n)}_w(\rho).
\end{equation*}}
\end{theorem}

 Theorems~\ref{T2.1} and \ref{T2.2} show that $E^{(n)}_w$ is an entanglement measure in the standard sense. It is also convex.

\begin{theorem}\label{T2.3} {\rm (Convexity)} {\it If $\rho=\sum_{j=1}^k p_j\rho_j$, then
\begin{equation*}
E^{(n)}_w(\rho)\leq \sum_{j=1}^kp_jE^{(n)}_w(\rho_j).
\end{equation*}}
\end{theorem}

For brevity, write ${\mathcal B}_1^+={\mathcal B}_1^+(H)$. Then
\begin{equation}
E_w^{(n)}(\rho)=\sup_{L\in{\mathcal B}_1^+}\ \sup_{g(L)\leq \lambda\leq\|L\|}
\max\{0,{\rm Tr}(L\rho)-\lambda\}.
\label{eq2.5}
\end{equation}
Optimizing over $\lambda$ gives a direct expression in terms of $g(L)$.
\begin{theorem}\label{T2.4} {\it For every $n$-partite state $\rho\in {\mathcal S}(H_1\otimes H_2\otimes\cdots\otimes H_n)$,}
\begin{equation}
E^{(n)}_w(\rho)= \sup_{L\in\mathcal{B}_1^+} \max\{0,{\rm Tr}(L\rho)-g(L)\}.
\label{eq2.6}
\end{equation}
\end{theorem}

{Subadditivity follows from the argument in Appendix~\ref{SecAA4}.}

\begin{theorem}\label{T2.5} {\rm (Subadditivity)} {\it If $\rho_1\in{\mathcal S}(H_1\otimes H_2\otimes \cdots\otimes H_n)$ and $\rho_2\in {\mathcal S}(K_1\otimes K_2\otimes\cdots\otimes K_m)$, then}
\begin{equation*}
E_w^{(n+m)}(\rho_1\otimes\rho_2)\leq E_w^{(n)}(\rho_1)+E_w^{(m)}(\rho_2).
\end{equation*}
\end{theorem}

\subsection{Sub-repartitions and the multipartite hierarchy}

The consistency of a multipartite entanglement measure across different descriptions of a many-body system is encoded by sub-repartitions and their hierarchy relation. This is the structural requirement that supplements faithfulness, symmetry, and LOCC monotonicity in the definition of a \emph{true} multipartite entanglement measure \cite{GY20,HLQ22}.

For $H=H_1\otimes\cdots\otimes H_n$, a sub-repartition $P=P_1|\cdots|P_m$ of $1|\cdots|n$ consists of nonempty, pairwise disjoint subsets $P_j\subseteq\{1,\ldots,n\}$. It specifies the { part system} and its reduced state,
\begin{equation*}
H_P=H_{P_1}\otimes\cdots\otimes H_{P_m},\qquad
\rho_P={\rm Tr}_{P^c}(\rho).
\end{equation*}
{ When $\bigcup_{j=1}^mP_j=\{1,\ldots,n\}$, $H_P$  reduces to an $m$-partition of the full system. The unification condition proposed in \cite{GY20,HLQ22} for a quantification of ME requires that considering $E^{(n)}$ alone is not enough, instead, we have to consider a sequence $\{E^{(m)}\}$ for every $m$ with the same construction to $E^{(n)}$ such that one can quantify entanglement of a state $\rho\in\mathcal S(H)$ in any part system $H_P$ as $E^{(m)}(\rho_P,P)$, by regarding $\rho_P$ as an $m$-partite state in $H_P$. It is clear that $E^{(n)}_w$ defined in Eq.(\ref{eq2.4}) satisfies the unification condition.}

Let $\mathcal{SP}^n$ be the set of all sub-repartitions of $1|\cdots|n$. For $P=P_1|\cdots|P_m$ and $Q=Q_1|\cdots|Q_r$ in $\mathcal{SP}^n$, three elementary relations describe how a lower {part system} can be obtained from a higher one. { We write $Q\preccurlyeq^a P$ if every block of $Q$ is a block of $P$, which means that $Q$ is obtained by  discarding some complete parties from $P$. We write $Q\preccurlyeq^b P$ if every block of $Q$ is a union of blocks of $P$ and $\bigcup_iQ_i=\bigcup_jP_j$, which means that $Q$ is obtained by grouping parties of $P$. Finally, $Q\preccurlyeq^c P$ if $r=m$ and every $Q_i$ is a nonempty subset of a distinct block of $P$, which means that  $Q$ is obtained by tracing out subsystems within the blocks of $P$. Illustrative examples are given in Appendix~\ref{SecWarmupDetails}. Their transitive closure defines a hierarchy relation $\preccurlyeq$ in $\mathcal{SP}^n$, which is a  partial order.}

\begin{definition}\label{D2.6} For $P,Q\in\mathcal{SP}^n$, we say that $Q$ has lower hierarchy than $P$, and write $Q\preccurlyeq P$, if there exist $R_1,R_2,\ldots,R_t\in\mathcal{SP}^n$ such that
\begin{equation*}
Q\preccurlyeq^{x_1}R_1\preccurlyeq^{x_2}R_2\preccurlyeq^{x_3}\cdots\preccurlyeq^{x_t}R_t\preccurlyeq^{x_{t+1}}P,
\end{equation*}
where $x_1,x_2,\ldots,x_t,x_{t+1}\in\{a,b,c\}$.
\end{definition}
\if false The relation $\preccurlyeq$ defines a partial order {after} {sub-repartitions that differ only by block permutations are identified}.\fi With this notation, the hierarchy condition can be stated compactly.

{ If $E^{(n)}$ is a quantification of ME for $n$-partite systems satisfying the unification condition, the hierarchy condition for $E^{(n)}$ is stated as, for any $Q=Q_1|Q_2\ldots|Q_r, P=P_1|P_2|\ldots|P_m\in\mathcal{SP}_n$ and any $\rho\in\mathcal S(H_1\otimes H_2\otimes\cdots\otimes H_n)$, $E^{(r)}(\rho_Q,Q)\leq E^{(m)}(\rho_P,P)$ whenever $Q\preccurlyeq P$. The following conclusion confirms that $E_w^{(n)}$ also obeys the hierarchy condition.}

\begin{theorem}\label{T2.7}  {\rm (Hierarchy condition)} {\it Consider an $n$-partite system $H=H_1\otimes H_2\otimes\cdots\otimes H_n$ with $n\geq 2$. For any sub-repartition $P=P_1|P_2|\ldots|P_m$ and $Q=Q_1|Q_2|\ldots |Q_r$ of $1|2|\ldots|n$, {with $2\leq r\leq m$,} if $Q\preccurlyeq P$, then
\begin{equation*}
E_w^{(r)}(\rho_Q,Q)\leq E_w^{(m)}(\rho_P,P)
\end{equation*}
holds for all states $\rho\in{\mathcal S}(H_1\otimes H_2\otimes\cdots\otimes H_n)$.}
\end{theorem}

Combining theorems \ref{T2.1}, \ref{T2.2} and \ref{T2.7}, we obtain the following

\begin{theorem}\label{T2.8} {\it  For every finite- or infinite-dimensional $n$-partite system $H_1\otimes H_2\otimes\cdots\otimes H_n$, $E^{(n)}_w$ is a true multipartite entanglement measure. Consequently, with fully separable states as free states and $n$-LOCCs as free operations, the multipartite entanglement is a many-body quantum resource.}
\end{theorem}

\section{Axiomatic framework and a true $k$-entanglement measure $E_w^{(k,n)}$}\label{Sec3}

We now extend the construction to general $k$-entanglement. The physical issue is that the same many-body state may be viewed at different resolutions by tracing out subsystems or grouping them into larger parties. A consistent resource measure must therefore compare not only different values of $k$ on a fixed system, but also the entanglement assigned to the resulting {part systems}. We first formulate these consistency requirements and then construct a witness-based measure satisfying them. In the standard formulation{~\cite{GY24}}, a $k$-entanglement measure $E^{(k,n)}$ is a nonnegative functional on ${\mathcal S}(H_1\otimes H_2\otimes\cdots\otimes H_n)$ satisfying the following three conditions.

($k$-EM1) (Faithfulness)
$E^{(k,n)}(\rho)=0$
if and only if $\rho$ is $k$-separable.

($k$-EM2) (Monotonicity under LOCC) For every $n$-partite LOCC operation $\Phi$ and every state $\rho$,
\begin{equation*}
E^{(k,n)}(\Phi(\rho))\leq E^{(k,n)}(\rho).
\end{equation*}

($k$-EM3) (Symmetry) $E^{(k,n)}$ is invariant under permutations of the subsystems.

{ As mentioned before, this axiomatic framework with the conditions ($k$-EM1)--($k$-EM3) does not characterize $k$-entanglement measures as a multipartite quantum  correlation since it does not regulate how to get the resource assigned to {part systems} and how to describe the requirement as a many-body resource that the resource possed by part system cannot exceed that possed by whole. Like the special case $k=n$ discussed in Section \ref{Sec2}, these lead to  the proposal of unification condition and hierarchy condition.}

\subsection{Unification and  hierarchy condition for quantifications of $k$-entanglement }

{ Let $E^{(k,n)}$ be a standard $k$-entanglement measure, that is, a quantification of $k$-entanglement for $n$-partite systems $H_1\otimes H_2\otimes\cdots\otimes H_n$ satisfying  ($k$-EM1)--($k$-EM3).
The unification condition is established as follows which defines how to get the strength of $s$-entanglement in $m$-partite part systems.}

($k$-EM4) (Unification condition) {A sequence
\begin{equation*}
\{E^{(s,m)}:2\leq s\leq\min\{k,m\},\ m=2,3,\ldots,n\}
\end{equation*}
of $s$-entanglement measures for any $m$-partite systems can be defined in the same way as $E^{(k,n)}$. For any sub-repartition $P=P_1|P_2|\cdots|P_m\in{\mathcal {SP}}^n$ and any state
$\rho\in{\mathcal S}(H_1\otimes H_2\otimes\cdots\otimes H_n),$
the reduced state $\rho_P$ of $\rho$ is regarded as an $m$-partite state on the part system
$H_P=H_{P_1}\otimes H_{P_2}\otimes\cdots\otimes H_{P_m},$
and $E^{(s,m)}(\rho_P,P)$ quantifies the $s$-entanglement of $\rho$ contained in the part system $H_P$.

{Note that, even if $Q\preccurlyeq P$ and $k\leq\{r,m\}$, we can not regard the $k$-entanglement in $\rho_Q$ with respect to $H_Q$ as a part of the $k$-entanglement in $\rho_P$ associated with $H_P$, because, by the definition of $k$-entanglement, the $k$-entanglement in $H_Q$ involves all $k$-partitions of $Q$ while the $k$-entanglement in $H_P$ involves all $k$-partitions of $P$. Hence, we can not simply say ``the $k$-entanglement possessed by a part system can not exceed the $k$-entanglement possessed in the entire system''. This is reasonable, for instance, consider the case of ME discussed in Section \ref{Sec2}, the hierarchy condition exactly says that ``the $r$-entanglement held in part system $H_Q$ does not exceed the $m$-entanglement held in part system $H_P$ if $Q\preccurlyeq P$''. So the reasonable statement of the hierarchy condition for $k$-entanglement should be ``the $s$-entanglement held in $H_Q$ does not larger than the $t$-entanglement held in $H_P$ for some $s,t$ if $Q\preccurlyeq P$''. In this case, we say that the hierarchy of $(s,Q)$ is lower than that of $(t,P)$.

Thus, to} describe how the part system possesses fewer resources, i.e., the hierarchy condition, we need first introduce a partial order in $\mathbb N\times \mathcal{SP}_n$. This involves a bit more technique than the case $k = n$ in Section \ref{Sec2}.

Let $Q=Q_1|\ldots|Q_r$ be associated with  $s$-entanglement and $P=P_1|\ldots|P_m$ with $t$-entanglement, denoted by $(s,Q)$ and $(t,P)$ respectively. A relation $(s,Q)\preccurlyeq'(t,P)$, defined below, expresses that $(s,Q)$ is lower in the hierarchy than $(t,P)$. Then, roughly speaking, the hierarchy condition means that the strength of $s$-entanglement assigned to the lower pair $(s,Q)$ does not   exceed the strength of $t$-entanglement assigned to the higher pair $(t,P)$. }

  Analogously to the partial order $\preccurlyeq$ on $\mathcal{SP}^n$ introduced in Section~\ref{Sec2}, we define three elementary relations for pairs consisting of an entanglement index and a sub-repartition. For a sub-repartition $P=P_1|\cdots|P_m$, let
$\mathcal P_t(P)$ denote the set of all $t$-partitions of $P$, namely, all partitions $R=R_1|\ldots|R_t$ obtained by grouping
the blocks of $P$. Similarly, $\mathcal P_s(Q)$ denotes the set of all $s$-partitions of $Q$.

\begin{definition}[Hierarchy relation for $k$-entanglement]\label{D3.1}
Let  $P=P_1|\cdots|P_m,Q=Q_1|\ldots|Q_r $ belong to $\mathcal{SP}^n$,  $2\le s\le r,  2\le t\le m. $

{\rm (a$'$)} $(s,Q)\preccurlyeq^{a'}(t,P)$ if $Q\preccurlyeq^a P$ and  $t-s\ge m-r.$

{\rm (b$'$)} $(s,Q)\preccurlyeq^{b'}(t,P)$ if $s\le t$, $Q\preccurlyeq^b P$, and, for every
$R\in\mathcal P_t(P)$, there exists $S\in\mathcal P_s(Q)$ such that $R$ is a refinement of $S$, equivalently,
$S\preccurlyeq^b R.$

{\rm (c$'$)} $(s,Q)\preccurlyeq^{c'}(t,P)$ if $ r=m, s\le t, Q\preccurlyeq^c P.$

We say that $(s,Q)$ has lower hierarchy than $(t,P)$, and write $(s,Q)\preccurlyeq'(t,P),$ if there exist sub-repartitions $R_1,\ldots,R_\alpha\in\mathcal{SP}^n$ and integers $k_1,\ldots,k_\alpha$, where $2\le k_j\le m_j$ whenever $R_j$ consists of $m_j$ parties, such that
\begin{equation}
\begin{split}
(s,Q)
&\preccurlyeq^{x_1}(k_1,R_1)
\preccurlyeq^{x_2}(k_2,R_2)
\preccurlyeq\cdots\\
&\preccurlyeq^{x_\alpha}(k_\alpha,R_\alpha)
\preccurlyeq^{x_{\alpha+1}}(t,P),
\end{split}
\label{eq:pair-hierarchy}
\end{equation}
where $x_1,\ldots,x_{\alpha+1}\in\{a',b',c'\}.$ Equivalently, $\preccurlyeq'$ is the transitive closure of $\preccurlyeq^{a'}$, $\preccurlyeq^{b'}$, and $\preccurlyeq^{c'}$.
\end{definition}

Generally,  $Q\preccurlyeq$ does not imply $(k,Q)\npreceq'(k,P)$ even if $k\leq \min\{r,m\}$ when $r<m$,

The additional conditions in $(a')$ and $(b')$ ensure the compatibility between the $t$-partition structures associated with $(t,P)$ and the $s$-partition structures associated with $(s,Q)$. This compatibility is required in order to compare consistently the $s$-entanglement contained in the part system $Q$ with the $t$-entanglement contained in the part system $P$.

For $Q\preccurlyeq^a P$, the sub-repartition $Q$ is obtained from $P$ by discarding exactly $m-r$ complete parties. Consider an arbitrary $t$-partition of $P$. After the discarding operation, at most $m-r$ of its blocks can disappear completely. Hence the induced partition on $Q$ contains at least $t-(m-r)$ nonempty blocks. Thus the condition $t-s\ge m-r$  is exactly the condition ensuring that every $t$-partition of $P$, after restriction to the remaining part system $Q$, contains sufficiently many nonempty blocks to be coarsened
to an $s$-partition of $Q$.

For $Q\preccurlyeq^b P$, no subsystem is discarded, but the blocks of $P$ are regrouped to form the blocks of $Q$. Compatibility therefore cannot in general be characterized merely by counting blocks. We require instead that every $t$-partition $R$ of $P$ refine some  $s$-partition $S$ of $Q$. Equivalently, $S\preccurlyeq^b R$. This condition guarantees that every $t$-partition structure relevant to
$(t,P)$ is compatible with an $s$-partition structure relevant to $(s,Q)$.

For $Q\preccurlyeq^c P$, the number of parties is preserved and, after a suitable ordering, each $Q_i$ is contained in the corresponding $P_i$. Hence every $t$-partition of $P$ naturally induces a $t$-partition of $Q$. Since $s\le t$, the induced $t$-partition can always be coarsened to an $s$-partition of $Q$. Thus no additional compatibility condition is needed in case $(c')$.

For ME in the two part systems, the entanglement indices coincide with the corresponding numbers of parties, namely, $s=r$ and $t=m$. In case $(a')$, the condition $t-s\ge m-r$ is then automatically satisfied whenever $Q\preccurlyeq^aP$. In case $(b')$, the only $m$-partition of $P$ is $P$ itself and the only $r$-partition of $Q$ is $Q$, so the refinement condition reduces to $Q\preccurlyeq^bP$. Case $(c')$ already reduces directly to $Q\preccurlyeq^cP$. Thus, for ME, $(r,Q)\preccurlyeq' (m,P)$ if and only if $Q\preccurlyeq P$. So, the hierarchy relation $\preccurlyeq'$ is a generation of the hierarchy relation $\preccurlyeq$ for ME in Section~\ref{Sec2} to $k$-entanglement.
Some illustrative examples and further details are given in Appendix~\ref{SecHierarchyDetails}.

{ Now the hierarchy condition for $k$-entanglement can be presented as

($k$-EM5) (Hierarchy condition) Let $E^{(k,n)}$ be a quantification of $k$-entanglement for $n$-partite system $H_1\otimes H_2\otimes\cdots\otimes H_n$ satisfying ($k$-EM4). For any sub-repartitions $P=P_1|P_2|\ldots|P_m$ and $Q=Q_1|Q_2|\ldots|Q_r\in \mathcal{SP}^n$ with $2\leq m,r\leq n$, the hierarchy relation $(s,Q)\preccurlyeq'(t,P)$ implies
\begin{equation*}
E^{(s,r)}(\rho_Q,Q)\leq E^{(t,m)}(\rho_P,P)
\end{equation*}
holds for all $n$-partite state
$\rho\in{\mathcal S}(H_1\otimes H_2\otimes\cdots\otimes H_n)$.

\begin{definition}\label{D3.2}
A quantification $E^{(k,n)}$ of $k$-entanglement is called a \emph{true} $k$-entanglement measure if it satisfies the rules ($k$-EM1)--($k$-EM5).
\end{definition}

\if false For the special case $k=n$, let $Q=Q_1|Q_2|\ldots|Q_r$ and $P=P_1|P_2|\ldots|P_m$. Then $s=r$ and $k=m$, and
the relation $(r,Q)\preccurlyeq'(m,P)$ holds if and only if
$Q\preccurlyeq P$.  Thus,\fi  When $k=n$, the unification and hierarchy conditions for $k$-entanglement reduce to that for ME described in Section~\ref{Sec2} (see also Ref.~\cite{GY20}). Hence a true $n$-entanglement measure $E^{(n)}=E^{(n,n)}$ is precisely a true multipartite entanglement measure.}

\subsection{ A true $k$-entanglement measure constructed based on witnesses}

{ Motivated by the definition of $E_w^{(n)}$ in Eq. (\ref{eq2.4}), we construct a true $k$-entanglement measure for arbitrary $2\leq k\leq n$. The construction is witness-based, avoids convex-roof extensions, and leads to a numerical evaluation scheme in Section~\ref{Sec4}.}

Following Refs.~\cite{GT,HGY}, a self-adjoint operator $W\in\mathcal{B}(H_1\otimes H_2\otimes\cdots\otimes H_n)$ is a $k$-entanglement witness ($k$-EW) if $\langle\psi|W|\psi\rangle\geq0$ for every $k$-separable pure state $|\psi\rangle$ and $W$ is not positive semidefinite.  The $k$-EW criterion asserts that a state $\rho$ is $k$-entangled {if and only if ${\rm Tr}(W\rho)<0$ for some $k$-EW $W$. It was shown in \cite{GHQH} that $\rho$ is $k$-entangled if and only if there exists $k$-EW of the form $\lambda I-L$ such that
${\rm Tr}((\lambda I-L)\rho)<0$, where $L$ is a positive operator  and $\lambda$ is a scalar with $0<\lambda\leq \|L\|$.
For a positive operator $L$ and  a scalar $\lambda>0$, the smallest admissible value of $\lambda$ such that ${\rm Tr}((\lambda I-L)\sigma)\geq 0$ for all $k$-separable state $\sigma$ is}
\begin{equation}
\begin{aligned}
\lambda_0 &= g_n^{(k)}(L) \\
&= \sup\bigl\{ \langle\phi|L|\phi\rangle : |\phi\rangle \in H_1\otimes H_2\otimes\cdots\otimes H_n, \\
&\qquad |\phi\rangle \text{ is a } k\text{-separable pure vector state}\bigr\}.
\end{aligned}
\label{eq3.3}
\end{equation}

Let ${\mathcal P}_n^k$ be the set of all $k$-partitions of $1|2|\ldots|n$. A pure state is $k$-separable exactly when it is a product vector with respect to at least one partition $P=P_1|P_2|\ldots|P_k\in{\mathcal P}_n^k$. Denote by ${\mathcal Pur}_n^k(P)$ the product vectors associated with $P$, and by ${\mathcal Pur}_n^k$ the set of all $k$-separable pure vectors. Then
\begin{equation}
{\mathcal Pur}_n^k = \bigcup_{P\in\mathcal{P}_n^k} {\mathcal Pur}_n^k(P).
\label{eq3.4}
\end{equation}
It follows from Eq.~(\ref{eq3.3}) that
\begin{equation}
g_n^{(k)}(L) = \max_{P\in\mathcal{P}_n^k} g(L,P),
\label{eq3.5}
\end{equation}
where, with $k$-partition  $P=P_1|P_2|\ldots|P_k$,
\begin{equation}
\begin{aligned}
& g(L,P)\\ =&\sup\bigl\{
\langle \phi_{P_1}\cdots\phi_{P_k}|L|
\phi_{P_1}\cdots\phi_{P_k}\rangle :
|\phi_{P_j}\rangle\in H_{P_j},\\
& \langle\phi_{P_j}|\phi_{P_j}\rangle=1,\quad
j=1,\ldots,k
\bigr\}.
\end{aligned}
\label{eq3.6}
\end{equation}

Hence the evaluation of $g_n^{(k)}(L)$ reduces to evaluating the multipartite quantity $g(L,P)$ for each inequivalent $k$-partition, using the same optimization as in Eq.~(\ref{eq2.2}).

Denote $
{\mathcal B}_1^+=\{L\in{\mathcal B}(H_1\otimes\cdots\otimes H_n):L\geq0,\ \|L\|\leq1\}$
and
${\mathcal {EW}}_1^{(k)}=\{\lambda I-L:L\in\mathcal B_1^+,\ g_n^{(k)}(L)\leq\lambda\leq\|L\|\}.
$ Every $W\in{\mathcal {EW}}_1^{(k)}$ is either a $k$-EW or a positive operator. For $\rho\in\mathcal{S}(H_1\otimes\cdots\otimes H_n)$, set
\begin{equation}
\begin{aligned}
E_w^{(k,n)}(\rho) &= \sup_{W\in\mathcal{EW}_1^{(k)}} \bigl|\min\{{\rm Tr}(W\rho),0\}\bigr| \\
&= \sup_{\substack{L\in \mathcal{B}_1^+ \\ g_n^{(k)}(L)\leq \lambda\leq \|L\|}} \max\{ {\rm Tr}(L\rho)-\lambda, 0\}.
\end{aligned}
\label{eq3.7}
\end{equation}

\if false The same witness arguments used in Section~\ref{Sec2} give the following properties.\fi

\begin{theorem}\label{T3.3} For any $\rho\in\mathcal{S}(H_1\otimes H_2\otimes\cdots\otimes H_n)$ and $2\leq k\leq n$, the following statements are true.

{\rm (1) (Faithfulness)} {\it $0\leq E_w^{(k,n)}(\rho)\leq 1$ and $\rho$ is $k$-separable if and only if $E_w^{(k,n)}(\rho)=0$. }

{\rm (2) (Monotonicity under LOCC)} {\it For any $n$-partite LOCC operation $\Phi$ and state $\rho\in {\mathcal S}(H_1\otimes H_2\otimes\cdots\otimes H_n)$, one has
\begin{equation*}
E_w^{(k,n)}(\Phi(\rho)) \leq E_w^{(k,n)}(\rho).
\end{equation*}
}

{\rm (3) (Symmetry)} {\it $E_w^{(k,n)}$ is invariant under permutations of the subsystems.}

{\rm (4) (Convexity)} {\it If $\rho=\sum_{j=1}^m p_j\rho_j$, then
\begin{equation*}
E^{(k,n)}_w(\rho)\leq \sum_{j=1}^mp_jE^{(k,n)}_w(\rho_j).
\end{equation*}}

\end{theorem}

Thus, $E_w^{(k,n)}$, satisfying ($k$-EW1)--($k$-EW3), is a $k$-entanglement measure in the standard sense \cite{GY24}.

Optimizing Eq.~(\ref{eq3.7}) over $\lambda$, as in Theorem~\ref{T2.4}, gives
\begin{equation}
E_w^{(k,n)}(\rho) = \sup_{L\in \mathcal{B}_1^+} \max\{ 0, {\rm Tr}(L\rho) - g_n^{(k)}(L) \}.
\label{eq3.8}
\end{equation}
\begin{theorem}\label{T3.4} {\rm (Subadditivity)} {\it If $\rho_1\in{\mathcal S}(H_1\otimes H_2\otimes \cdots\otimes H_n)$ and $\rho_2\in {\mathcal S}(K_1\otimes K_2\otimes\cdots\otimes K_m)$, then, for $2\leq k_1\leq n$ and $2\leq k_2\leq m$,}
\begin{equation*}
E_w^{(k_1+k_2,n+m)}(\rho_1\otimes\rho_2)\leq E_w^{(k_1,n)}(\rho_1)+E_w^{(k_2,m)}(\rho_2).
\end{equation*}
\end{theorem}

Because the same construction defines $E_w^{(k,n)}$ for every admissible pair $(k,n)$, the unification condition ($k$-EM4) follows directly.

{The  following result is crucial, non-trivial and technical.}

\begin{theorem}[Hierarchy condition]\label{T3.5}
Let $P=P_1|\ldots|P_m,$ $ Q=Q_1|\ldots|Q_r$ be sub-repartitions in $\mathcal{SP}^n$, with
$2\le s\le r,$ $ 2\le t\le m.$ For any state $\rho\in\mathcal{S}(H_1\otimes\cdots\otimes H_n),$
let \[
\rho_P=\operatorname{Tr}_{P^c}(\rho),
\qquad
\rho_Q=\operatorname{Tr}_{Q^c}(\rho)
\]
be regarded as an $m$-partite state on $H_P$ and an
$r$-partite state on $H_Q$, respectively. If $(s,Q)\preccurlyeq'(t,P),$
then
$$E_w^{(s,r)}(\rho_Q,Q)
\le
E_w^{(t,m)}(\rho_P,P).
$$
\end{theorem}

The proofs of Theorems~\ref{T3.4} and \ref{T3.5} are given in Appendix~\ref{SecAB}. Combining Theorems~\ref{T3.3} and \ref{T3.5} yields the main  result in this section.

\begin{theorem}\label{T3.6} {\it For every $n$-partite system and every $2\leq k\leq n$, $E_w^{(k,n)}$ defined in Eq.(\ref{eq3.7}) is a true $k$-entanglement measure. Consequently, $k$-entanglement is a many-body quantum resource with the $k$-separable states as free states and $n$-partite LOCCs as free operations. In particular, both ME and GME are many-body quantum resources.}
\end{theorem}

\section{Evaluation of $E_w^{(k,n)}$}
\label{Sec4}

The true $k$-entanglement measure $E_w^{(k,n)}$
can be evaluated numerically in finite-dimensional systems.
According to Eq.~(\ref{eq3.8}), its evaluation reduces to
computing
$g_n^{(k)}(L)$ for positive operators
$L\in{\mathcal B}_1^+$. By Eq.~(\ref{eq3.5}),
$g_n^{(k)}(L)$ is obtained by maximizing $g(L,P)$ over
$k$-partitions $P$. For a fixed partition $P$, the quantity
$g(L,P)$ is determined by the product-state optimization in
Eq.~(\ref{eq3.6}), which was established in
Ref.~\cite{SV3} and further discussed in Ref.~\cite{GHQH}.
Accordingly, the numerical evaluation consists of two parts:
the values of $g(L,P)$ are first computed and stored for a
finite set of positive operators $L$, and the resulting
database is then used to evaluate $E_w^{(k,n)}$ for given
states.

The partition optimization in Eq.~(\ref{eq3.5}) contains
redundant terms. Two $k$-partitions
$P=P_1|\cdots|P_k$ and $Q=Q_1|\cdots|Q_k$ are said to be
equivalent, denoted by $P\sim Q$, if their blocks differ only
by a permutation. For equivalent partitions $Q\backsim P$,
$g(L,P)=g(L,Q)$ holds for every positive operator $L$.
Thus it is enough choose one representative from each equivalence class and
call it a valid $k$-partition. The set of all valid
$k$-partitions is denoted by ${\mathcal P}_n^{vk}$. Equation
(\ref{eq3.5}) can then be written as
\begin{equation}
g_n^{(k)}(L)
=
\max_{P\in{\mathcal P}_n^{vk}}g(L,P).
\label{eq4.1}
\end{equation}
Restricting the maximization to ${\mathcal P}_n^{vk}$ removes
the redundant evaluations caused by permutations of partition
blocks without changing $g_n^{(k)}(L)$. Examples of
${\mathcal P}_n^{vk}$ and further details of this partition
reduction are given in Appendix~\ref{SecComputationalDetails}.

{ The point is that, for $2\leq k<n$, values $g(L,P)$ for every valid
$k$-partition $P\in{\mathcal P}_n^{vk}$ is computable numerically, of which the algorithms are presented in Appendix \ref{SecAC}. \if false These values determine
$g_n^{(k)}(L)$ through Eq.~(\ref{eq4.1}).\fi Storing the resulting
quantities for all $k=2,3,\ldots,n$ enable us to construct the general database
used below.}

{
\subsection{Evaluation of $E_w^{(n)}$}
To make the idea clear, we first consider the case $k=n$, that is, the computation of $E_w^{(n)}=E^{(n,n)}_w$.
 In this case, there is only one valid
$n$-partition, $1|2|\ldots|n\in \mathcal P_n^{vn}$, so
\begin{equation*}
g_n^{(n)}(L)=g(L, 1|2|\ldots|n)=g(L)
\end{equation*}
and
$$E_w^{(n)}(\rho)= \sup_{L\in\mathcal{B}_1^+} \max\{0,{\rm Tr}(L\rho)-g(L)\}
$$ for every $\rho\in\mathcal S(H)$. $E_w^{(n)}$ is continuous on $\mathcal S(H)$.}

Let $\widetilde{\mathcal B}_1^+$ be a finite subset of
${\mathcal B}_1^+$. Computing $g(L)$ for each
$L\in\widetilde{\mathcal B}_1^+$ gives the finite database
\begin{equation*}
\widetilde{\mathcal{EW}}_{ME}
=
\left\{
(L,g(L)):
L\in\widetilde{\mathcal B}_1^+
\right\}.
\end{equation*}
For any state $\rho$ in the given system, this database leads to
a value
\begin{equation}
\widetilde E_{w,\widetilde{\mathcal B}_1^+}^{(n)}(\rho)
=
\max_{(L,g(L))\in\widetilde{\mathcal{EW}}_{ME}}
\max\left\{
{\rm Tr}(L\rho)-g(L),0
\right\}.
\label{eq4.2}
\end{equation}
Since the maximization in Eq.~(\ref{eq4.2}) is taken over only
a finite subset of ${\mathcal B}_1^+$, one has
\begin{equation*}
\widetilde E_{w,\widetilde{\mathcal B}_1^+}^{(n)}(\rho)
\leq
E_w^{(n)}(\rho).
\end{equation*}

{Let $$\Omega=\{\widetilde{\mathcal B}_1^+: \widetilde{\mathcal B}_1^+\subset \mathcal B_1^+(H) \ \mbox{\rm is finite subset} \}.$$
With ``$\subseteq$'' as the partial order, $\Omega$ is a direct set and $\{\widetilde E_{w,\widetilde{\mathcal B}_1^+}^{(n)}(\rho)\}_\Omega$ is a net. Consequently,
\begin{equation*}\label{eq4.0}
\begin{array}{rl}& \lim_\Omega \|E_w^{(n)}-\widetilde E_{w,\widetilde{\mathcal B}_1^+}^{(n)}\|\\
=&\lim_\Omega [\max_{\rho\in\mathcal S(H)} |E_w^{(n)}(\rho)-\widetilde E_{w,\widetilde{\mathcal B}_1^+}^{(n)}(\rho)|]\\
=&0.
\end{array}
\end{equation*}
 Therefore, if $\widetilde{\mathcal B}_1^+$ is big enough, $\widetilde E_{w,\widetilde{\mathcal B}_1^+}^{(n)}(\rho)\approx E_w^{(n)}(\rho)$ for all $\rho\in\mathcal S(H)$. In the sequel, we simply denote $\widetilde E_{w,\widetilde{\mathcal B}_1^+}^{(n)}$ by $\widetilde E_{w}^{(n)}$.}

\subsection{General evaluation scheme}\label{Sec4.1}

Now we describe the general evaluation scheme of $E_{w}^{(k,n)}$.

Let $H=H_1\otimes\cdots\otimes H_n$ be a finite-dimensional $n$-partite system and
\begin{equation*}
{\mathcal B}_1^+=\{L\in{\mathcal B}(H):L\geq0,\ \|L\|\leq1\}.
\end{equation*}
For each $L\in{\mathcal B}_1^+$, denote
\begin{equation*}
\mathcal G_k(L)=\{g(L,P):P\in{\mathcal P}_n^{vk}\},
\qquad 2\leq k\leq n-1.
\end{equation*}
The complete database can then be written compactly as
\begin{equation}
\begin{aligned}
&\mathcal{EW}(H)=\mathcal{EW}(H_1\otimes\cdots\otimes H_n)\\
&=\bigl\{[L,g(L),\mathcal G_{n-1}(L),\ldots,\mathcal G_2(L)]
:L\in{\mathcal B}_1^+\bigr\}.
\end{aligned}
\label{eq4.3}
\end{equation}
Then, for any state $\rho\in{\mathcal S}(H)$, one gets  values   $E_w^{(k,n)}(\rho)$ for all $k$ as a vector
\begin{equation}
\begin{aligned}
 \mathcal{E}^{(n)}_w(\rho)
= \bigl(E_w^{(n,n)}(\rho),E_w^{(n-1,n)}(\rho),\ldots,E_w^{(3,n)}(\rho),E_w^{(2,n)}(\rho)\bigr),
\end{aligned}
\label{eq4.4}
\end{equation}
where $E_w^{(n,n)}=E_w^{(n)}$ and, for $2\leq k< n$,
\begin{equation}
E_w^{(k,n)}(\rho)=\max_{L\in{\mathcal B}_1^+}
\max\left\{{\rm Tr}(L\rho)-\max_{P\in{\mathcal P}_n^{vk}}g(L,P),0\right\}.
\label{eq4.5}
\end{equation}

In practice, only a finite subset $\widetilde{\mathcal B}_1^+\subset{\mathcal B}_1^+$ can be sampled. Then the corresponding finite database is
\begin{equation}
\begin{aligned}
&\widetilde{\mathcal{EW}}(H_1\otimes\cdots\otimes H_n)\\
&=\bigl\{[L,g(L),\mathcal G_{n-1}(L),\ldots,\mathcal G_2(L)]:L\in\widetilde{\mathcal B}_1^+\bigr\}.
\end{aligned}
\label{eq4.6}
\end{equation}
It yields
\begin{equation}
\begin{aligned}
 \widetilde{\mathcal E}^{(n)}_{w, \widetilde{B}_1^+}(\rho)
=\bigl(\widetilde E_w^{(n,n)}(\rho),\widetilde E_w^{(n-1,n)}(\rho),\ldots, \widetilde E_w^{(3,n)}(\rho),\widetilde E_w^{(2,n)}(\rho)\bigr),
\end{aligned}
\label{eq4.7}
\end{equation}
with
\begin{equation}
\begin{array}{rl}\widetilde E_w^{(k,n)}(\rho)=&\max_{L\in\widetilde{\mathcal B}_1^+}
\max\left\{0,{\rm Tr}(L\rho)-\max_{P\in{\mathcal P}_n^{vk}}g(L,P)\right\} \\
\leq & E_w^{(k,n)}(\rho).
\end{array}\label{eq4.8}
\end{equation}
{Since the maximization is performed over a finite subset $\widetilde{\mathcal B}_1^+\subset\mathcal B_1^+$, one has
\begin{equation*}
\widetilde E_w^{(k,n)}(\rho)\leq E_w^{(k,n)}(\rho).
\end{equation*}
{In particular, $\widetilde E_w^{(k,n)}(\rho)>0$ ensures that $\rho$ is $k$-entangled.

Since
$$\lim_\Omega \|\widetilde E_w^{(k,n)}- E_w^{(k,n)}\|=0$$
for each $k=2,3,\ldots,n$,
increasing the size and coverage of $\widetilde{\mathcal B}_1^+$ can improve the finite-database approximation.

We remark that $\widetilde E_w^{(k,n)}(\rho)=0$ is generally inconclusive, since $\rho$ may be $k$-entangled but the
database may not contain a witness capable of detecting the given state $\rho$. This does not hinder the practical use of
$\widetilde E_w^{(k,n)}$, since a positive value still certifies $k$-entanglement. Also, adding new samples while retaining the
existing ones in $\widetilde{\mathcal B}_1^+$ may improve the approximation accuracy.}

\subsection{Accuracy analysis}

The error between approximate value $\widetilde E_w^{(k,n)}$ and true value $E_w^{(k,n)}$ mainly comes from two sources: The error comes from the computation of $g(L,P)$ and the error caused by the size of $\widetilde{\mathcal B}_1^+$.

Give a tolerable error  $\varepsilon>0$.  Preset the error of computing $g(L,P)$ as $\varepsilon/2$, which can be by the algorithm, and take $\widetilde{\mathcal B}_{1,0}^+\in\Omega$ such that $\widetilde{\mathcal B}_{1,0}^+\subseteq \widetilde{\mathcal B}_{1}^+$ implies that $\|\widetilde E_w^{(k,n)}- E_w^{(k,n)}\|<\varepsilon/2$ for all $k=2,3,\ldots,n$.
Denote the value we practically obtained is $\hat{E}_w^{(k,n)}$. Then
$$\|\hat{E}_w^{(k,n)}-\widetilde E_w^{(k,n)}\|<\varepsilon/2,
$$
and consequently,
$$\begin{array}{rl} & \|\hat{E}_w^{(k,n)}- E_w^{(k,n)}\| \\ \leq &\|\hat{E}_w^{(k,n)}-\widetilde E_w^{(k,n)}\|+ \|\widetilde E_w^{(k,n)}- E_w^{(k,n)}\| \\ < &\varepsilon \end{array}$$
for each $k$ with $2\leq k\leq n$
Therefore, by above approach we obtain an approximation of $E_w^{(k,n)}$ with error less than given tolerable error $\varepsilon$.

As we do not know the true value of $E_w^{(k,n)}$, it is somewhat difficult in the above approach  to find the finite sample set $\widetilde{\mathcal B}_{1,0}^+$ for  given tolerable error $\varepsilon$. We provide a possible method as follows: (1) Generate randomly a finite subset $\Lambda_1$ of $\mathcal B_1^+$, where  $\#(\Lambda_1)=N$ is a rule-of-thumb estimate, and obtain $\widetilde{\mathcal  E}_{w,\Lambda_1}^{(n)}$ as in Eq.(\ref{eq4.7}). (2) Add randomly generated $N/2$ other samples into $\Lambda _1$, form $\Lambda_2$ with scale $(1+\frac{1}{2})N$ and get $\widetilde{\mathcal  E}_{w,\Lambda_2}^{(n)}$. (3) Compute $\|\widetilde{\mathcal  E}_{w,\Lambda_1}^{(n)}-\widetilde{\mathcal  E}_{w,\Lambda_2}^{(n)}\|_\infty=\max_k\|\widetilde{E}_{w,\Lambda_2}^{(k,n)}-\widetilde{E}_{w,\Lambda_1}^{(k,n)}\|
=\delta$. If $\delta<\varepsilon/2$, then stop and take $\widetilde{\mathcal B}_{1,0}^+=\Lambda_2$; if $\delta\geq \varepsilon/2$, repeat step (2)  and step (3) by taking $\Lambda _2$ as $\Lambda_1$. Stop above operations until get $\delta <\varepsilon/2$.

\if false
Apart from factors related to sample size and distribution, errors in other aspects when computing the approximate value $\tilde E_w^(k,n)(\rho)$ arise from database errors generated in calculating $g(L,P)$. This also requires improving the accuracy of calculating $g(L,P)$ as much as possible.

With exact values of $g(L,P)$, the maximization in Eq.~(\ref{eq4.8}) is restricted to a finite subset of admissible operators, so
\begin{equation}
0\leq\tilde E_w^{(k,n)}(\rho)\leq E_w^{(k,n)}(\rho).
\label{eq4.lowerbound}
\end{equation}
Thus, for every $2\leq k\leq n$, a positive value of $\tilde E_w^{(k,n)}(\rho)$ certifies that $\rho$ is $k$-entangled.

The numerical error contains contributions from finite sampling of $L$, approximation of the product-state maxima $g(L,P)$, and floating-point arithmetic. These contributions enter at different stages of the evaluation. If every stored coefficient has absolute error at most $\epsilon$, the finite-database value changes by at most $\epsilon$, excluding trace-evaluation errors. The outer sampling error remains even when all product-state maxima are evaluated exactly.\fi

Constructing the database is the computationally demanding stage because the underlying product-state optimizations are generally hard. Once the database has been generated for a fixed multipartite system, Eq.~(\ref{eq4.8}) reduces the evaluation of each  state to finitely many trace calculations and maximizations. This separation between database construction and state evaluation underlies the software described below.

\subsection{Evaluation for $m$-qubit systems}\label{Sec4.2}

For implementation, we specialize to an $m$-qubit Hilbert space
$H=H_1\otimes\cdots\otimes H_m, H_j=\mathbb C^2$, { by applying the general scheme in previous subsection.
\if false The restriction is computational rather than conceptual; the preceding construction applies to arbitrary finite local dimensions.\fi Note that, an} $m$-qubit system can be grouped into an $n$-partite system for any $2\leq n\leq m$. We write $E_{w,m}^{(k,n)}$ to display both the number of qubits and the selected multipartite structure. A database constructed for the $m$ individual qubits contains the entries required for all coarser groupings, provided that the optimization is restricted to partitions compatible with the chosen grouping.

The explicit formulas for the $n$-partite $n$-qubit and $n$-partite $m$-qubit cases, including Eqs.~(\ref{eq4.9})--(\ref{eq4.14}), are collected in Appendix~\ref{SecQubitDetails}.

\section{Implementation for two to four qubits}\label{Sec5}

{ To check the feasibility of the computing scheme  of $E_{w}^{(k,n)}$ proposed in Section~\ref{Sec4}, we implement the scheme for  $E_{w,m}^{(k,n)}$ with $m=2,3,4$ in three stages.}

First, { a finite set $\widetilde{\mathcal B}_1^+$ of operators $L\in\mathcal B_1^+$ are sampled and, for each $L\in \widetilde{\mathcal B}_1^+$, then the product-state optimization $g_m(L,P)$ is carried out for every valid partition $P$ using the general iterative algorithm in Appendix~\ref{SecAC}. There are  one valid full partition and three valid bipartitions for $m=3$; one four-partition, six tripartitions and seven bipartitions for $m=4$. The resulting records form the reusable database $\widetilde{\mathcal{EW}}^{(m)}$ as in Eq.(\ref{eq4.6}).The sampling procedure is  random on $\mathcal B_1^+$, and every computed $g_m(L,P)$ is stored with randomly sampled positive operator $L$  and each valid partition $P$. \if false {For each sampled $L$ and partition $P$, an operator $c_m(L,P)I-L$ is nonnegative on all product states of $P$ if and only if $c_m(L,P)\geq g_m(L,P)$. This condition specifies the coefficient accuracy required for entanglement certification.}\fi  The explicit database entries, valid partitions, and case-by-case formulas are given in Appendix~\ref{SecImplDetails}; the iterative product-state algorithms are given in Appendix~\ref{SecAC}.}

Second, the same record is used for both the natural $m$-partite structure and every coarser grouping. If $Q=Q_1|\cdots|Q_n$ is an $n$-partition of the $m$ qubits, only the $k$-partitions compatible with $Q$ are retained, and { an approximation of  $E_{w,m,Q}^{(k,n)}(\rho)$ at an $m$-qubit state $\rho$  with respect to $Q$ and the established  finite-database can be obtained as}
\begin{equation}
\begin{aligned}
\tilde E_{w,m,Q}^{(k,n)}(\rho)= \max_{L\in\widetilde{\mathcal B}_1^+}\max\biggl\{0,\,{\rm Tr}(L\rho)
-\max_{\substack{R\in\mathcal P_m^{vk}\\ R\preccurlyeq^b Q}}g_m(L,R)\biggr\}.
\end{aligned}
\label{eq5.main}
\end{equation}
For the natural grouping $Q=1|2|\cdots|m$, Eq.~(\ref{eq5.main}) reduces to $\tilde E_{w,m}^{(k,m)}$. For example, one four-qubit database gives the $4$-, $3$-, and $2$-entanglement values of the four-partite state, the ME and GME values of each compatible tripartite grouping, and the entanglement value of each compatible bipartite grouping.

Third, these evaluations are assembled into a software used in Section~\ref{Sec6}.

\section{Database, software, and numerical tests}\label{Sec6}

{ A software~\cite{Software} is developed to evaluate $\tilde{E}_{w,m,Q}^{(k,n)}$ in Eq.(\ref{eq5.main})  numerically for $2\leq k\leq n\leq m\leq4$, which is an approximated values of $ E_{w,m,Q}^{(k,n)}$, the $k$-entanglement measure of $n$-partite $m$-qubit system determined by $n$-partite partition $Q$ of $1|2|\ldots|m$}.

For each $m=2,3,4$, our database contains more than $3\times10^4$ randomly generated positive semi-definite  matrices $L\in\mathcal B_1^+$. The coefficients $g_m(L,P)$ are obtained using the iterative algorithm in Appendix~\ref{SecAC} in validation errors of order $10^{-4}$. This error characterizes the tested coefficient calculations separately from the finite sampling of $L$.

{ The numerical tests include a comparison with know criteria of $k$-entanglement to check the reliability and accuracy of the developed software. And then used to discover new $k$-entangled states and visualize the change in its intensity.
The software can also examine different subsystem groupings and detect  PPT-entangled states. Several supplementary numerical curves are collected in Appendix~\ref{SecNumericalDetails}.}

\subsection{{Comparison with entanglement negativity for case $m=2$}}

For a bipartite state $\rho$, the entanglement negativity is $$
{\mathcal N}(\rho)=\frac{\|\rho^{T_A}\|_1-1}{2},$$
where $T_A$ denotes partial transposition with respect to one subsystem. In a two-qubit system, ${\mathcal N}(\rho)=0$ if and only if $\rho$ is separable by PPT criterion. {The positivity of negativity therefore provides an exact entanglement criterion for two-qubit states.}

{
We establish randomly five sets of sizes 1,000, 2,000, 3,000, 4,000, and 5,000 two-qubit states to compare the values obtained by the software with the values obtained by the entanglement negativity $\mathcal N$. For each input, a matrix $A=X+\mathrm{i}Y\in\mathbb C^{4\times4}$ is generated with NumPy, with the entries of $X$ and $Y$ sampled independently and uniformly from $[0,1)$. We set
$B=A^\dagger A$ and define the normalized density matrix by $\rho=B/{\rm Tr}B$. Since $B=A^\dagger A$ is positive semidefinite, $\rho$ is a valid quantum state with unit trace. It is represented as a QuTiP~\cite{QuTiP2012} object with subsystem dimensions $(2,2)$. Then
$\tilde{E}_{w,2}^{(2,2)}(\rho)$ and ${\mathcal N}(\rho)$ are compared.
Table~\ref{tab1} gives the numbers of positive and zero values. We use $N_{\rm diff}=N_{\mathcal N}^{+}-N_{\rm sw}^{+}$ and
$r_{\rm diff}=100N_{\rm diff}/N$ to measure the difference between the two positive counts, where $N$ is the sample size,
$N_{\mathcal N}^{+}$ counts the positive values of ${\mathcal N}(\rho)=(\|\rho^{T_A}\|_1-1)/2$, and $N_{\rm sw}^{+}$ counts the positive values obtained by the software. The quantity $r_{\rm diff}$ measures the difference between the aggregate positive counts. Table~\ref{tab1} shows that the relative differences between the two positive counts range from $1.30\%$ to $0.68\%$ as the sample size increases from 1,000 to 5,000.}

\begin{table*}[t]
\caption{Comparison of the negativity ${\mathcal N}(\rho)$ and $\widetilde E_{w,2}^{(2,2)}(\rho)$ for randomly generated two-qubit states. Here $N_{\rm diff}$ is the difference between the numbers of positive values obtained by the two methods and $r_{\rm diff}=100N_{\rm diff}/N$.}
\label{tab1}
\centering
\begin{tabular}{rcc ccrr}
\toprule
& \multicolumn{2}{c}{Negativity ${\mathcal N}(\rho)$}& \multicolumn{2}{c}{$\tilde E_{w,2}^{(2,2)}(\rho)$}& & \\
\cline{2-3}\cline{4-5}
Sample size $N$ & {Positive} & {Zero} & {Positive} & {Zero} & $N_{\rm diff}$ & $r_{\rm diff}$ (\%) \\
\hline
1000 & 957  & 43  & 944  & 56  & 13 & 1.30 \\
2000 & 1924 & 76  & 1908 & 92  & 16 & 0.80 \\
3000 & 2861 & 139 & 2837 & {163} & 24 & 0.80 \\
4000 & 3818 & 182 & 3790 & 210 & 28 & 0.70 \\
5000 & 4773 & 227 & 4739 & {261} & 34 & 0.68 \\
\bottomrule
\end{tabular}
\end{table*}

{
The two entanglement measures ${E}_{w,2}^{(2,2)}$ and ${\mathcal N}$ do not induce the same ordering. For two density matrix inputs, denoted by $\rho_1$ and $\rho_2$, the numerical values are ${\mathcal N}(\rho_1)=0.12924261200412$ and $\tilde E_{w,2}^{(2,2)}(\rho_1)=0.2129571233$, whereas ${\mathcal N}(\rho_2)=0.0624506547894968$ and $\tilde  E_{w,2}^{(2,2)}(\rho_2)=0.2279157623$. We have ${\mathcal N}(\rho_1)>{\mathcal N}(\rho_2)$ but $\tilde E_{w,2}^{(2,2)}(\rho_1)
< \tilde E_{w,2}^{(2,2)}(\rho_2)$.}

{For a comparison on physical states with a known separability boundary, consider}
$\rho(p)=p|\psi^-\rangle\langle\psi^-|+(1-p)\frac{I_4}{4}$
and
$\sigma(p)=p|\psi^+\rangle\langle\psi^+|+(1-p)\frac{I_4}{4},$ where $
|\psi^-\rangle=\frac{|01\rangle-|10\rangle}{\sqrt2},$ $
|\psi^+\rangle=\frac{|01\rangle+|10\rangle}{\sqrt2}.$
Both families are entangled exactly when $p>1/3$. The two Bell states are related by a local unitary, so the exact negativity and the exact witness-based measure take the same values on the two families, that is, $\tilde{E}_{w,2}^{(2)}(\rho(p))=\tilde{E}_{w,2}^{(2)}(\sigma(p))$ and ${\mathcal N}(\rho(p))={\mathcal N}(\sigma(p))$. \if false Numerically, the corresponding software curves coincide within the reported resolution.\fi  As shown in Fig.~\ref{fuxing}, both quantities vanish for $p\leq0.3333333$ and are positive for $p\geq0.3333334$, {giving} {a} {numerical} {sign-change} interval of width $10^{-7}$ {containing the exact threshold $1/3$}. Both curves increase monotonically with $p$.

{ The numerical tests above show that the developed software is sufficiently reliable and accurate for the two-qubit system with the database size of $3\times 10^4$ samples.}

\begin{figure}[!htbp]
\centering
\includegraphics[width=0.48\textwidth,height=0.2\textheight]{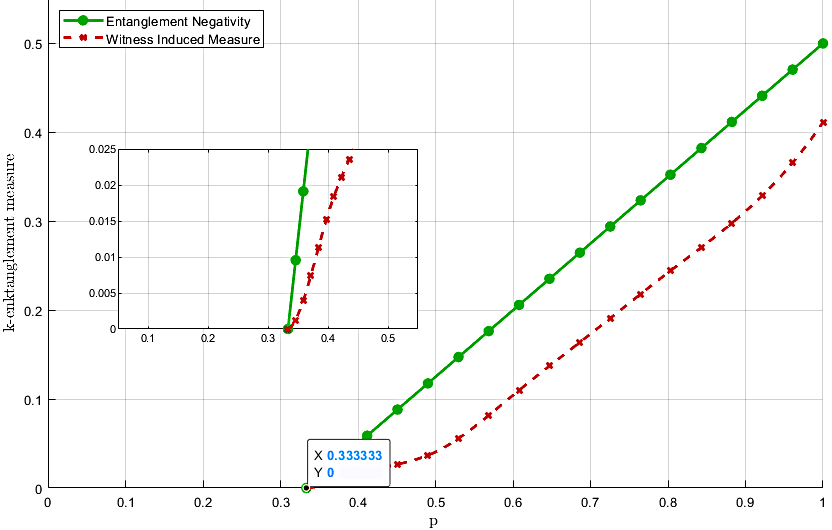}
\caption{Plots of $\tilde E_{w,2}^{(2,2)}(\rho(p))$
(red, $-+-$) and the negativity ${\mathcal N}(\rho(p))$
(green, $-\!\cdot\!-$) as functions of $p$ for two families
of Bell states mixed with white noise.}
\label{fuxing}
\end{figure}

\subsection{$n$-partite $n$-qubit systems}\label{SecRepresentativeExamples}

{ Since there are no computable $k$-entanglement measures for $n$-partite $n$-qubit system with $n >2$, we  utilize some known results in the identification of $k$-entanglement to perform reliability testing for our software.
 Here several families of pure states mixed with white noise, that is,
$$\varrho(p)=p|\psi\rangle\langle\psi|+(1-p)\frac{I}{2^n},$$
are considered,} where {$0\leq p\leq1$ is the pure-state weight,} $I$ is the identity operator on the $n$-qubit Hilbert space and $|\psi\rangle$ denotes the representative multipartite pure state specified in each of Examples~6.1--6.6. The resulting white-noise admixtures include the Werner-type, W-type, linear-cluster, Dicke-type, and singlet-type families considered below, { which  are several classes of quantum states that have been extensively studied and applied. Here a smaller  pure-state weight $p$ means greater white-noise tolerance.} \if false Comparisons below use the same state family, partition, and entanglement index; a software sign change is distinguished from a proven separability boundary. Examples~6.1 and 6.2 examine the dependence on the number of parties, Examples~6.3--6.5 compare inequivalent four-qubit entanglement structures, and Example~6.6 studies the effect of changing the {subsystem} grouping while keeping the underlying qubit state fixed.\fi

\textbf{Example 6.1 (Werner states).}
{ For $n=2$, let $|\psi_2\rangle=\frac{|01\rangle-|10\rangle}{\sqrt2},$
and, for $n=3,4$, let $
|\psi_n\rangle=\frac{|0\rangle^{\otimes n}+|1\rangle^{\otimes n}}{\sqrt2}.$
The Werner states are states of the form
$$\varrho_n(p)=p|\psi_n\rangle\langle\psi_n|+(1-p)\frac{I}{2^n}.$$
The graphs of  $\tilde E_{w,n}^{(k,n)}(\varrho(p))$ for $n=2,3,4$ as functions of $p\in[0,1]$ are presented in {\color{blue} Fig. ~\ref{fig22}.}
%\begin{widetext}

\begin{figure}[htpb]
\centering
\includegraphics[width=0.48\textwidth,height=0.22\textheight]{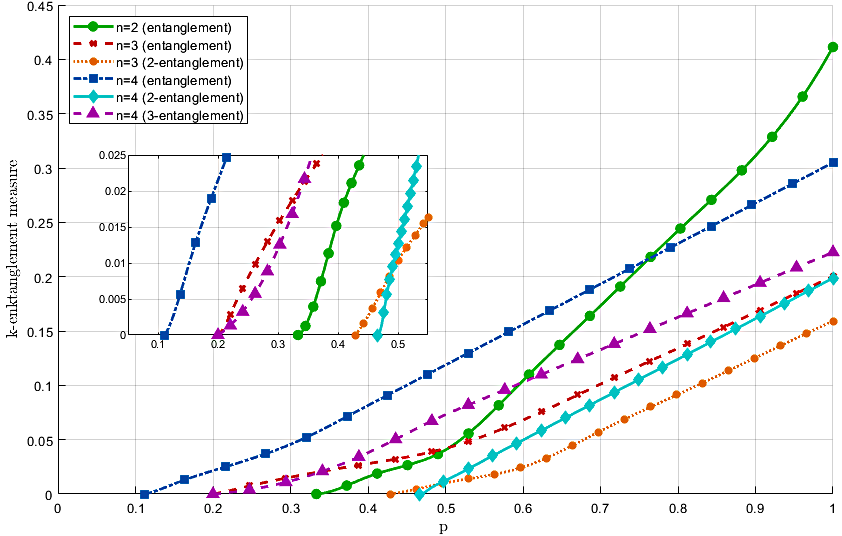}
\caption{The measures $\tilde E_{w,n}^{(k,n)}$ for the Werner-type families of Example~6.1 with $n=2,3,4$.}
\label{fig22}
\end{figure}

%\end{widetext}
It is known that, for $n=2$, $\varrho(p)$ is entangled if and only if $p> 1/3$; for $n=3$, $\varrho(p)$ is ME if and only if $p>0.2$ and is GME if $p\geq 0.4285714$; for $n=4$, $\varrho(p)$ is respectively ME, 3-entangled and GME if and only if $p>1/9$, $p>1/5$ and $p>7/15$ ~{\cite{F-sep1,F-sep2,2-sep,K-sep}}. Fig.~\ref{fig22} reveals that
all these known thresholds are rediscovered with high precision of order $10^{-8}$ by our software. Moreover, the software  visualizes the $k$-entanglement strength of the Werner states.
}

\if false
the {numerical} {onset agrees with} the exact threshold $p=1/3$. For $n=3$, {the} {numerical curves reproduce} the known {ME transition} at $p=0.2$ and {the} GME {transition near $p=0.4285714$}~{\cite{F-sep1,F-sep2,2-sep,K-sep}}. For $n=4$, the {numerical} {onsets} {for} $k=4,3,2$ {occur} at approximately $0.11111111$, $0.2$, and $7/15=0.46666667$, respectively. {The} {ME} {and} {GME} {onsets} {agree} {with} the {analytical} {thresholds~\cite{F-sep1,F-sep2,2-sep}}. The corresponding curves are shown in Fig.~\ref{fig2} of Appendix~\ref{SecNumericalDetails}. {In the benchmark comparison with an exact value, the final numerical result has an absolute error of order $10^{-8}$.}\fi

\textbf{Example 6.2 (W states).}
Let $\varrho(p)=p|W_n\rangle\langle W_n|+(1-p)\frac{I}{2^n} $, where
$$
|W_n\rangle=\frac{1}{\sqrt n}\bigl(|10\ldots0\rangle+|010\ldots0\rangle+\cdots+|00\ldots1\rangle\bigr)$$
is the W state.
Unlike the GHZ-type family in Example~6.1, the W family carries a single delocalized excitation and therefore tests a different multipartite correlation structure. For $n=2$, the {numerical onset of $\tilde E_{w,2}^{(k,2)}(\varrho(p))$ by the software agrees with the} exact known threshold $p=1/3$. For $n=3$, $\tilde E_{w,3}^{(3,3)}(\varrho(p))$ (ME) {and} $\tilde E_{w,3}^{(2,3)}(\varrho(p))$ (GME) become positive respectively from $p=0.26882292$ and {$p=0.4790001$. For $n=4$, {the} {computed} {ME}, $3$-entanglement, and GME {values become positive} from $p=0.20001$, $0.2326897$, and $0.3027026$, respectively (See Fig.~\ref{fig2} in Appendix~\ref{SecNumericalDetails}). {Related} criteria and {numerical} {benchmarks} for {noisy} {W} {states} are {discussed} {in Refs.~\cite{GHQH,JMG,2-sep-w,CLW23}} { and, our test results recover the known exact thresholds,  improve the known criteria $p\geq 0.30000752$ and $p\geq 0.5555556$ on  $3$-entanglement and GME respectively for 4-qubit case.}

\textbf{Example 6.3 (linear-cluster state).}
Consider $\varrho(p)=p|Cl_4\rangle\langle Cl_4|+(1-p)\frac{I}{2^4} $ with the cluster state $$
|Cl_4\rangle=\frac{|0000\rangle+|0011\rangle+|1100\rangle-|1111\rangle}{2}.$$
For the noisy cluster family $\varrho(p)$, {the} {computed GME value $\tilde E_{w,4}^{(2,4)}(\varrho(p))$ becomes positive near $p=0.385$, in agreement with} the known transition~{\cite{JMG}}. In addition, we find new results that $\varrho(p)$ is $3$-entangled and multipartite entangled respectively  for $p\geq 0.0973904$ and $p\geq 0.0769231$. The numerical curves are given in Fig.~\ref{fig3} of Appendix~\ref{SecNumericalDetails}.

\textbf{Example 6.4 (two-excitation Dicke state).}
 Consider the 4-qubit Dicke state $ |D_{24}\rangle $ mixed with white noise $\varrho(p)=p |D_{24}\rangle \langle D_{24}| + (1-p) \frac{{I}}{2^4}$  \cite{JMG, GO07}, where
\begin{equation*}\begin{aligned}
|D_{2,4}\rangle=\frac{1}{\sqrt6}\bigl(&|0011\rangle+|1100\rangle+|0101\rangle\\
&+|0110\rangle+|1001\rangle+|1010\rangle\bigr).\end{aligned}\end{equation*}
The {computed} value of  $\tilde{E}_{w,4}^{(2,4)}(\varrho(p))$ reveals that $\varrho(p)$ is GME if  $p\geq 0.320059$, improving the GME criterion  $p\geq 0.461$ obtained in {Ref.}~{\cite{JMG}}. We also get two new observations that  $\varrho(p)$ is $3$-entangled if $p\geq 0.0914763$ and multipartite entangled if $p\geq 0.05974273$. The corresponding curves for this permutation-symmetric two-excitation family are shown in Fig.~\ref{fig4} of Appendix~\ref{SecNumericalDetails}.

\textbf{Example 6.5 (four-qubit singlet-type state).} Consider the 4-qubit state $ |\psi_{S,4}\rangle $ \cite{JMG, GO07} mixed with white noise, denoted as $\varrho(p)=p |\psi_{S,4}\rangle \langle \psi_{S,4}| + (1-p) \frac{{I}}{2^4}$, where
\begin{equation*}\begin{aligned}
|\psi_{S,4}\rangle=\frac{1}{\sqrt3}\biggl[&|0011\rangle+|1100\rangle\\
&-\frac12\bigl(|0101\rangle+|0110\rangle+|1001\rangle+|1010\rangle\bigr)\biggr].\end{aligned}\end{equation*}

For this family, the software finds that $\varrho(p)$ is GME if $p\geq 0.3512337$, {improving the result with $p\geq 0.447$ in {Ref.}~\cite{JMG} {and an earlier criterion $p\geq 0.683$ in Ref.~\cite{GO07}}. In addition, the software discovers that $\varrho(p)$ is $3$-entangled if $p\geq 0.1175803$ and is multipartite entangled if $p\geq 0.08993755$. The graphs of $\tilde{E}_{w,4}^{(k,4)}(\varrho(p))$ for $k=2,3,4$ are displayed in Fig.~\ref{fig5} of Appendix~\ref{SecNumericalDetails}.

\if false

{The comparison with Ref.~\cite{JMG} is summarized in Table~\ref{tab:onsets}. That reference reports white-noise tolerances, which are converted here to pure-state weights through $p=1-p_{\rm tol}$. The software onsets are lower by approximately $0.1713$, $0.1409$, and $0.0958$ for the four-qubit W, Dicke, and singlet families, respectively.

\begin{table*}[t]
\caption{{Numerical GME onsets for the natural four-qubit partition ($k=2$), compared with the pure-state weights required by the sufficient criteria in Ref.~\cite{JMG}. The comparisons do not establish exact separability boundaries.}}
\label{tab:onsets}
\centering
\begin{tabular}{lcc}
\toprule
State family & Software onset & Reference pure-state weight \\
\midrule
$W_4$ & $0.3027026$ & $0.474$ \\
$D_{2,4}$ & $0.320059$ & $0.461$ \\
$|\psi_{S,4}\rangle$ & $0.3512337$ & $0.447$\\
\bottomrule
\end{tabular}
\end{table*}
\fi

{
Examples~6.1--6.5 show that, for $n$-partite $n$-qubit systems, the $k$-entanglement detection by the software  highly aligns with the known sufficient and necessary criteria or improves the known sufficient criteria.
This leads us to believe that, with the database size $3\times 10^4$ samples for $3$- and $4$-qubit systems, the $k$-entanglement calculation software is already sufficiently reliable and accurate for detecting $k$-entanglement and its intensity in quantum states.}

\subsection{$n$-partite $m$-qubit systems}\label{SecGroupedExample}

The preceding examples use the natural $n$-partite structure in which each qubit is one party. We now provide an example for $n$-partite $m$-qubit systems with $2\leq n<m\leq 4$.

\textbf{Example 6.6 (The same qubit state under different groupings).} { Since no existing criteria are available for all
cases considered here, we test the software using randomly
generated $k$-separable states. The states are generated
according to the prescribed multipartite structures: random
density matrices on the individual groups are generated using
QuTiP~\cite{QuTiP2012}, and their tensor products are used to
construct $k$-separable states. For each $(k,n,m)$ and
multipartite structure listed in Table~\ref{tab:sep-test},
1,000 $k$-separable states are generated and evaluated.
A zero output is counted as correct, whereas a positive output
is counted as a false positive. The accuracy is the percentage
of zero outputs. }

\begin{table*}[!t]
\caption{Numerical tests on randomly generated $k$-separable $n$-partite $m$-qubit states. Each case contains 1,000 samples.
The columns $\widetilde E=0$ and $\widetilde E>0$ denote the numbers of zero outputs and false positives, respectively.}
\label{tab:sep-test}
\centering
\begin{tabular}{ccccccccc}
\toprule
$m$ & $n$ & $k$ & Dimension
& Multipartite structure
& Sample size
& $\widetilde E=0$
& $\widetilde E>0$
& Accuracy (\%) \\
\midrule
3 & 2 & 2 & 8
& $4\times2$
& 1000 & 1000 & 0 & 100.0 \\
3 & 2 & 2 & 8
& $2\times4$
& 1000 & 999 & 1 & 99.9 \\
4 & 2 & 2 & 16
& $4\times4$
& 1000 & 1000 & 0 & 100.0 \\
4 & 2 & 2 & 16
& $2\times8$
& 1000 & 1000 & 0 & 100.0 \\
4 & 2 & 2 & 16
& $8\times2$
& 1000 & 1000 & 0 & 100.0 \\
4 & 3 & 2 & 16
& $4\times2\times2$
& 1000 & 1000 & 0 & 100.0 \\
4 & 3 & 2 & 16
& $2\times4\times2$
& 1000 & 1000 & 0 & 100.0 \\
4 & 3 & 2 & 16
& $2\times2\times4$
& 1000 & 1000 & 0 & 100.0 \\
4 & 3 & 3 & 16
& $4\times2\times2$
& 1000 & 1000 & 0 & 100.0 \\
4 & 3 & 3 & 16
& $2\times4\times2$
& 1000 & 1000 & 0 & 100.0 \\
4 & 3 & 3 & 16
& $2\times2\times4$
& 1000 & 1000 & 0 & 100.0 \\
\bottomrule
\end{tabular}
\end{table*}
  For the structure $\mathbb C^2\otimes\mathbb C^4$, 999 of the 1,000 $2$-separable states give zero outputs, corresponding to an  accuracy of $99.9\%$. In the remaining ten tested cases, all 1,000 sampled states give zero outputs, yielding an accuracy of $100\%$ in each case. The single positive output is a numerical false positive, whose magnitude lies within the numerical error observed in the validation of $g(L,P)$. {These ensures that the detecting results of our software is credible for these $n$-partite $m$-qubit systems and thus is safe to be used.}

Consider  the Werner  states $\varrho(p)$ in Example 6.1. One can detect the $k$-entanglement of $\varrho(p)$ by  applying $\tilde E_{w,m}^{(k,n)}(\varrho(p))$ as an $n$-partite state for $2\leq n<m\leq 4$.

For $m=3$, we regard $\varrho(p)$ as bipartite states under the grouping
$(\mathbb C^2\otimes\mathbb C^2)\otimes\mathbb C^2,$
and evaluate $\tilde E_{w,3,12|3}^{(2,2)}$.
It follows that $\tilde E_{w,3,12|3}^{(2,2)}$ vanishes for $p\leq0.12000456$ and becomes positive for $p\geq0.12000457$, witnessing the entanglement of $\varrho(p)$ as a state in $(\mathbb C^2\otimes\mathbb C^2)\otimes\mathbb C^2$ (see Fig~\ref{fig7}).

 For $m=4$,
we consider both the tripartite grouping
$(\mathbb C^2\otimes\mathbb C^2)\otimes\mathbb C^2\otimes\mathbb C^2$
and the bipartite grouping $
(\mathbb C^2\otimes\mathbb C^2)\otimes(\mathbb C^2\otimes\mathbb C^2).$
It follows that $\tilde E_{w,4, 12|3|4}^{(3,3)}$ vanishes for $p\leq0.1174894$ and becomes positive for $p\geq0.1174895$;  $\tilde E_{w,4, 12|3|4}^{(2,3)}$ vanishes for $p\leq0.18410412$ and becomes positive for $p\geq0.18410413$; while $\tilde E_{w,4, 12|34}^{(2,2)}$ vanishes for $p\leq0.13533319$ and becomes positive for $p\geq0.13533320$ as shown in Fig~\ref{fig7}.

\begin{figure}[htpb]
\centering
\includegraphics[width=0.45\textwidth,height=0.24\textheight]{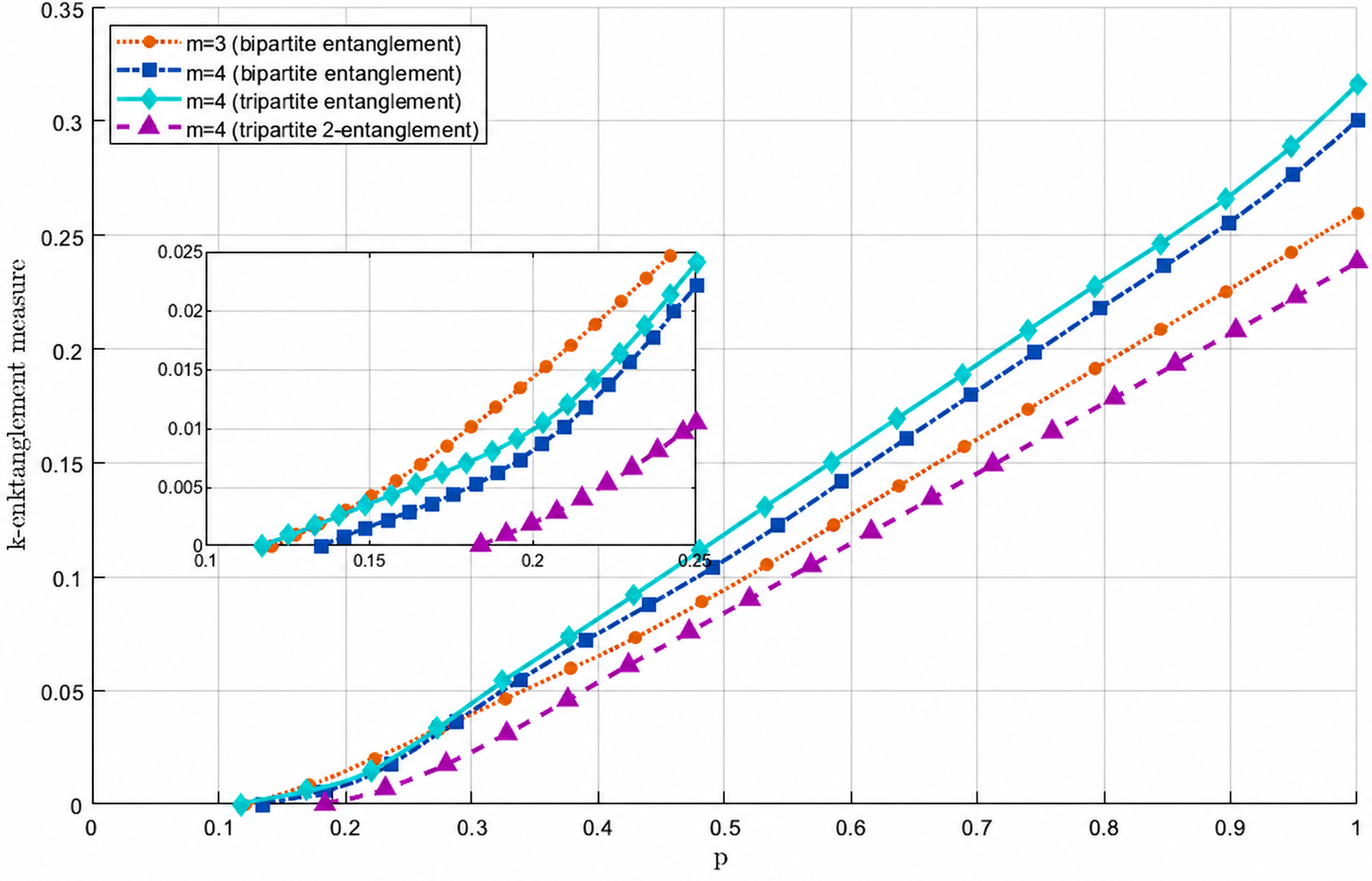}
\caption{ Finite-database measures $\tilde E_{w,n}^{(k,m)}$ for the Werner-type family under coarser part-system groupings with $n<m$. Each curve corresponds to a different grouping and hence to a different free-state set.}
\label{fig7}
\end{figure}

\subsection{Detection beyond the PPT criterion}

\textbf{Example 6.7.}
To illustrate the sensitivity of the software, we try to use the software to detect  entangled   PPT states. Consider bipartite four-qubit system
$
(\mathbb C^2\otimes\mathbb C^2)\otimes(\mathbb C^2\otimes\mathbb C^2)$.
 In the computational basis $\{|1\rangle,|2\rangle,|3\rangle,|4\rangle\}$ of $\mathbb C^2\otimes\mathbb C^2$, define
$|\omega_1\rangle=\frac12\sum_{i=1}^4|i,i'\rangle,|\omega_2\rangle=\frac12\sum_{i=1}^4|i,i{+}1'\rangle,$
where the second index is taken modulo $4$, and let $\rho_j=|\omega_j\rangle\langle\omega_j|$ for $j=1,2$. Let
$\rho_3=\frac14\sum_{i=1}^4|i,i{+}2'\rangle\langle i,i{+}2'|,
\rho_4=\frac14\sum_{i=1}^4|i,i{+}3'\rangle\langle i,i{+}3'|.$
and consider the state
\begin{equation*}
\rho(p_1,p_2,p_3,p_4)=\sum_{j=1}^4p_j\rho_j,
\qquad
p_j\geq0,\quad \sum_{j=1}^4p_j=1.
\end{equation*}
It was shown in Ref.~\cite[Example~4.3]{QH-bipatite}.
$\rho(p_1,p_2,p_3,p_4)$ is a PPT-entangled state if and only if $p_1=\frac{1}{2}p_2$ or $p_3=p_4\geq4p_1$. { Our software recovers this result numerically. For instance, by above criterion,
$\rho(0.07,0.14,0.395,0.395)$ is PPT entangled and thus cannot be recognized by the PPT criterion.
The software gives
$\tilde E_{w,4,12|34}^{(2,2)}(\rho)=0.0118878187>0$, witnessing the entanglement.}

{ From the numerical tests, we see that, the software using about $3\times10^4$ sampled positive operators for each $m$-qubit system, $m=2,3,4$,  achieves high reliability and accuracy. In addition, typical evaluation times are below 50 seconds for two qubits, about 80 seconds for three qubits, and below 200 seconds for four qubits. The experiments are conducted on a workstation equipped with an Intel Xeon Gold processor and 32\,GB RAM, running Windows~10. These timings refer to state evaluation after database construction. Once a finite database is available, each evaluation requires only finitely many trace calculations and maximizations.}

\section{Conclusion and discussion}

{ By the structure of $k$-entanglement, including ME and GME, for multipartite quantum systems and requirements from many-body resource theory, quantifying $k$-entanglement requires a consistent description of both the separability hierarchy and the entanglement assigned to different  part systems. We have addressed these issues by formulating an axiomatic framework that supplements the usual faithfulness, LOCC monotonicity, and permutation symmetry with unification and hierarchy conditions that regulate how to measure the entanglement content in part systems and how to describe the rule that the entanglement content possessed by a part system cannot exceed that possesses by entire system.  The quantifications obeying the new axiomatic framework are called true $k$-entanglement measures. From the perspective of multipartite quantum resource theory, the proposed framework provides a basis for investigating further structural properties of $k$-entanglement measures.
The existence of such a true $k$-entanglement measure is proved by  constructing a witness-based measure $E_w^{(k,n)}$ with $2\leq k\leq n$ for any $n$-partite systems, which also satisfies convexity and subadditivity.

The witness formulation also leads to a finite-dimensional evaluation procedure for $E_w^{(k,n)}$. The idea is that, for a given $n$-partite system, establish  a database of  $k$-entanglement witnesses generated by finitely many  positive operators in advance and then utilize the database to calculate the approximate value of $E_w^{(k,n)}(\rho)$ for any $k\leq n$ and any state $\rho$ in the system. As an illustration,  we have implemented a computation software for systems of up to four qubits with database containing $3\times10^4$ randomly generated positive operators. Numerical tests show that this software can detect $k$-entanglement as well as the strength of $k$-entanglement in states up to 4 qubits  efficiently with high accuracy.

\if false Calculations for representative state families reproduce selected established thresholds and give numerical GME onsets below those of the cited sufficient criteria for noisy four-qubit W, Dicke, and singlet states. The resulting positive intervals identify parameter regions detected by the proposed witness beyond the ranges reported by the cited sufficient criteria. A positive value is also obtained for a known PPT-entangled state, demonstrating detection beyond the PPT criterion. These results are based on finite-database numerical evaluations and should be interpreted as numerical witness detections rather than exact determinations of separability boundaries.\fi

Though the computation approach of  the true $k$-entanglement measure $E_w^{(k,n)}(\rho)$ applies to arbitrary finite  dimensional $n$-partite systems,  extending the implementation to larger systems calls for more efficient operator sampling and product-state optimization, together with improved control of finite-sampling and computational errors. }

{ To the best of our knowledge, $E_w^{(k,n)}$ is currently the only true $k$-entanglement measure. It is also interesting to construct more  true $k$-entanglement measures, and to determine which previously proposed $k$-entanglement measures satisfy  $k$-entanglement measures that satisfy the unification and the hierarchy conditions. \if false The GME quantifier of Ref.~\cite{JMG} is defined relative to outer approximations such as PPT mixtures and therefore can vanish on {genuinely multipartite} entangled states, so it does not satisfy faithfulness in the present sense. Earlier witness-based measures such as Ref.~\cite{FGS} impose general spectral constraints on admissible witnesses but do not address hierarchical consistency across prescribed subsystem partitions. These distinctions motivate the explicit construction above and the numerical evaluation developed next.\fi

 Another  theoretical question is whether $E_w^{(k,n)}$ satisfies certain types of multipartite monogamy relations that are compatible with the proposed hierarchy condition.}

\section*{Acknowledgments} { This work was supported by the National Natural Science Foundation of China (Grant Nos. 12571138, 12071336, 12171290, and 12271394; Key program No. 62636009).}

\section*{Conflict of Interest} The authors declare that they have no conflict of interest.

\section*{Data Availability Statement} The {software} {is} available {via} the repository cited in Ref.~\cite{Software}.

\begin{widetext}
\appendix
\renewcommand{\thesubsection}{\Alph{section}\arabic{subsection}}

\section{Warm-up: details for the case $k=n$ }\label{SecWarmupDetails}

\subsection{Equivalent characterizations of $g(L)$}

The properties in Eq.~(\ref{eq2.3}) hold whenever $\alpha>0$ and $D,L,L_1,L_2$ { and $M$} are positive operators. Since every fully separable state is a convex combination of product pure states, or a trace-norm limit of such combinations, it follows that
\begin{equation*}
\begin{aligned}
g(L)=\sup\bigl\{{\rm Tr}(L\sigma):
\sigma\in{\mathcal S}_{\rm sep}(H_1\otimes\cdots\otimes H_n)\bigr\},=\sup\bigl\{{\rm Tr}[L(\sigma_1\otimes\cdots\otimes\sigma_n)]:\sigma_j\in{\mathcal S}(H_j),\ j=1,2,\ldots,n\bigr\}.
\end{aligned}
\end{equation*}
where ${\mathcal S}_{\rm sep}(H_1\otimes H_2\otimes\cdots\otimes H_n)={\mathcal S}_n$, the set of all fully separable states. Moreover,  $W=\lambda I-L\in\mathcal{EW}_1$ if and only if $g(L)\leq \lambda\leq\|L\|\leq 1$.

\if false
\subsection{Convexity and witness optimization}

Let $E:{\mathcal S}(H)\to\mathbb R_+$ be an entanglement measure. Recall that $E$ is convex if
$E(\sum_jp_j\rho_j)\leq\sum_jp_jE(\rho_j)$ for every ensemble $\{p_j,\rho_j\}$. { The  result Theorem~\ref{T2.3} establishes this property for $E_w^{(n)}$. Moreover, the proof of Theorem~\ref{T2.2} in Appendix~\ref{SecAA} shows that, for every $L\in{\mathcal B}_1^+$ and every LOCC operation $\Phi$, $\Phi^\dagger (L)\in\mathcal B_1$ and} $g(\Phi^\dagger(L))\leq g(L).$
\fi

\subsection{Examples of the elementary hierarchy relations}

The three elementary relations defined in Section~\ref{Sec2} correspond, respectively, to discarding complete parties, grouping parties, and tracing out subsystems within the parties. The following examples illustrate these operations without repeating their definitions. Let $n\geq5$ and $P=13|25|4$. Then $13|4\preccurlyeq^a P$, because the second block of $P$ is discarded; $13|245\preccurlyeq^b P$, because the blocks $25$ and $4$ are grouped; and $3|2|4\preccurlyeq^c P$, because subsystems are traced out within the first two blocks. These elementary operations generate the partial order $\preccurlyeq$ through Definition~\ref{D2.6}.

\section{Supplementary remarks on the hierarchy condition}\label{SecHierarchyDetails}

We give some further details on the compatibility conditions appearing in Definition~3.1. These conditions ensure that the partition structures associated with the two pairs $(s,Q)$ and $(t,P)$ can be compared consistently. We first consider case $(a')$. Let
$P=P_1|\cdots|P_m,Q=Q_1|\cdots|Q_r,$ and suppose that $Q\preccurlyeq^a P$. By the definition of $\preccurlyeq^a$, $Q$ is obtained from $P$ by discarding exactly $m-r$ complete parties. Take an arbitrary $R=R_1|\cdots|R_t\in\mathcal P_t(P).$ After the discarded parties are removed, delete the empty blocks from $R$ and denote the resulting partition of $Q$ by $R_Q=T_1|\cdots|T_\ell.$
A block {\color{red} $T_j$} can disappear only if it contains at least one of the $m-r$ discarded parties. Hence at most $m-r$ blocks of $R$ can disappear, and $\ell\ge t-(m-r).$ Under the condition $t-s\ge m-r,$  we have $\ell\ge s$, so $R_Q$ can be coarsened to an $s$-partition of $Q$. For the converse, assume that $t-s<m-r.$ In this case, $m-r\ge t-s+1.$ Choose $t-s+1$ discarded parties and let each of them form a distinct block of a $t$-partition of $P$. Arrange the retained parties, together with any remaining discarded parties, into the other $s-1$ nonempty blocks. After all discarded parties are removed, at most $s-1$ nonempty blocks remain, which is insufficient to form an  $s$-partition of $Q$. The condition $t-s\ge m-r$ therefore characterizes the compatibility required in case $(a')$: every $t$-partition of $P$, after restriction to $Q$,  can be coarsened to an $s$-partition of $Q$. For case $(b')$, compatibility cannot be characterized by the number of blocks alone, since the blocks of $Q$ are obtained by grouping blocks of $P$. We impose the structural condition directly: for every $R\in\mathcal P_t(P)$, there exists $S\in\mathcal P_s(Q)$ such that $R$ refines $S$, equivalently, $ S\preccurlyeq^b R. $ For case $(c')$, $r=m$ and, after a suitable ordering, $Q_i\subseteq P_i$ for every $i$. Each $t$-partition of $P$ then induces a $t$-partition of $Q$. Since $s\le t$, the induced partition can always be coarsened to an $s$-partition of $Q$, and no additional compatibility condition is required.

For example, if $Q\preccurlyeq^aP$ and $(s,t)\in\{(r-1,m-1),(r,m)\},$ then $t-s=m-r$, so $(s,Q)\preccurlyeq^{a'}(t,P)$. {While $(t,Q)\preccurlyeq^{b'}(t,P)$ holds if and only if $Q=P$, and thus, if $Q\not =P$ and $(s,Q)\preccurlyeq^{b'}(t,P)$, then $s<t$}.

\section{Partition reduction---valid partitions}\label{SecComputationalDetails}
The number of $k$-partitions in $\mathcal P_n^{(k)}$ grows rapidly with $n$; for example, $\#(\mathcal P_n^2)=(n-1)n!$. Nevertheless, the maximization defining $g_n^{(k)}(L)$ need not be performed over all $k$-partitions. We say that two partitions $P=P_1|\cdots|P_k$ and $Q=Q_1|\cdots|Q_k$ are equivalent, denoted by $P\thicksim Q$, if their blocks differ only by a permutation: $Q_j=P_{\eta(j)}$ { as sets} for some permutation $\eta$ of $(1,\ldots,k)$. $\thicksim $ is an equivalence relation on ${\mathcal P}_n^k$, and $g(L,P)=g(L,Q)$ whenever $P\thicksim Q$. For each equivalence class $[P]\in\mathcal P_n^{(k)}/\thicksim$, {choose one representative $P^v=P^v_{[P]}$, called a valid $k$-partition, and define {
$$\mathcal P_n^{vk}=\{P^v_{[P]}:[P]\in\mathcal P_n^{(k)}/\sim\}.$$
be  the set by taking one representative element $P^v=P^v_{[P]}$ from each $[P]\in \mathcal P_n^{(k)}/\thicksim$. } Restricting the optimization to ${\mathcal P}_n^{vk}$ removes redundant evaluations without changing the maximum. This yields Eq.~(\ref{eq4.1}) from Eq.~(\ref{eq3.5}). For example, when $n=3$ and $k=2$,
$$\begin{array}{rl}
{\mathcal P}_3^{2}=\{1|23,12|3,1|32,13|2,2|13,21|3,
2|31,23|1,3|12,31|2,3|21,32|1\},
\end{array}$$
whereas ${\mathcal P}_3^{v2}=\{12|3,1|23,13|2\}.$}

\if false
{ To compute $E_w^{(k,n)}$, we first consider the case $k=n$, for which $E_w^{(n)}=E_w^{(n,n)}$.} For each sampled operator $L\in\mathcal B_1^+(H_1\otimes\cdots\otimes H_n)$, the iterative procedure of Appendix~\ref{SecAC}, motivated by Ref.~\cite{SV3}, numerically evaluates $g(L)$. A finite sample $\widetilde{\mathcal B}_1^+\subset\mathcal B_1^+$ then generates a database
$$\widetilde{\mathcal{EW}}_{ME}(H)=\{(L,g(L)):L\in\tilde{\mathcal B}_1^+\}.$$
Maximizing over this finite database gives $\tilde E_w^{(n)}(\rho)$ in Eq.~(\ref{eq4.2}),  a lower bound on $E_w^{(n)}(\rho)$: a positive value certifies multipartite entanglement, whereas a zero value is inconclusive. { Keeping existing entries  and adding new operators into the database will increase the value $\tilde E_w^{(n)}(\rho)$, which theoretically approaches the true value $E_w^{(n)}(\rho)$. The error of computing $\tilde E_w^{(n)}(\rho)$ mainly comes from   computational errors of $g(L)$ and ${\rm Tr}(L\rho)$. If the calculation error is controlled within $\delta$, then the error of $\tilde E_w^{(n)}(\rho)$ will not exceed $2\delta$ since both $\operatorname{Tr}(L\rho)$ and $g(L)$ are 1-Lipschitz in $L$. }

The same construction applies to every $2\leq k\leq n$ through Eq.~(\ref{eq3.8}):
$$E_w^{(k,n)}(\rho)=\sup_{L\in\mathcal B_1^+}\max\{0,{\rm Tr}(L\rho)-g_n^{(k)}(L)\}.$$

{ In fact, for each $L\in\tilde{\mathcal B}_1^+\subset\mathcal B_1^+$, we compute $\mathcal G_k(L)=\{g(L,P): P\in\mathcal P^{vk}\}$, $k=2,3,\ldots, n$, and
construct the database
\begin{equation}
\begin{aligned}
\widetilde{\mathcal{EW}}(H_1\otimes\cdots\otimes H_n)=\bigl\{[L,g(L),\mathcal G_{n-1}(L),\ldots,\mathcal G_2(L)]:L\in\widetilde{\mathcal B}_1^+\bigr\}.
\end{aligned}
\label{eqC1}
\end{equation}
Using the database, one obtains, for every $\rho\in\mathcal S(H)$,
\begin{equation}
\begin{aligned}
 \tilde{\mathcal{E}}^{(n)}_{w}(\rho)
= \bigl( \tilde{E}^{(n)}_{w}(\rho), \tilde{E}_{w}^{(n-1,n)}(\rho), \ldots, \tilde{E}_{w}^{(3,n)}(\rho), \tilde{E}_{w}^{(2,n)}(\rho) \bigr)
\end{aligned}
\label{eqC2}
\end{equation}
by
$$\tilde{E}_w^{(k,n)}(\rho)=\sup_{L\in\tilde{\mathcal B}_1^+}\max\{0,{\rm Tr}(L\rho)-g_n^{(k)}(L)\}$$
with $g_n^{(k)}(L)=\max\{\beta: \beta\in\mathcal G_k(L)\} $.}\fi

\section{Detailed evaluation formulas for $m$-qubit systems}\label{SecQubitDetails}

We now specialize the general formulas to an $m$-qubit Hilbert space $H=H_1\otimes\cdots\otimes H_m$, with $H_j=\mathbb C^2$. This restriction is made only to present the formulas used in the implementation; the framework itself allows arbitrary local dimensions. Because the same $m$-qubit system may be grouped into $n$ parties with $2\leq n\leq m$, we write $E_{w,m}^{(k,n)}$ to make both the number of qubits and the chosen multipartite structure explicit.

\subsection{$n$-partite $n$-qubit system}\label{Sec4.2.1}
When $m=n$, each party is a single qubit. Using the database $\mathcal{EW}^{(n)}$ defined in Eq.~(\ref{eq4.3}), the complete tuple of $k$-entanglement measures is
\begin{equation}
\begin{aligned}
 \mathcal{E}^{(n)}_{w,n}(\rho)
= \bigl( E^{(n)}_{w,n}(\rho), E_{w,n}^{(n-1,n)}(\rho), \ldots, E_{w,n}^{(3,n)}(\rho), E_{w,n}^{(2,n)}(\rho) \bigr)
\end{aligned}
\label{eq4.9}
\end{equation}
as in Eq.~(\ref{eq4.4}). Replacing $\mathcal{EW}^{(n)}$ by the finite subset $\widetilde{\mathcal{EW}}^{(n)}$ of Eq.~(\ref{eq4.6}) gives the corresponding numerical tuple
\begin{equation}
\begin{aligned}
 \tilde{\mathcal{E}}^{(n)}_{w,n}(\rho)
= \bigl( \tilde{E}^{(n)}_{w,n}(\rho), \tilde{E}_{w,n}^{(n-1,n)}(\rho), \ldots, \tilde{E}_{w,n}^{(3,n)}(\rho), \tilde{E}_{w,n}^{(2,n)}(\rho) \bigr)
\end{aligned}
\label{eq4.10}
\end{equation}
as specified in Eq.~(\ref{eq4.8}).

 \subsection{$n$-partite $m$-qubit systems}\label{Sec4.2.2}
Assume that the $m$-qubit database $\mathcal{EW}^{(m)}$ of Eq.~(\ref{eq4.3}) has been constructed. An $n$-partite grouping, with $2\leq n<m$, is specified by an $n$-partition $Q=Q_1|\cdots|Q_n$ of $1|2|\ldots|m$ and the associated system is  $H_Q=H_{Q_1}\otimes\cdots\otimes H_{Q_n}$. To evaluate the $k$-entanglement measure $E_{w,m,Q}^{(k,n)}$ for $2\leq k\leq n$ in system $H_Q$, choose the valid representative $P=P_1|\cdots|P_n\in\mathcal P_m^{vn}$ satisfying $P\thicksim Q$. The corresponding partition optimizations are invariant under this relabeling, and hence
\begin{equation}
E_{w,m,Q}^{(k,n)} = E_{w,m,P}^{(k,n)}.
\label{eq4.11}
\end{equation}
The required entries can therefore be read from $\mathcal{EW}^{(m)}$ using the representative $P$. A valid $k$-partition $R\in\mathcal P_m^{vk}$ is compatible with this grouping precisely when it is a $k$-partition of $P$, equivalently $R\preccurlyeq^b P$. Consequently, for every $\rho\in\mathcal S(H_{P_1}\otimes\cdots\otimes H_{P_n})$,
\begin{equation}
\begin{aligned}
 E_{w,m,P}^{(k,n)}(\rho)
= \max_{L\in\mathcal{B}_1^+} \max\Bigl\{0,\ {\rm Tr}(L\rho) - \max_{R\in\mathcal{P}_m^{vk},\ R\preccurlyeq^b P} g(L,R)\Bigr\}.
\end{aligned}
\label{eq4.12}
\end{equation}
Collecting these quantities for $k=n,n-1,\ldots,2$ gives
\begin{equation}
\begin{aligned}
\mathcal{E}^{(n)}_{w,m,P}(\rho) = \bigl( E^{(n,n)}_{w,m,P}(\rho), E_{w,m,P}^{(n-1,n)}(\rho), \ldots, E_{w,m,P}^{(3,n)}(\rho), E_{w,m,P}^{(2,n)}(\rho) \bigr),
\end{aligned}
\label{eq4.13}
\end{equation}
Replacing the full database by the finite subset $\widetilde{\mathcal{EW}}^{(m)}$ in Eq.~(\ref{eq4.6}) yields the corresponding numerical tuple
\begin{equation}
\begin{aligned}
\tilde{\mathcal{E}}^{(n)}_{w,m,P}(\rho) = \bigl( \tilde{E}^{(n,n)}_{w,m,P}(\rho), \tilde{E}_{w,m,P}^{(n-1,n)}(\rho),  \ldots, \tilde{E}_{w,m,P}^{(3,n)}(\rho), \tilde{E}_{w,m,P}^{(2,n)}(\rho) \bigr),
\end{aligned}
\label{eq4.14}
\end{equation}
The tuple $\tilde{\mathcal E}_{w,m,Q}^{(n)}(\rho)$ for the original grouping $Q$ is then obtained through Eq.~(\ref{eq4.11}).
\section{Detailed implementation for two to four qubits}\label{SecImplDetails}
This section specifies the implementation of Section~\ref{Sec4} for $m=2,3,4$ qubits. The construction proceeds in three stages.

\textbf{Step 1: Construct the finite databases.} For each $m$, sample a finite set $\widetilde{\mathcal B}_1^+\subset\mathcal B_1^+$ and evaluate the partition-dependent quantities required to form a finite subset $\widetilde{\mathcal{EW}}^{(m)}\subset\mathcal{EW}^{(m)}$. The corresponding full database structures are
\begin{equation}
\mathcal{EW}^{(2)} = \{ (L, g_2(L)) : L \in \mathcal{B}_1^+(\mathbb{C}^2 \otimes \mathbb{C}^2) \},
\label{eq5.1}
\end{equation}
\begin{equation}
\begin{aligned}
\mathcal{EW}^{(3)} = \{  (L, g_3(L, 1|2|3), g_3(L, 12|3), g_3(L, 1|23),  g_3(L, 13|2)) : L \in \mathcal{B}_1^+(\mathbb{C}^2 \otimes \mathbb{C}^2 \otimes \mathbb{C}^2) \},
\end{aligned}
\label{eq5.2}
\end{equation}
and
\begin{equation}
\begin{aligned}
\mathcal{EW}^{(4)}=\bigl\{(L,g_4(L,1|2|3|4),
\{g_4(L,P):P\in\mathcal P_4^{v3}\},
\{g_4(L,P):P\in\mathcal P_4^{v2}\}):
L\in\mathcal B_1^+((\mathbb C^2)^{\otimes4})\bigr\}.
\end{aligned}
\label{eq5.3}
\end{equation}
%and
%$$\begin{array}{rl}\mathcal{EW}^{(5)}_{(0,1)}= & \{(L, g_5(L, 1|2|3|4|5),\{g_5(L, P): P\in{\mathcal P}_4^{v4}\}; \{g_5(L, P): P\in{\mathcal P}_4^{v3}\} ,\\
%& \{g_5(L, P): P\in{\mathcal P}_4^{v2}\})  :   L\in {\mathcal B}_1^+(\mathbb C^2\otimes\mathbb C^2\otimes \mathbb C^2\otimes\mathbb C^2\otimes\mathbb C^2)\}
% \end{array} \eqno(5.4)
%$$
where
\begin{equation}
\begin{aligned}
\mathcal P_4^{v2}=\{1|234,\ 2|134,\ 3|124,\ 123|4,12|34,\ 13|24,\ 14|23\}.
\end{aligned}
\label{eq5.4}
\end{equation}
and
\begin{equation}
\begin{aligned}
\mathcal P_4^{v3}=\{12|3|4,\ 1|23|4,\ 1|2|34,13|2|4,\ 14|2|3,\ 1|3|24\}.
\end{aligned}
\label{eq5.5}
\end{equation}
%and
%$$\begin{array}{rl}{\mathcal P}_5^{v2}=&\{1|2345, 2|1345, 3|1245, 4|1235, 1234|5, 12|345, 13|245, 14|235\\
%& 15|234, 23|145,24|135, 25|134, 34|125, 35|124, 45|123\},
%\end{array}\eqno(5.7)$$
%$$\ldots\ldots$$

The algorithms used to compute the parameters $g_m(L,P)$ are presented in Appendix~\ref{SecAC}.
The positive operators $L$ in $\widetilde{\mathcal{EW}}^{(m)}$ are sampled { randomly}. \if false according to the distribution specified by the implementation; no uniform measure on $\mathcal B_1^+$ is assumed.\fi The stored values of $g_m(L,P)$ are numerical approximations accompanied by the prescribed optimization tolerance.

\textbf{Step 2: Evaluate the measures.} For every $m$-qubit state $\rho$, the $k$-entanglement measure of $m$-partite system is computed by
\begin{equation}
\begin{aligned}
 \tilde{E}_{w,m}^{(k,m)}(\rho)
= \max_{L\in\tilde{\mathcal B}_1^+} \max\bigl\{0,\ {\rm Tr}(L\rho) - \max_{P\in\mathcal{P}_m^{vk}} g_m(L,P)\bigr\}.
\end{aligned}
\label{eq5.6}
\end{equation}
For each $n$-partition $Q=Q_1|Q_2|\ldots|Q_n$ of $1|2|\ldots|m$, { take $P\in \mathcal P_m^{vn}$ such that $P\thicksim Q$, and then the $k$-entanglement measure with respect to $n$-partite system $H_Q$ is computed by}
\begin{equation}
\begin{aligned}
\tilde{E}_{w,m,Q}^{(k,n)}(\rho) = \max_{L\in\tilde{\mathcal{B}}_1^+} \max\bigl\{0,\ {\rm Tr}(L\rho) - \max_{R\in\mathcal{P}_m^{vk},\ R\preccurlyeq^b P} g_m(L,R) \bigr\}.
\end{aligned}
\label{eq5.7}
\end{equation}

The cases $m=2,3,4$ are detailed below.

(1) In the case $m=2$,  we only need to compute the entanglement
measure using the database $\widetilde{\mathcal{EW}}^{(2)}$:
$$\begin{array}{rl} \tilde{E}_{w,2}(\rho)= \tilde{E}_{w,2}^{(2,2)}(\rho)
=  \max_{L\in\tilde{\mathcal B}^+_1 (\mathbb C^2\otimes\mathbb C^2)}\max\{0,\ {\rm Tr}(L\rho)-g_2(L)\}.
\end{array}$$

(2) In the case $m=3$, we use the database $\widetilde{\mathcal{EW}}^{(3)}$ to compute the ME measure for the natural tripartite structure
 $$\begin{array}{rl}  \tilde{E}_{w,3}^{(3)}(\rho)= \tilde{E}_{w,3}^{(3,3)}(\rho) = \max_{L\in\tilde{\mathcal B}^+_1(\mathbb C^2\otimes\mathbb C^2\otimes \mathbb C^2)}\max\{0,\ {\rm Tr}(L\rho)-g_3(L,1|2|3)\},\end{array}
$$
and the  GME measure for the tripartite system
$$ \tilde{E}_{w,3}^{(2,3)}(\rho)=\max_{L\in\tilde{\mathcal B}^+_1(\mathbb C^2\otimes\mathbb C^2\otimes \mathbb C^2)}\max\{0,\ {\rm Tr}(L\rho)-g_3^{(2)}(L)\},
$$
with $g_3^{(2)}(L)=\max\{g_3(L, 12|3), g_3(L, 1|23), g_3(L, 13|2)\} $.

When the 3-qubit system is regarded as bipartite, we illustrate  two cases, that is, $(\mathbb C^2\otimes\mathbb C^2)\otimes\mathbb C^2$ and $\mathbb C^2\otimes(\mathbb C^2\otimes\mathbb C^2)$.  The entanglement measures  are respectively
$$\begin{array}{rl}
\tilde{E}_{w,3,12|3}^{(2)}(\rho)=\tilde{E}_{w,3,12|3}^{(2,2)}(\rho)=\max_{L\in\tilde{\mathcal B}_1^+(\mathbb C^2\otimes\mathbb C^2\otimes \mathbb C^2)} \max\{0,\  {\rm Tr}(L\rho)-g_3 (L, 12|3)\},
\end{array}$$
and
$$
\begin{array}{rl} \tilde{E}_{w,3,1|23}^{(2)}(\rho)=\tilde{E}_{w,3,1|23}^{(2,2)}(\rho) =\max_{L\in\tilde{\mathcal B}_1^+(\mathbb C^2\otimes\mathbb C^2\otimes \mathbb C^2)} \max\{0,\  {\rm Tr}(L\rho)-g_3 (L, 1|23)\}.\end{array}
$$

 (3) In the case $m=4$, use the database $\widetilde{\mathcal{EW}}^{(4)}$, and we have the following cases:

 (3.1) When the four-qubit system is regarded as four-partite, we compute the ME measure by
$$\begin{array}{rl}  \tilde{E}_{w,4}^{(4)}(\rho)=\tilde{E}_{w,4}^{(4,4)}(\rho)= \max_{L\in\tilde{\mathcal B}_1^+(\mathbb C^2\otimes\mathbb C^2\otimes \mathbb C^2\otimes\mathbb C^2)} \max\{0,\  {\rm Tr}(L\rho)-g_4(L,1|2|3|4)\},\end{array}
$$
compute the 3-entanglement measure by
$$\begin{array}{rl}  \tilde{E}_{w,4}^{(3,4)}(\rho) =  \max_{L\in\tilde{\mathcal B}_1^+(\mathbb C^2\otimes\mathbb C^2\otimes \mathbb C^2\otimes\mathbb C^2)} \max\{0,\ {\rm Tr}(L\rho)-\max_{P\in{\mathcal P}_4^{v3}}g_4(L,P)\}\end{array}
$$
with ${\mathcal P}_4^{v3}$ as presented in Eq.(\ref{eq5.5}),
and compute the GME measure by
$$\begin{array}{rl}  \tilde{E}_{w,4}^{(2,4)}(\rho) = \max_{L\in\tilde{\mathcal B}_1^+(\mathbb C^2\otimes\mathbb C^2\otimes \mathbb C^2\otimes\mathbb C^2)} \max\{0,\  {\rm Tr}(L\rho)-\max_{P\in{\mathcal P}_4^{v2}}g_4(L,P)\} \end{array}
$$
with ${\mathcal P}_4^{v2}$ as presented in Eq.(\ref{eq5.4}).

(3.2) When the four-qubit system is regarded as tripartite, we consider every valid grouping $P\in\mathcal P_4^{v3}$ and compute the ME measure as
$$\begin{array}{rl}
\tilde{E}_{w,4,P}^{(3,3)}(\rho)= \max_{L\in\tilde{\mathcal B}_1^+(\mathbb C^2\otimes\mathbb C^2\otimes \mathbb C^2\otimes\mathbb C^2)} \max\{0,\  {\rm Tr}(L\rho)-g_4 (L, P)\},\end{array}
$$
the GME measure as
$$\begin{array}{rl}
\tilde{E}_{w,4,P}^{(2,3)}(\rho)= \max_{L\in\tilde{\mathcal B}_1^+(\mathbb C^2\otimes\mathbb C^2\otimes \mathbb C^2\otimes\mathbb C^2)} \max\{0,\   {\rm Tr}(L\rho)-\max_{R\in{\mathcal P}_4^{v2}, R\preccurlyeq^b P}g_4 (L, R)\}.\end{array}
$$
For example, if $P=12|3|4$, then $R\in\mathcal P_4^{v2}$ and $R\preccurlyeq^bP$ if and only if $R\in\{12|34,123|4,124|3\}$. Hence
\begin{equation*}
\begin{aligned}
\widetilde E_{w,4,12|3|4}^{(2,3)}(\rho)
 &=\max_{L\in\widetilde{\mathcal B}_1^+}
   \max\{0,{\rm Tr}(L\rho)-G_{12|3|4}(L)\},\\
G_{12|3|4}(L)
 &=\max\bigl\{g_4(L,12|34),g_4(L,123|4),g_4(L,124|3)\bigr\}.
\end{aligned}
\end{equation*}

(3.3) When the four-qubit system is regarded as bipartite, the relevant groupings are all $P\in\mathcal P_4^{v2}$, and the implemented entanglement measure is
\begin{equation*}
\begin{aligned}
\widetilde E_{w,4,P}^{(2)}(\rho)
=\widetilde E_{w,4,P}^{(2,2)}(\rho)=\max_{L\in\widetilde{\mathcal B}_1^+}
   \max\{0,{\rm Tr}(L\rho)-g_4(L,P)\}.
\end{aligned}
\end{equation*}

%The case of $n=5$ is dealt with similarly.

\textbf{Step 3: Combine the above databases and computation into a software  for evaluating the ME, GME and $k$-entanglement measures.}

The algorithms employed in constructing the above databases are provided  in Appendix \ref{SecAC}.

\section{Supplementary numerical figures}\label{SecNumericalDetails}
\setcounter{figure}{0}
\renewcommand{\thefigure}{F\arabic{figure}}
\renewcommand{\theHfigure}{F.\arabic{figure}}

This appendix collects the curves supporting Examples~6.2--6.5. The physical systems and the hierarchy levels being tested are stated in Section~\ref{SecRepresentativeExamples} and are not repeated here.

\begin{figure}[htpb]
\centering
\includegraphics[width=0.58\textwidth,height=0.22\textheight]{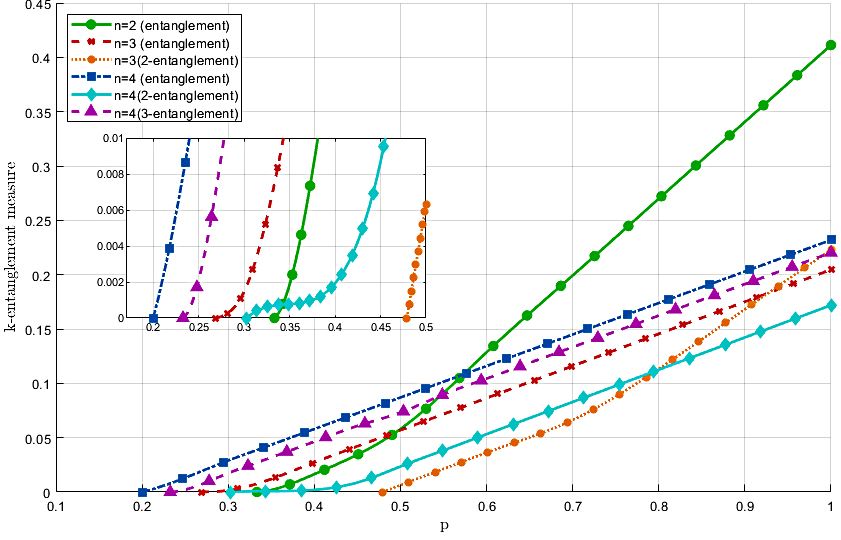}
\caption{The measures $\tilde E_{w,n}^{(k,n)}$ for the noisy W families of Example~6.2 with $n=2,3,4$.}
\label{fig2}
\end{figure}

\begin{figure}[htpb]
\centering
\includegraphics[width=0.58\textwidth,height=0.22\textheight]{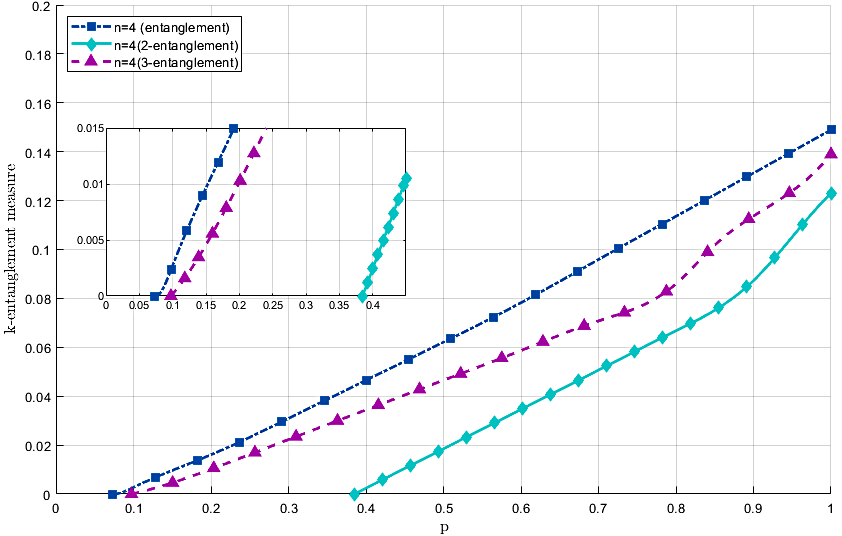}
\caption{The measures $\tilde E_{w,4}^{(k,4)}$, $k=2,3,4$, for the noisy four-qubit linear-cluster state of Example~6.3.}
\label{fig3}
\end{figure}

\begin{figure}[htpb]
\centering
\includegraphics[width=0.58\textwidth,height=0.22\textheight]{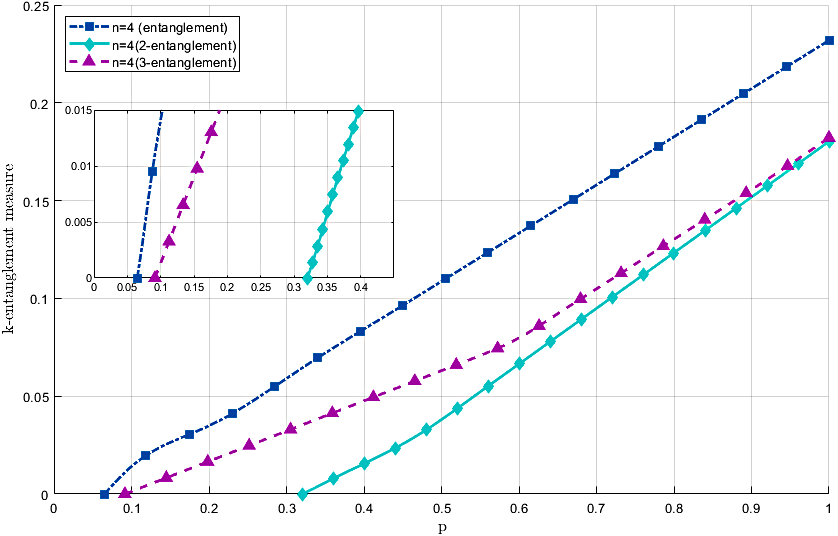}
\caption{The measures $\tilde E_{w,4}^{(k,4)}$, $k=2,3,4$, for the noisy two-excitation Dicke state of Example~6.4.}
\label{fig4}
\end{figure}

\begin{figure}[htpb]
\centering
\includegraphics[width=0.58\textwidth,height=0.22\textheight]{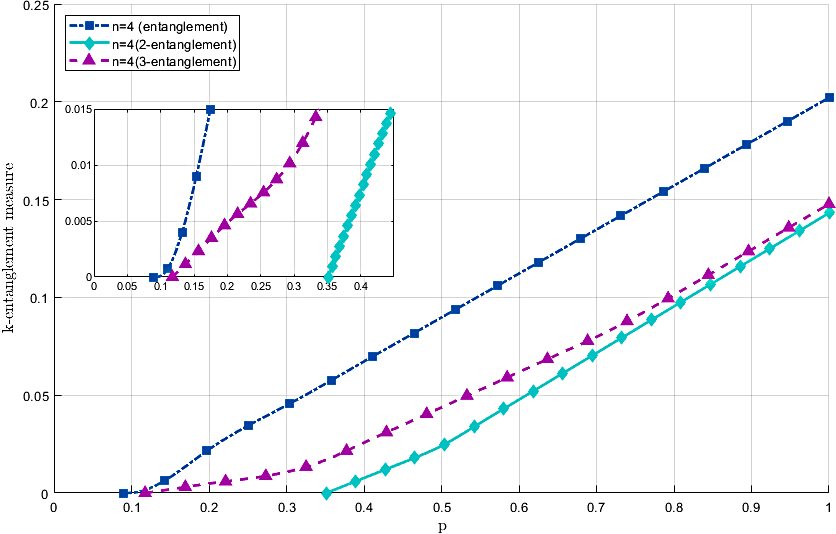}
\caption{The measures $\tilde E_{w,4}^{(k,4)}$, $k=2,3,4$, for the noisy four-qubit singlet-type state of Example~6.5.}
\label{fig5}
\end{figure}

\clearpage

\section{Proofs of the results in Section~\ref{Sec2}} \label{SecAA}

\subsection{Proof of Theorem  \ref{T2.2}.} \label{SecAA1}
%{\bf A1. Proof of Theorem  \ref{T2.2}.}
{\bf Proof.} Note that, for any quantum channel $\Phi:{\mathcal T}(H)\to{\mathcal T}(H)$ (i.e., completely positive linear map preserving trace, or briefly, CPTP map), $\Phi^\dagger: {\mathcal B}(H)\to{\mathcal B}(H)$ is completely positive and unital, and thus  $\|\Phi\|=1$. If $\Phi$ is a LOCC, then $\Phi$ transforms (fully) separable states into separable ones.
For any $W=\lambda I -L\in\mathcal{EW}_1$ and any separable state $\sigma$, we have
$${\rm Tr}(\Phi^\dagger(W)\sigma)={\rm Tr}(W\Phi(\sigma)) \geq 0
$$
and $\Phi^\dagger(W)=\lambda I-\Phi^\dagger(L)$ as $\Phi^\dagger(I)=I$. Consequently, as $0\leq \Phi^\dagger(L)\leq I$, $\Phi^\dagger(W)\in\mathcal{EW}_1$ whenever $W\in\mathcal{EW}_1$.

Now let $\rho\in\mathcal S(H_1\otimes\cdots\otimes H_n)$. If $\rho$ is separable, then
\begin{equation*}
E_w^{(n)}(\Phi(\rho))=0=E_w^{(n)}(\rho).
\end{equation*}
If $\rho$ is entangled, then
\begin{equation*}
\begin{aligned}
E_w^{(n)}(\Phi(\rho))
&=\sup_{W\in\mathcal{EW}_1}
\bigl|\min\{{\rm Tr}[W\Phi(\rho)],0\}\bigr|\\
&=\sup_{W\in\mathcal{EW}_1}
\bigl|\min\{{\rm Tr}[\Phi^\dagger(W)\rho],0\}\bigr|\\
&\leq\sup_{W'\in\mathcal{EW}_1}
\bigl|\min\{{\rm Tr}(W'\rho),0\}\bigr|
=E_w^{(n)}(\rho).
\end{aligned}
\end{equation*}
Hence,
$$E^{(n)}_w(\Phi(\rho))\leq E^{(n)}_w(\rho)$$
holds for all states $\rho$. \hfill$\Box$

\subsection{Proof of Theorem  \ref{T2.3}.} \label{SecAA2}
%{\bf A2. Proof of Theorem \ref{T2.3}.}
{\bf Proof.} Assume $\rho=\sum_{j=1}^k p_j\rho_j$. By the definition of $E^{(n)}_w$ in Eq.(\ref{eq2.4}),
 $$\begin{array}{rl}
 E^{(n)}_w(\rho)=& \sup_{W\in\mathcal{EW}_1} |\min\{\sum_j p_j{\rm Tr}(W\rho_j),0\}|\\
 \leq & \sup_{W\in\mathcal{EW}_1} \sum_j p_j|\min\{{\rm Tr}(W\rho_j),0\}|\\
 \leq & \sum_j p_j\sup_{W\in\mathcal{EW}_1} |\min\{{\rm Tr}(W\rho_j),0\}|\\
 = & \sum_{j=1}^kp_jE^{(n)}_w(\rho_j).
\end{array}$$

\hfill$\Box$

\subsection{Proof of Theorem  \ref{T2.4}.} \label{SecAA3}
%{\bf A3. Proof of Theorem \ref{T2.4}.}
{\bf Proof.} Let $ \mathcal{EW}_{(1,0)}=\{ (L, g(L)): L\in\mathcal{B}_1^+\}.$
Then $\{ g(L)-L: (L, g(L))\in{\mathcal{EW}}_{(1,0)}\} $ is a subset of $\mathcal{EW}_1$. It follows that, for any state $\rho$,
$$\begin{array}{rl} & \sup_{L\in\mathcal{B}_1^+} \max\{0, {\rm Tr}(L\rho)-g(L)\}  = \sup_{(L,g(L))\in\mathcal{EW}_{(1,0)}}\max\{0, -{\rm Tr}((g(L) I-L)\rho)\}
\leq  E^{(n)}_w(\rho). \end{array}
$$

If $E^{(n)}_w(\rho)>0$, for any $\varepsilon>0$ small enough, by Theorem  \ref{T2.1}, there exists $W=\lambda I-L\in\mathcal{EW}_1$ such that
$$ \begin{array}{rl}  0< E^{(n)}_w(\rho)-\varepsilon\leq -{\rm Tr}(W\rho) = &{\rm Tr}(L\rho)-\lambda
\leq  {\rm Tr}(L\rho)-g(L)\leq E^{(n)}_w(\rho) \end{array}
$$
as $g(L)\leq \lambda$, which gives
$$E^{(n)}_w(\rho)-\varepsilon\leq \sup_{L\in\mathcal{B}_1^+} \max\{0, {\rm Tr}(L\rho)-g(L)\}\leq E^{(n)}_w(\rho).$$
Let $\varepsilon\to 0$, we get the equality.\hfill$\Box$

\subsection{Proof of Theorem  \ref{T2.5}.} \label{SecAA4}
%{\bf A4. Proof of Theorem  \ref{T2.5}.}
To begin with, we present the following Lemma  \ref{LA4.1}.

\begin{lemma}\label{LA4.1}  {\it Let $L\in{\mathcal B}^+(H_1\otimes H_2\otimes\cdots\otimes H_n)$ be a state. Then}
{\it \begin{equation*}
\begin{aligned}
g_{\max}({\rm Tr}_n(L)) = \sup\bigl\{ {\rm Tr}[(\eta_1\otimes \eta_2\otimes\cdots\otimes \eta_{n-1}\otimes I_{H_n})L] \qquad : \eta_j\in {\mathcal Pur}(H_j), j=1,2,\ldots, n-1 \bigr\},
\end{aligned}
\end{equation*}
}
{\it and}
$$ g_{\rm max}(L)\leq g_{\rm max}({\rm Tr}_n(L))\leq d_n g_{\rm max}(L),
$$
{where $d_n=\dim H_n$.}
\end{lemma}

{\bf Proof.} As $\eta_1\otimes \eta_2\otimes\cdots\otimes \eta_n\leq \eta_1\otimes \eta_2\otimes\cdots\otimes \eta_{n-1}\otimes I_n$, where $I_n$ is the identity of $H_n$, we have

\begin{align*}
g_{\max}(L) &= \sup\bigl\{ {\rm Tr}[(\eta_1\otimes \eta_2\otimes\cdots\otimes \eta_n)L] :\eta_j\in{\mathcal Pur}(H_j), j=1,2,\ldots, n \bigr\} \\
&\leq \sup\bigl\{ {\rm Tr}[(\eta_1\otimes \eta_2\otimes\cdots\otimes \eta_{n-1}\otimes I_{H_n})L] :\eta_j\in{\mathcal Pur}(H_j), j=1,2,\ldots, n-1 \bigr\} \\
&= \sup\bigl\{ {\rm Tr}[(\eta_1\otimes \eta_2\otimes\cdots\otimes \eta_{n-1}){\rm Tr}_n(L)] : \eta_j\in{\mathcal Pur}(H_j), j=1,2,\ldots, n-1 \bigr\} \\
&= g_{\max}({\rm Tr}_n(L)).
\end{align*}

So, $g_{\rm max}({\rm Tr}_n(L))=\sup\{{\rm Tr}[(\eta_1\otimes \eta_2\otimes\cdots\otimes \eta_{n-1}\otimes I_{H_n})L]: \eta_j\in{\mathcal Pur}(H_j), j=1,2,\ldots, n-1\}
$ and $ g_{\rm max}(L)\leq g_{\max}({\rm Tr}_n(L))$.

To check the  inequality $g_{\max}({\rm Tr}_n(L))\leq d_ng_{\max}(L)$, write $I_n=\sum_{r=1}^{d_n} \omega_r$, where $\{\omega_r\}_{r=1}^{d_n}\subset{\mathcal Pur}(H_n)$ is an orthogonal set of pure states. Without loss of generality, we may assume $d_n=\dim H_n<\infty$. Then
%\begin{widetext}
\begin{align*}
g_{\max}({\rm Tr}_n(L))
&= \sup\bigl\{ {\rm Tr}[(\eta_1\otimes \eta_2\otimes\cdots\otimes \eta_{n-1} \otimes (\sum_{r=1}^{d_n} \omega_r))L] :  \eta_j\in{\mathcal Pur}(H_j), j=1,2,\ldots, n-1 \bigr\}\\
&= d_n \sup\bigl\{ {\rm Tr}[(\sum_{r=1}^{d_n}\frac{1}{d_n}(\eta_1\otimes \eta_2\otimes\cdots\otimes \eta_{n-1}\otimes  \omega_r))L] :  \eta_j\in{\mathcal Pur}(H_j), j=1,2,\ldots, n-1 \bigr\}\\
&\leq d_n \max \bigl\{ {\rm Tr}((\eta_1\otimes \eta_2\otimes\cdots\otimes \eta_{n-1}\otimes \eta_n)L) :  \eta_j\in{\mathcal Pur}(H_j), j=1,2,\ldots, n \bigr\}\\
&= d_n g_{\max}(L).
\end{align*}
%\end{widetext}
Hence, the proposition is true.\hfill$\Box$

Note that, for  operators $A\in\mathcal B(H)$, $B\in\mathcal B(K)$, and $C\in \mathcal B(H\otimes K)$, the following trace
identities are true:
 $$
{\rm Tr}\!\left[(A\otimes B)C\right]
= {\rm Tr}_1\!\left[A\,{\rm Tr}_2\!\left((I\otimes B)C\right)\right],
 $$
  $$
{\rm Tr}\!\left[C(A\otimes B)\right]
= {\rm Tr}_1\!\left[{\rm Tr}_2\!\left(C(I\otimes B)\right)A\right].
 $$
Consequently,
 $$
{\rm Tr}_1\!\left[A\,{\rm Tr}_2\!\left((I\otimes B)C\right)\right]
=
{\rm Tr}_1\!\left[{\rm Tr}_2\!\left(C(I\otimes B)\right)A\right].
 $$

Now, we are in a position to provide a proof of Theorem  \ref{T2.5}.

{\bf Proof.}
Let $H=H_1\otimes H_2\otimes \cdots\otimes H_n$ and $K=K_1\otimes K_2\otimes\cdots\otimes K_m$.

Given $L\in{\mathcal B}_1^+(H\otimes K)$, for any $\rho_1\in{\mathcal S}(H)$ and $\rho_2\in{\mathcal S}(K)$, we have
$${\rm Tr}(L(\rho_1\otimes\rho_2))={\rm Tr}_1[{\rm Tr}_2(L(I\otimes \rho_2))\rho_1]\leq 1.
$$
Fixing $\rho_2$ and letting $\rho_1$ runs over all of ${\mathcal S}(H)$, the above equation ensures that ${\rm Tr}_K(L(I_H\otimes \rho_2))\in {\mathcal B}_1^+(H)$ holds for every $\rho_2\in{\mathcal S}(K)$. Symmetrically, ${\rm Tr}_H(L(\rho_1\otimes I_K))\in {\mathcal B}_1^+(K)$ holds for every $\rho_1\in{\mathcal S}(H)$. These observations will be used frequently below.

 Thus, for vector states $|\phi_j\rangle\in H_j$, $j=1,,\ldots,n$, and $|\psi_i\rangle\in K_i$, $i=1,2,\ldots,m$,

 %\begin{widetext}
$$\begin{array}{rl} & {\rm Tr}[L(|\phi_1\rangle\langle\phi_1|\otimes\cdots \otimes|\phi_n\rangle\langle\phi_n|\otimes  |\psi_1\rangle\langle\psi_1|\otimes\cdots \otimes|\psi_m\rangle\langle\psi_m|)]\\
= & {\rm Tr}_H[{\rm Tr}_K(L(I_H\otimes ( |\psi_1\rangle\langle\psi_1|\otimes\cdots \otimes|\psi_m\rangle\langle\psi_m|))(|\phi_1\rangle\langle\phi_1|\otimes\cdots \otimes|\phi_n\rangle\langle\phi_n|))],
\end{array}$$
which gives  that
$$\begin{array}{rl}
g(L)=& \sup_{|\phi_j\rangle\in H_j, |\psi_i\rangle\in K_i} {\rm Tr}[L(|\phi_1\rangle\langle\phi_1|\otimes\cdots \otimes|\phi_n\rangle\langle\phi_n|\otimes  |\psi_1\rangle\langle\psi_1|\otimes\cdots \otimes|\psi_m\rangle\langle\psi_m|)]\\
=& \sup_{ |\psi_i\rangle\in K_i} g({\rm Tr}_K(L(I_H\otimes ( |\psi_1\rangle\langle\psi_1|\otimes\cdots \otimes|\psi_m\rangle\langle\psi_m|))).
\end{array}$$
%\end{widetext}
Symmetrically,
$$g(L)=\sup_{|\phi_j\rangle\in H_j}g({\rm Tr}_H(L((|\phi_1\rangle\langle\phi_1|\otimes\cdots\otimes|\phi_n\rangle\langle\phi_n|)\otimes I_K))).
$$
Then, by applying Lemma \ref{LA4.1}, we have
%\begin{widetext}
\begin{align*}
&\operatorname{Tr}(L(\rho_1\otimes \rho_2))-g(L) \\
= &\operatorname{Tr}_H[\operatorname{Tr}_K(L(I\otimes\rho_2))\rho_1] - g[\operatorname{Tr}_K(L(I\otimes\rho_2))]+ \sup_{|\phi_j\rangle\in H_j} \operatorname{Tr}_H[\operatorname{Tr}_K(L(I\otimes\rho_2))(|\phi_1\rangle\langle\phi_1|\otimes\cdots\otimes|\phi_n\rangle\langle\phi_n|)] - g(L) \\
\leq &\operatorname{Tr}_H[\operatorname{Tr}_K(L(I\otimes\rho_2))\rho_1] - g[\operatorname{Tr}_K(L(I\otimes\rho_2))] + \sup_{|\phi_j\rangle\in H_j} \Bigl\{ \operatorname{Tr}_K\bigl[ \operatorname{Tr}_H\bigl( L((|\phi_1\rangle\langle\phi_1|\otimes\cdots\otimes|\phi_n\rangle\langle\phi_n|)\otimes I ) \rho_2 \bigr) \bigr] \\
   &\qquad - g\bigl( \operatorname{Tr}_H( L((|\phi_1\rangle\langle\phi_1|\otimes\cdots\otimes|\phi_n\rangle\langle\phi_n|)\otimes I ) ) \bigr) \Bigr\} \\
\leq & \sup_{C\in\mathcal{B}_1^+(H)} \{ \operatorname{Tr}_H(C\rho_1) - g(C) \} + \sup_{D\in\mathcal{B}_1^+(K)} \{ \operatorname{Tr}_K(D\rho_2) - g(D) \}.
\end{align*}
%\end{widetext}
Therefore, we achieve that
$$\begin{array}{rl}
 E_w^{(n+m)}(\rho_1\otimes\rho_2)  =& \sup_{L\in{\mathcal B}_1^+(H\otimes K)}\max\{0, {\rm Tr}(L(\rho_1\otimes \rho_2))-g(L)\} \\\leq  &  \sup_{C\in{\mathcal B}_1^+(H)}\max\{0,{\rm Tr}_H(C\rho_1)-g(C)\} +\sup_{D\in{\mathcal B}_1^+(K)}\max\{0,{\rm Tr}_K(D\rho_2)-g(D)\} \\
= & E_w^{(n)}(\rho_1)+E_w^{(m)}(\rho_2).
\end{array}$$
This completes the proof. \hfill$\Box$

\subsection{Proof of Theorem  \ref{T2.7}.} \label{SecAA5}
%{\bf A5. Proof of Theorem 2.7.}
{\bf Proof.} We first consider the special cases $\preccurlyeq^x$, $x\in\{a,b,c\}$.

(a). For the case $\preccurlyeq^a$, without loss of generality, it is sufficient to prove that, for any $\rho_{1,2,\ldots,n}\in{\mathcal S}(H_1\otimes H_2\otimes\cdots\otimes H_n)$ and any subset $\{j_1,j_2,\ldots,j_r\}\subseteq\{1,2,\ldots,n\}$, we have $$E_w^{(r)}(\rho_{j_1,\ldots, j_r})\leq E_w^{(n)}(\rho_{1,2,\ldots,n}),$$
 where $\rho_{j_1,\ldots, j_r}={\rm Tr}_{\{j_1,j_2,\ldots,j_r\}^c}(\rho_{1,2,\ldots,n})$ is the reduced state of $\rho_{1,2,\ldots,n}$ to the subsystem $j_1|j_2|\ldots|j_r$. We can also assume that
 $j_1|j_2|\ldots|j_r=1|2|\ldots|r$ with $r<n$. Note that, for any positive operator $D$, we have $g(D\otimes I)=g(D)$. By the definition,
 %\begin{widetext}
$$\begin{array}{rl}
E_w^{(r)}(\rho_{1,2,\ldots,r})=&\sup_{D\in{\mathcal B}_1^+(H_1\otimes H_2\otimes\cdots\otimes H_r)}\max\{0,{\rm Tr}(D\rho_{1,2,\ldots,r})-g(D)\}\\
=& \sup_{D\otimes I\in{\mathcal B}_1^+(H_1\otimes H_2\otimes\cdots\otimes H_n)}\max\{0,{\rm Tr}((D\otimes I)\rho_{1,2,\ldots,n})-g(D\otimes I)\}\\
\leq & \sup_{L\in{\mathcal B}_1^+(H_1\otimes H_2\otimes\cdots\otimes H_n)}\max\{0,{\rm Tr}(L\rho_{1,2,\ldots,n})-g(L)\}
= E_w^{(n)}(\rho_{1,2,\ldots,n}),
\end{array}$$
%\end{widetext}
as desired.

(b). For the case $\preccurlyeq^b$, it suffices to check that, for any $r$-partition $Q=Q_1|Q_2|\ldots|Q_r$ of $1|2|\ldots|n$ and for any $\rho=\rho_{1,2,\ldots,n}\in{\mathcal S}(H_1\otimes H_2\otimes\cdots\otimes H_n)$,   regarding also $\rho\in{\mathcal S}(H_{Q_1}\otimes H_{Q_2}\otimes\cdots\otimes H_{Q_r})$ as an $r$-partite state, we have
$$ E_w^{(r)}(\rho,Q)\leq E_w^{(n)}(\rho).$$
Note that, in this case, $\rho_Q=\rho$.

 Observe that, for any $L\in{\mathcal B}_1^+(H_1\otimes H_2\otimes\cdots\otimes H_n)={\mathcal B}_1^+(H_{Q_1}\otimes H_{Q_2}\otimes\cdots\otimes H_{Q_r})$,
 %\begin{widetext}
$$\begin{array}{rl}
& g(L,Q) \\ = & \sup\{{\rm Tr}[L(|\omega_1\rangle\langle\omega_1|\otimes|\omega_2\rangle\langle\omega_2|\otimes\cdots \otimes |\omega_r\rangle\langle\omega_r|)] : |\omega_j\rangle\in H_{Q_j}, \langle\omega_j|\omega_j\rangle=1, j=1,2,\ldots, r\}\\
\geq & \sup\{{\rm Tr}[L(|\phi_1\rangle\langle\phi_1|\otimes|\phi_2\rangle\langle\phi_2|\otimes\cdots \otimes |\phi_n\rangle\langle\phi_n|)] : |\phi_i\rangle\in H_{i}, \langle\phi_i|\phi_i\rangle=1, j=1,2,\ldots, n\}\\
= & g(L).
\end{array}
$$
% \end{widetext}
So, we have
$$\begin{array}{rl}
 E_w^{(r)}(\rho,Q)=& \sup_{L\in{\mathcal B}_1^+}\max\{0, {\rm Tr}(L\rho)-g(L,Q)\}\\
 \leq &  \sup_{L\in{\mathcal B}_1^+}\max\{0, {\rm Tr}(L\rho)-g(L)\}\\
 =& E_w^{(n)}(\rho).
\end{array}$$
This completes the proof of case $\preccurlyeq^b$.

(c). Assume $Q\preccurlyeq^c P$. Then we have $r=m$.  Without loss of generality, suppose $P$ is a $m$-partition of $\{1|2|\ldots|n\}$, that is, $H_{P_1}\otimes H_{P_2}\otimes\cdots\otimes H_{P_m}=H= H_1\otimes H_2\otimes\cdots\otimes H_n$, and we may assume that $Q_i\subseteq P_i$ for each $i=1,2,\ldots,m$.  For any $\rho=\rho_{1,2,\ldots,n}\in{\mathcal S}(H_1\otimes H_2\otimes\cdots\otimes H_n)$,  we also regard $\rho_P=\rho$   and $\rho_Q={\rm Tr}_{Q^c}(\rho)$ as   $m$-partite states in $H_P$ and $H_Q$, respectively. We have to check
$$ E_w^{(m)}(\rho_Q,Q)\leq E_w^{(m)}(\rho,P).$$

 Denote $Q_j^{c_j}=P_j\setminus Q_j$. For any $D\in{\mathcal B}_1^+(Q)={\mathcal B}_1^+(H_{Q_1}\otimes H_{Q_2}\otimes\cdots\otimes H_{Q_m})$,
 %\begin{widetext}
$$\begin{array}{rl}
g (D,Q)=&\sup\{{\rm Tr}(D(\eta_1\otimes \eta_2\otimes\cdots\otimes \eta_m)): \eta_j\in{\mathcal Pur}(H_{Q_j}), j=1,2,\ldots,m\}\\
= & \sup\{{\rm Tr}[(D\otimes I_{Q^c})((\eta_1\otimes\xi_1)\otimes (\eta_2\otimes\xi_2)\otimes\cdots\otimes (\eta_m\otimes\xi_m))]: \eta_j\in{\mathcal Pur}(H_{Q_j}), \xi_j\in{\mathcal Pur}(H_{Q_j^{c_j}}), j=1,2,\ldots,m\}\\
\leq & \sup\{{\rm Tr}((D\otimes I_{Q^c})(\varphi_1\otimes \varphi_2\otimes\cdots\otimes \varphi_m)): \varphi_j\in{\mathcal Pur}(H_{P_j}), j=1,2,\ldots,m\}
=  g(D\otimes I_{Q^c},P).
\end{array}$$
%\end{widetext}
On the other hand, as
$$\begin{array}{rl}
 {\rm Tr}((D\otimes I_{Q^c})(\varphi_1\otimes \varphi_2\otimes\cdots\otimes \varphi_m))
= {\rm Tr}[D{\rm Tr}_{Q^c}((\varphi_1\otimes \varphi_2\otimes\cdots\otimes \varphi_m)]
=  {\rm Tr}[D({\rm Tr}_{Q_1^{c_1}}(\varphi_1)\otimes {\rm Tr}_{Q_2^{c_2}}(\varphi_2)\otimes
\cdots\otimes {\rm Tr}_{Q_m^{c_m}}(\varphi_m))],
\end{array}$$
we see that
\begin{align*}
g(D\otimes I_{Q^c},P) &= \sup\bigl\{ \operatorname{Tr}((D\otimes I_{Q^c})(\varphi_1\otimes \varphi_2\otimes\cdots\otimes \varphi_m)) : \varphi_j\in\mathcal{Pur}(H_{P_j}), j=1,2,\ldots,m \bigr\} \\
&\leq \sup\bigl\{ \operatorname{Tr}(D(\sigma_1\otimes \sigma_2\otimes\cdots\otimes \sigma_m))  : \sigma_j\in\mathcal{S}(H_{Q_j}), j=1,2,\ldots,m \bigr\} \\
&= g(D,Q).
\end{align*}
This establishes that
$g(D,Q)= g(D\otimes I_{Q^c},P).$
It follows that
$$\begin{array}{rl}
& E_w^{(m)}(\rho_Q,Q)\\ =& \sup_{D\in{\mathcal B}_1^+(Q)}\max\{0, {\rm Tr}(D\rho_Q)-g(D,Q)\}\\
= &  \sup_{D\otimes I_{Q^c}\in{\mathcal B}_1^+({P})}\max\{0, {\rm Tr}(D\rho_Q)-g(D\otimes I_{Q^c},P)\}\\
\leq &  \sup_{L\in{\mathcal B}_1^+({P})}\max\{0, {\rm Tr}(L\rho)-g(L,P)\}\\
=& E_w^{(m)}(\rho,P),
\end{array}$$
 where $\mathcal{B}^+_1( {P})=\mathcal{B}^+_1(H_{ {P}_1}\otimes H_{ {P}_2}\otimes \cdots H_{ {P}_m})$.

By now we have shown that, for $x\in\{a,b,c\}$, $Q\preccurlyeq^x P$ will imply that $E_w^{(r)}(\rho_Q,Q)\leq E_w^{(m)}(\rho_P,P)$ for any $\rho\in\mathcal H$.
Assume $Q\preccurlyeq P$. By Definition \ref{D2.6}, there are $R_i\in \mathcal{SP}^n$ and $x_i\in\{a,b,c\}$ such that
$Q\preccurlyeq^{x_1} R_1\preccurlyeq^{x_2} R_2\preccurlyeq^{x_3}\cdots \preccurlyeq^{x_t} R_t \preccurlyeq^{x_{t+1}} P$. Denote by $r_i$ the partite of $R_i$, $i=1,2,\ldots,t$. Then, by what we have proved above,
$$\begin{array}{rl}
 E_w^{(r)}(\rho_Q,Q)\leq E_w^{(r_1)}(\rho_{R_1},R_1)\leq E_w^{(r_2)}(\rho_{R_2},R_2)  \leq \cdots  \leq  E_w^{(r_t)}(\rho_{R_t},R_t)\leq E_w^{(m)}(\rho_P,P) \end{array}
$$
for all $\rho\in{\mathcal S}(H_1\otimes H_2\otimes\cdots\otimes H_n)$. \hfill$\Box$

\section{ Proofs of the results in Section~\ref{Sec3}}\label{SecAB}

\subsection{Proof of Theorem  \ref{T3.3}.} \label{SecAB1}
%{\bf B1. Proof of Theorem \ref{T3.3}.}

{\bf Proof.}

(1) If $\rho$ is $k$-entangled, then there exists a $k$-entanglement witness $W=\lambda I-L\in {\mathcal {EW}}_1^{(k)}$ such that ${\rm Tr}(W\rho)<0$. Consequently, ${\rm Tr}(L\rho)-\lambda>0$ and $E_w^{(k,n)}(\rho)>0.$
Conversely, if $E_w^{(k,n)}(\rho)>0$, then ${\rm Tr}(L\rho)-\lambda>0$ for some $\lambda I-L\in \mathcal{EW}_1^{(k)}$, which implies that $\rho$ is $k$-entangled. We can conclude that $\rho$ is $k$-entangled if and only if $E_w^{(k,n)}(\rho)>0$. Therefore, $\rho$ is $k$-separable if and only if $E_w^{(k,n)}(\rho)=0$.

(2) Let $\Phi$ be an $n$-partite LOCC. For any $\rho\in\mathcal{S}(H_1\otimes H_2\otimes\cdots\otimes H_n)$, if $\sigma$ is $k$-separable,
then $\Phi(\sigma)$ is $k$-separable. If $W=\lambda I-L\in \mathcal{EW}_1^{(k)}$, then
$${\rm Tr}(\Phi^\dagger(W)\sigma)={\rm Tr}(W\Phi(\sigma))\geq0$$
holds for all $k$-separable states $\sigma$.
Thus, as $\Phi^\dagger(I)=I$, $0\leq \Phi^\dagger(L)\leq I$, we have $\Phi^\dagger(W)\in \mathcal{EW}_1^{(k)}$,  which implies that $\Phi^\dagger({\mathcal {EW}}_1^{(k)})\subseteq {\mathcal {EW}}_1^{(k)}$.

Based on the above discussion, if $\rho$ is $k$-separable, then we have
$$ \begin{array}{rl} E^{(k,n)}_w(\Phi(\rho)) &=\sup_{W\in\mathcal{EW}_1^{(k)}}|\min\{{\rm Tr}(W\Phi(\rho)),0\}|=0 =E^{(k,n)}_w(\rho); \end{array}$$
 if
$\rho$ is $k$-entangled, then
$$ \begin{array}{rl}
& E^{(k,n)}_w(\Phi(\rho)) \\ =& \sup_{W\in\mathcal{EW}_1}|\min\{{\rm Tr}(W\Phi(\rho)),0\}|\\
=&  \sup_{W\in\mathcal{EW}_1^{(k)}}|\min\{{\rm Tr}(\Phi^\dagger(W)\rho),0\}|\\
\leq & \sup_{W'\in\mathcal{EW}_1^{(k)}}|\min\{{\rm Tr}(W'\rho),0\}|=E^{(k,n)}_w(\rho).
\end{array}$$
Hence, $E^{(k,n)}_w(\Phi(\rho))\leq E^{(k,n)}_w(\rho)$ holds for all state $\rho$.

(3)  Obvious.

(4) Assume $\rho=\sum_{j=1}^m p_j\rho_j$, then
that
 $$\begin{array}{rl}
 E^{(k,n)}_w(\rho)=& \sup_{W\in\mathcal{EW}_1^{(k)}} |\min\{\sum_j p_j{\rm Tr}(W\rho_j),0\}|\\
 \leq & \sup_{W\in\mathcal{EW}_1^{(k)}} \sum_j p_j|\min\{{\rm Tr}(W\rho_j),0\}|\\
 \leq & \sum_j p_j\sup_{W\in\mathcal{EW}_1^{(k)}} |\min\{{\rm Tr}(W\rho_j),0\}|\\
 = & \sum_{j=1}^mp_jE^{(k,n)}_w(\rho_j).
\end{array}$$
 \hfill$\Box$

\subsection{Proof of Theorem  \ref{T3.4}.} \label{SecAB2}
%{\bf B2. Proof of Theorem \ref{T3.4}.}
{\bf Proof.} It follows from Eqs.(\ref{eq3.4}) and (\ref{eq3.8}) that we get
$$\begin{array}{rl} & E_w^{(k_1+k_2,n+m)}(\rho_1\otimes\rho_2)
= \sup_{L\in \mathcal{B}_1^+(H\otimes K)}\max\{0, {\rm Tr}(L\rho_1\otimes \rho_2)-g_{n+m}^{(k_1+k_2)}(L)\}.
\end{array}$$
 Denote $H_{n+i}=K_i$, $i=1,2,\ldots, m$. Then the set of all $(k_1+k_2)$-partition ${\mathcal P}_{n+m}^{k_1+k_2}(\tilde{H})$ of $(n+m)$-partite system $\tilde{H}=H_1\otimes H_2\otimes\cdots\otimes H_n\otimes K_1\otimes K_2\otimes\cdots\otimes K_m$ is
$$\begin{array}{rl}
&  {\mathcal P}_{n+m}^{k_1+k_2}(\tilde{H})
= \{ R=R_1|R_2|\ldots|R_{k_1+k_2} :R_j\subset\{1,2,\ldots, n+m\},  R_j\cap R_i=\emptyset, \bigcup _{j=1}^{k_1+k_2} R_j=\{1,2,\ldots, n+m\}\}.
\end{array}$$
 It is clear that,
\begin{equation*}
\begin{aligned}
\bigl\{P|Q:\;&P\in\mathcal P_n^{k_1}(H_1\otimes\cdots\otimes H_n),Q\in\mathcal P_m^{k_2}(K_1\otimes\cdots\otimes K_m)\bigr\}
\subset\mathcal P_{n+m}^{(k_1+k_2)}(\widetilde H).
\end{aligned}
\end{equation*}
It follows that
%\begin{widetext}
$$\begin{array}{rl}
&g_{n+m}^{(k_1+k_2)}(L)
\\=&
\max_{R\in{\mathcal P}^{k_1+k_2}_{n+m}(\tilde{H})}\sup_{|\phi_{R_i}\rangle\in{H_{R_i}}}
\{  {\rm Tr}[L(|\phi_{R_1}\rangle\langle\phi_{R_1}|\otimes
|\phi_{R_2}\rangle\langle\phi_{R_2}| \otimes\cdots\otimes
|\varphi_{R_{k_1+k_2}}\rangle\langle\varphi_{R_{k_1+k_2}}|)]\}\\
\geq &\max_{P\in{\mathcal P}^{k_1}_{n}(H)}\max_{Q\in{\mathcal P}^{k_2}_{m}(K)}\sup_{|\phi_{P_i}\rangle\in{H_{P_i}}}\sup_{|\varphi_{Q_{j}}\rangle\in{K_{Q_{j}}}}
\{{\rm Tr}[L(|\phi_{P_1}\rangle\langle\phi_{P_1}|
\otimes\cdots \\ & \otimes
|\phi_{P_{k_1}}\rangle\langle\phi_{P_{k_1}}|\otimes |\varphi_{Q_{1}}\rangle\langle\varphi_{Q_{1}}|\otimes\cdots\otimes
|\varphi_{Q_{k_2}}\rangle\langle\varphi_{Q_{k_2}}|
)]\} \\
=&\max_{P\in{\mathcal P}^{k_1}_{n}(H)}\max_{Q\in{\mathcal P}^{k_2}_{m}(K)}\sup_{|\phi_{P_i}\rangle\in{H_{P_i}}}\sup_{|\varphi_{Q_{j}}\rangle\in{K_{Q_{j}}}}
\{{\rm Tr}_K[{\rm Tr}_H(L(|\phi_{P_1}\rangle\langle\phi_{P_1}| \\
&\otimes\cdots\otimes
|\phi_{P_{k_1}}\rangle\langle\phi_{P_{k_1}}|\otimes I_{K})) |\varphi_{Q_{1}}\rangle\langle\varphi_{Q_{1}}|\otimes\cdots\otimes
|\varphi_{Q_{k_2}}\rangle\langle\varphi_{Q_{k_2}}|
]\} \\
=&\max_{P\in{\mathcal P}^{k_1}_{n}(H)}\sup_{|\phi_{P_i}\rangle\in{H_{P_i}}}g^{k_2}_m({\rm Tr}_H[L(|\phi_{P_1}\rangle\langle\phi_{P_1}|\otimes\cdots\otimes
|\phi_{P_{k_1}}\rangle\langle\phi_{P_{k_1}}|\otimes I_K)])
\end{array}$$
%\end{widetext}
with $i\in\{1,2,\ldots,k_1\}$ and $j\in\{1,2,\ldots,k_2\}$.
Therefore, if ${\mathcal B}_1^+(H)=\{ L\in{\mathcal B}(H_1\otimes H_2\otimes\cdots\otimes H_n): L\geq 0, \|L\|\leq 1\}$ and ${\mathcal B}_1^+(K)=\{ L\in{\mathcal B}(K_1\otimes K_2\otimes\cdots\otimes K_m): L\geq 0, \|L\|\leq 1\}$, then we have that
%\begin{widetext}
$$\begin{array}{rl}
&{\rm Tr}(L(\rho_1\otimes \rho_2))-g^{(k_1+k_2)}_{n+m}(L) \\
=&{\rm Tr}_H[{\rm Tr}_K(L(I\otimes\rho_2))\rho_1]-g_n^{(k_1)}[{\rm Tr}_    K(L(I\otimes \rho_2))]-g^{(k_1+k_2)}_{n+m}(L)\\
&+\max_{P\in{\mathcal P}^{k_1}_{n}(H)}\sup_{|\phi_{P_i}\rangle\in{H_{P_i}}}{\rm Tr}_H[{\rm Tr}_K(L(I\otimes \rho_2))(|\phi_{P_1}\rangle\langle\phi_{P_1}|\otimes\cdots\otimes
|\phi_{P_{k_1}}\rangle\langle\phi_{P_{k_1}}|)]\\
\leq &{\rm Tr}_H[{\rm Tr}_K(L(I\otimes\rho_2))\rho_1]-g_n^{(k_1)}[{\rm Tr}_    K(L(I\otimes \rho_2))] \\ &-\max_{P\in{\mathcal P}^{k_1}_{n}(H)}\sup_{|\phi_{P_i}\rangle\in{H_{P_i}}}g^{(k_2)}_m({\rm Tr}_H(L[(|\phi_{P_1}\rangle\langle\phi_{P_1}|\otimes\cdots\otimes
|\phi_{P_{k_1}}\rangle\langle\phi_{P_{k_1}}|)\otimes I_K])\\
&+\max_{P\in{\mathcal P}^{k_1}_{n}(H)}\sup_{|\phi_{P_i}\rangle\in{H_{P_i}}}{\rm Tr}_K[{\rm Tr}_H(L((|\phi_{P_1}\rangle\langle\phi_{P_1}|\otimes\cdots\otimes
|\phi_{P_{k_1}}\rangle\langle\phi_{P_{k_1}}|)\otimes I_K)\rho_2]\\
\leq&\sup_{C\in\mathcal{B}_1^+(H)}\{{\rm Tr}_H(C\rho_1)-g_n^{(k_1)}(C)\}+\sup_{D\in\mathcal{B}_1^+(K)}\{{\rm Tr}_K(D\rho_2)-g_m^{(k_2)}(D)\},
\end{array}$$
%\end{widetext}
because ${\rm Tr}_K(L(I_H\otimes \rho_2))\in{\mathcal B}_1^+(H)$ whenever $L\in{\mathcal B}_1^+(H\otimes K)$ and $\rho_2\in{\mathcal S}(K)$ (see the proof of Theorem  \ref{T2.5} in Appendix \ref{SecAA}).  Hence
$$\begin{array}{rl}
&E_w^{(k_1+k_2,n+m)}(\rho_1\otimes\rho_2) \\
=&\sup_{L\in \mathcal{B}_1^+(H\otimes K)}\max\{ {\rm Tr}(L\rho_1\otimes \rho_2)-g_{n+m}^{(k_1+k_2)}(L), 0\} \\
\leq&\sup_{C\in\mathcal{B}_1^+(H)}\max\{{\rm Tr}_H(C\rho_1)-g_n^{(k_1)}(C),0\} \\ & +\sup_{D\in\mathcal{B}_1^+(K)}\max\{{\rm Tr}_K(D\rho_2)-g_m^{(k_2)}(D),0\}\\
=&E_w^{(k_1,n)}(\rho_1)+E_w^{(k_2,m)}(\rho_2).
\end{array}$$
The proof is completed.\hfill$\Box$

\subsection{Proof of Theorem~\ref{T3.5}.}
\label{SecAB3}

By Definition~\ref{D3.1}, to prove Theorem~\ref{T3.5}, it suffices to prove Lemmas~\ref{LB3.1}--\ref{LB3.3}.

\begin{lemma}\label{LB3.1}
{\it
If $(s,Q)\preccurlyeq^{a'}(t,P)$, then
$$
E_w^{(s,r)}(\rho_Q,Q)
\leq
E_w^{(t,m)}(\rho_P,P)
$$
holds for all $\rho\in\mathcal S(H_1\otimes H_2\otimes\cdots\otimes H_n)$.
}
\end{lemma}

{\bf Proof.}
Since $Q\preccurlyeq^aP$, $Q$ is obtained from $P$ by discarding some complete parties. For $\rho\in\mathcal S(H_1\otimes H_2\otimes\cdots\otimes H_n)$, regarding $\rho_P$ as an $m$-partite state on
$H_{P_1}\otimes\cdots\otimes H_{P_m}$ and relabeling these $m$ parties if necessary, we may assume that
$P=P_1|P_2|\cdots|P_m,$ $Q=P_1|P_2|\cdots|P_r.$ Set $K=H_{P_1}\otimes H_{P_2}\otimes\cdots\otimes H_{P_r},$ and
$K^\perp=H_{P_{r+1}}\otimes H_{P_{r+2}}\otimes\cdots\otimes H_{P_m}.$
Thus $K=H_Q$ and $H_P=K\otimes K^\perp$. Let $P'=P'_1|P'_2|\cdots|P'_t\in\mathcal P_t(P)$ be an arbitrary $t$-partition of $P$. For each $j=1,2,\ldots,t$, let $R_j=P'_j\cap\left(\bigcup_{i=1}^rP_i\right).$ Some of the $R_j$ may be empty. Suppose that exactly $l$ of them
are nonempty, denoted by $R_{j_1},R_{j_2},\ldots,R_{j_l}$. Then $R=R_{j_1}|R_{j_2}|\cdots|R_{j_l}$ is an $l$-partition of $Q$.
Since $Q$ is obtained from $P$ by discarding exactly $m-r$ complete parties, a block $P'_j$ can disappear after the
discarding operation only if it contains at least one of these $m-r$ discarded parties. As the blocks $P'_1,\ldots,P'_t$ are mutually disjoint, at most $m-r$ blocks can disappear. Hence $l\geq t-(m-r).$

By the definition of $\preccurlyeq^{a'}$, $t-s\geq m-r,$ and therefore $l\geq t-(m-r)\geq s.$ Thus the $l$ blocks of $R$ can be grouped into $s$ nonempty blocks. Let $Q'=Q'_1|Q'_2|\cdots|Q'_s$ be such an $s$-partition. Then $ Q'\in\mathcal P_s(Q). $

For $D\in\mathcal B_1^+(K)$, consider $D\otimes I_{K^\perp}$. For the fixed partition $P'$, we have

%\begin{widetext}
\begin{equation}
\begin{aligned}
g(D\otimes I_{K^\perp},P')={}&\sup\Bigl\{{\rm Tr}\bigl[(D\otimes I_{K^\perp})(|\phi_1\rangle\langle\phi_1|\otimes|\phi_2\rangle\langle\phi_2|\otimes\cdots\otimes|\phi_t\rangle\langle\phi_t|)\bigr]:
\\&\hspace{12em}|\phi_j\rangle\in H_{P'_j},\
\langle\phi_j|\phi_j\rangle=1,\
j=1,2,\ldots,t\Bigr\}\\
={}&
\sup\Bigl\{{\rm Tr}_{K}\bigl[D\,{\rm Tr}_{K^\perp}(|\phi_1\rangle\langle\phi_1|\otimes|\phi_2\rangle\langle\phi_2|\otimes\cdots\otimes
|\phi_t\rangle\langle\phi_t|)\bigr]:\\
&\hspace{12em}
|\phi_j\rangle\in H_{P'_j},\
\langle\phi_j|\phi_j\rangle=1,\
j=1,2,\ldots,t\Bigr\}\\
\leq{}&
\sup\Bigl\{{\rm Tr}\bigl[D(\sigma_1\otimes\sigma_2\otimes\cdots\otimes\sigma_s)\bigr]:\sigma_i\in\mathcal S(H_{Q'_i}),\
i=1,2,\ldots,s
\Bigr\}
\\
={}&
\sup\Bigl\{
{\rm Tr}\bigl[
D(
|\omega_1\rangle\langle\omega_1|\otimes|\omega_2\rangle\langle\omega_2|\otimes\cdots\otimes|\omega_s\rangle\langle\omega_s|
)\bigr]:
\\
&\hspace{12em}
|\omega_i\rangle\in H_{Q'_i},\
\langle\omega_i|\omega_i\rangle=1,\
i=1,2,\ldots,s
\Bigr\}
\\
={}&g(D,Q')
\\
\leq{}&g_r^{(s)}(D,Q).
\end{aligned}
\tag{H1}
\label{eqB1}
\end{equation}
%\end{widetext}

Here the fourth equality follows from the equivalent characterization of $g$ by product mixed states given in Appendix A.

Since $P'\in\mathcal P_t(P)$ was arbitrary, Eq.~(\ref{eqB1}) gives
\begin{equation}
\begin{aligned}
g_m^{(t)}(D\otimes I_{K^\perp},P)
&=
\max_{P'\in\mathcal P_t(P)}
g(D\otimes I_{K^\perp},P')
\\
&\leq
g_r^{(s)}(D,Q).
\end{aligned}
\tag{H2}
\label{eqB2}
\end{equation}

Moreover,
$$
{\rm Tr}(D\rho_Q)
=
{\rm Tr}[(D\otimes I_{K^\perp})\rho_P].
$$
Since $D\in\mathcal B_1^+(K)$ implies $D\otimes I_{K^\perp}\in\mathcal B_1^+(H_P)$, Eq.~(\ref{eq3.8}) and Eq.~(\ref{eqB2}) give

$$
\begin{array}{rl}
E_w^{(s,r)}(\rho_Q,Q)
=&
\displaystyle
\sup_{D\in\mathcal B_1^+(K)}
\max\left\{
0,
{\rm Tr}(D\rho_Q)-g_r^{(s)}(D,Q)
\right\}
\\
\leq&
\displaystyle
\sup_{D\in\mathcal B_1^+(K)}
\max\left\{
0,
{\rm Tr}[(D\otimes I_{K^\perp})\rho_P]
-
g_m^{(t)}(D\otimes I_{K^\perp},P)
\right\}
\\
\leq&
\displaystyle
\sup_{L\in\mathcal B_1^+(H_P)}
\max\left\{
0,
{\rm Tr}(L\rho_P)-g_m^{(t)}(L,P)
\right\}
\\
=&
E_w^{(t,m)}(\rho_P,P).
\end{array}
$$
This completes the proof of Lemma~\ref{LB3.1}.
\hfill$\Box$

\begin{remark}
In the particular case: $t-s=m-r,$ the inequality in Eq.~(\ref{eqB2}) becomes an equality. Indeed, Eq.~(\ref{eqB2}) gives
\begin{equation*}
g_m^{(t)}(D\otimes I_{K^\perp},P)
\leq
g_r^{(s)}(D,Q).
\end{equation*}
Conversely, for any $S=S_1|\cdots|S_s\in\mathcal P_s(Q)$, since $t-s=m-r$, the partition $R=S_1|\cdots|S_s|P_{r+1}|\cdots|P_m$
belongs to $\mathcal P_t(P)$. Every product pure state with respect to $S$ admits a product pure-state extension with
respect to $R$. Therefore,
\begin{equation*}
g(D,S)
\leq
g(D\otimes I_{K^\perp},R).
\end{equation*}
Taking the maximum over all $S\in\mathcal P_s(Q)$ yields \begin{equation*}
g_r^{(s)}(D,Q)
\leq
g_m^{(t)}(D\otimes I_{K^\perp},P).
\end{equation*}
Hence,
\begin{equation*}
g_m^{(t)}(D\otimes I_{K^\perp},P)
=
g_r^{(s)}(D,Q).
\end{equation*}
\end{remark}

\begin{lemma}\label{LB3.2}
{\it
If $(s,Q)\preccurlyeq^{b'}(t,P)$, then, for any $\rho\in\mathcal S(H_1\otimes H_2\otimes\cdots\otimes H_n)$, we have
$$
E_w^{(s,r)}(\rho_Q,Q)
\leq
E_w^{(t,m)}(\rho_P,P).
$$
}
\end{lemma}

{\bf Proof.}
As $Q\preccurlyeq^bP$, the two sub-repartitions have the same underlying total system. Therefore, without loss of generality, we may assume $H_P$ and $H_Q$ are respectively an $m$-partition and an $r$-partition of the entire system $H$. Then we have $\rho_P=\rho_Q=\rho.$ Regard $\rho$ as an $m$-partite state of $H_{P_1}\otimes  H_{P_2}\otimes\cdots\otimes H_{P_m},$ and also as an $r$-partite state of $H_{Q_1}\otimes H_{Q_2}\otimes\cdots\otimes H_{Q_r}.$
Let $\mathcal P_m^{t,P}$ denote the set of all $t$-partitions of $P=P_1|P_2|\cdots|P_m$, that is,
\begin{equation*}
\begin{aligned}
\mathcal P_m^{t,P}
=
\bigl\{
R=R_1|R_2|\cdots|R_t :
R_j\text{ is a union of some }P_i,\
j=1,2,\ldots,t
\bigr\}.
\end{aligned}
\end{equation*}

Similarly, let $\mathcal P_r^{s,Q}$ denote the set of all $s$-partitions of $Q=Q_1|Q_2|\cdots|Q_r$, i.e.,

\begin{equation*}
\begin{aligned}
\mathcal P_r^{s,Q}
=
\bigl\{
S=S_1|S_2|\cdots|S_s :
S_j\text{ is a union of some }Q_i,\
j=1,2,\ldots,s
\bigr\}.
\end{aligned}
\end{equation*}

If $S=S_1|S_2|\cdots|S_s\in\mathcal P_r^{s,Q},$ then, since $s\leq t\leq m$ and $Q\preccurlyeq^bP$, the blocks of $S$ can be refined by the blocks of $P$ until one obtains a $t$-partition $R=R_1|R_2|\cdots|R_t\in\mathcal P_m^{t,P},$ such that every $S_i$ is a union of some $R_j$. In this case, $R$ is a refinement of $S$.

Conversely, let $R=R_1|R_2|\cdots|R_t\in\mathcal P_m^{t,P}.$ Since $(s,Q)\preccurlyeq^{b'}(t,P),$ Definition~\ref{D3.1} implies directly that there exists $S=S_1|S_2|\cdots|S_s\in\mathcal P_r^{s,Q}$ such that $R$ is a refinement of $S$, that is, $S\preccurlyeq^bR.$ Now let $L\in\mathcal B_1^+(H).$ For $S=S_1|S_2|\cdots|S_s\in\mathcal P_r^{s,Q}$ and a refinement $R=R_1|R_2|\cdots|R_t\in\mathcal P_m^{t,P}$ of $S$, every product pure state with respect to $R$ is also a product pure state with respect to $S$. Hence,

%\begin{widetext}
$$
\begin{array}{rl}
g(L,S)=&
\displaystyle
\sup
\left\{
{\rm Tr}
\left[
L(
|\omega_1\rangle\langle\omega_1|
\otimes\cdots\otimes
|\omega_s\rangle\langle\omega_s|
)
\right]:
|\omega_j\rangle\in H_{S_j},
\
\langle\omega_j|\omega_j\rangle=1,
\
j=1,\ldots,s
\right\}
\\
\geq&
\displaystyle
\sup
\left\{
{\rm Tr}
\left[
L(
|\phi_1\rangle\langle\phi_1|
\otimes\cdots\otimes
|\phi_t\rangle\langle\phi_t|
)
\right]:
|\phi_j\rangle\in H_{R_j},
\
\langle\phi_j|\phi_j\rangle=1,
\
j=1,\ldots,t
\right\}
\\
=&g(L,R).
\end{array}
$$

Since, by Definition~\ref{D3.1}, every $R\in\mathcal P_m^{t,P}$ is a refinement of some $S\in\mathcal P_r^{s,Q}$, we obtain
$$
\begin{array}{rl}
g_r^{(s)}(L,Q)=&
\displaystyle
\max_{S\in\mathcal P_r^{s,Q}}g(L,S)
\\
\geq&
\displaystyle
\max_{R\in\mathcal P_m^{t,P}}g(L,R)
\\
=&
g_m^{(t)}(L,P).
\end{array}
$$
%\end{widetext}

Since $H_{Q_1}\otimes\cdots\otimes H_{Q_r}=H_{P_1}\otimes\cdots\otimes H_{P_m}=H,$ the corresponding sets $\mathcal B_1^+$ are identical. Therefore,
$$
\begin{array}{rl}
E_w^{(s,r)}(\rho,Q)
=&
\displaystyle
\sup_{L\in\mathcal B_1^+}
\max
\left\{
0,
{\rm Tr}(L\rho)-g_r^{(s)}(L,Q)
\right\}
\\
\leq&
\displaystyle
\sup_{L\in\mathcal B_1^+}
\max
\left\{
0,
{\rm Tr}(L\rho)-g_m^{(t)}(L,P)
\right\}
\\
=&
E_w^{(t,m)}(\rho,P).
\end{array}
$$

This completes the proof of Lemma~\ref{LB3.2}.
\hfill$\Box$

If $Q\preccurlyeq^bP$ and $2\leq t\leq r$, then every
$t$-partition of $Q$ is also a $t$-partition obtainable by
grouping blocks of $P$. Thus $\mathcal P_r^{t,Q}\subseteq\mathcal P_m^{t,P}.$
Consequently,
\begin{equation*}
\begin{aligned}
g_m^{(t)}(L,P)
&=
\max_{R\in\mathcal P_m^{t,P}}g(L,R)
\\
&\geq
\max_{S\in\mathcal P_r^{t,Q}}g(L,S)
\\
&=
g_r^{(t)}(L,Q),
\end{aligned}
\end{equation*}

and hence $E_w^{(t,r)}(\rho,Q)\geq E_w^{(t,m)}(\rho,P).$

If $(t,Q)\preccurlyeq^{b'}(t,P),$ Lemma~\ref{LB3.2} gives the reverse inequality. Therefore the two quantities are equal. As discussed in
Appendix~\ref{SecHierarchyDetails}, the relation $(t,Q)\preccurlyeq^{b'}(t,P)$ holds if and only if $Q=P$. This illustrates why the structural refinement condition in the definition of $\preccurlyeq^{b'}$ is required.

\begin{lemma}\label{LB3.3}
{\it
If $(s,Q)\preccurlyeq^{c'}(t,P)$, then, for any $\rho\in\mathcal S(H_1\otimes H_2\otimes\cdots\otimes H_n)$,
we have
$$
E_w^{(s,m)}(\rho_Q,Q)
\leq
E_w^{(t,m)}(\rho_P,P).
$$
}
\end{lemma}

{\bf Proof.}
As $(s,Q)\preccurlyeq^{c'}(t,P),$
we have
$$
Q\preccurlyeq^cP,
\qquad
r=m,
\qquad
s\leq t.
$$

Since the inequality only concerns the reduced states
$\rho_P$ and $\rho_Q$, we may regard $H_P$ as the ambient $m$-partite system and write $\rho=\rho_P$. After a suitable ordering, let $P=P_1|P_2|\cdots|P_m$ and $Q=Q_1|Q_2|\cdots|Q_m,$ where $Q_j\subseteq P_j,$ $j=1,2,\ldots,m.$ Then $\rho_Q={\rm Tr}_{Q^c}(\rho).$ It remains to prove
$$
E_w^{(s,m)}(\rho_Q,Q)
\leq
E_w^{(t,m)}(\rho,P).
$$

Denote $Q_j^{c_j}=P_j\backslash Q_j,$ $j=1,2,\ldots,m.$ Let $\mathcal P_m^{t,P}$ be the set of all $t$-partitions of $P=P_1|P_2|\cdots|P_m$, and let $\mathcal P_m^{s,Q}$ be the set of all $s$-partitions of $Q=Q_1|Q_2|\cdots|Q_m$.
If $S=S_1|S_2|\cdots|S_s\in\mathcal P_m^{s,Q},$ then every $S_j$ is a union of some $Q_{j_1},Q_{j_2},\ldots,Q_{j_{l_j}}$, i.e.,
$S_j=\bigcup_{\alpha=1}^{l_j}Q_{j_\alpha}.$
Define $V_j=\bigcup_{\alpha=1}^{l_j}P_{j_\alpha},j=1,2,\ldots,s.$ Then $V=V_1|V_2|\cdots|V_s$ is an $s$-partition of $P$. Since $s\leq t$, there exists a $t$-partition $R=R_1|R_2|\cdots|R_t\in\mathcal P_m^{t,P}$ such that $R$ refines $V$. Consequently, every $S_i$ is a union of some of the sets $R_j\cap Q$. Conversely, let $R=R_1|R_2|\cdots|R_t \in\mathcal P_m^{t,P}.$ Since $Q\preccurlyeq^cP$ and each $Q_i$ is contained in the corresponding $P_i$, there exists $V=V_1|V_2|\cdots|V_t\in\mathcal P_m^{t,Q}$ such that $V_j\subseteq R_j, j=1,2,\ldots,t.$ More explicitly, if $R_j$ is the union of $P_{i_1},\ldots,P_{i_l}$, then $V_j$ is the union of the corresponding $Q_{i_1},\ldots,Q_{i_l}$. Since $s\leq t$, the blocks of $V$ can be grouped into $s$ nonempty blocks. Let $S=S_1|S_2|\cdots|S_s$ be such an $s$-partition of $V$. Then $S\in\mathcal P_m^{s,Q},$ and each $S_i$ is a union of some of the sets $R_j\cap Q$. Thus the $t$-partition structure of $P$ and the $s$-partition structure of $Q$ are compatible in this case.

Now fix $R=R_1|R_2|\cdots|R_t\in\mathcal P_m^{t,P},$ and let $V=V_1|V_2|\cdots|V_t\in\mathcal P_m^{t,Q}$ be the corresponding $t$-partition of $Q$ described above. For each $j$, define $V_j^c=R_j\backslash V_j.$ Then $H_{R_j}=H_{V_j}\otimes H_{V_j^c}.$ For any $\varphi_j\in{\mathcal Pur}(H_{R_j}), j=1,2,\ldots,t,$
we have
\begin{equation}
\begin{aligned}
{\rm Tr}
\left[(D\otimes I_{Q^c})(\varphi_1\otimes\varphi_2\otimes\cdots\otimes\varphi_t)
\right]
={}&
{\rm Tr}\left[D\,{\rm Tr}_{Q^c}(\varphi_1\otimes\varphi_2\otimes\cdots\otimes\varphi_t)\right]
\\
={}&
{\rm Tr}\left[D\left({\rm Tr}_{V_1^c}(\varphi_1)\otimes{\rm Tr}_{V_2^c}(\varphi_2)\otimes\cdots\otimes{\rm Tr}_{V_t^c}(\varphi_t)\right)
\right].
\end{aligned}
\tag{H3}
\label{eqB3}
\end{equation}

For every $R\in\mathcal P_m^{t,P}$, the above construction gives a $t$-partition $V\in\mathcal P_m^{t,Q}$ by grouping the corresponding $Q_i$ in exactly the same way as the $P_i$ are grouped in $R$. Conversely, every $t$-partition of $Q$ determines a
$t$-partition of $P$ by grouping the corresponding $P_i$ in the same way. Thus, applying Eq.~(\ref{eqB3}), we obtain

%\begin{widetext}
\begin{equation}
\begin{aligned}
g_m^{(t)}(D\otimes I_{Q^c},P)
={}&
\max_{R\in\mathcal P_m^{t,P}}
\sup
\Bigl\{
{\rm Tr}
\left[(D\otimes I_{Q^c})(\varphi_1\otimes\varphi_2\otimes\cdots\otimes\varphi_t)\right]:
\varphi_j\in{\mathcal Pur}(H_{R_j}),
\
j=1,2,\ldots,t
\Bigr\}
\\
={}&
\max_{V\in\mathcal P_m^{t,Q}}
\sup
\Bigl\{
{\rm Tr}
\left[
D
\left(
{\rm Tr}_{V_1^c}(\varphi_1)
\otimes
{\rm Tr}_{V_2^c}(\varphi_2)
\otimes\cdots\otimes
{\rm Tr}_{V_t^c}(\varphi_t)
\right)
\right]:
\varphi_j\in{\mathcal Pur}(H_{R_j}),
\
j=1,2,\ldots,t
\Bigr\}
\\
\leq{}&
\max_{V\in\mathcal P_m^{t,Q}}
\sup
\Bigl\{
{\rm Tr}
\left[
D(
\sigma_1\otimes\sigma_2\otimes\cdots\otimes\sigma_t
)
\right]:
\sigma_j\in{\mathcal S}(H_{V_j}),
\
j=1,2,\ldots,t
\Bigr\}
\\
={}&
g_m^{(t)}(D,Q).
\end{aligned}
\tag{H4}
\label{eqB4}
\end{equation}

Furthermore, since $s\leq t$, for every
$V=V_1|V_2|\cdots|V_t
\in\mathcal P_m^{t,Q},$
one can choose an $s$-partition
$S=S_1|S_2|\cdots|S_s
\in\mathcal P_m^{s,Q}$
such that every $S_i$ is a union of some $V_j$.
Therefore every product pure state with respect to $V$ is
also a product pure state with respect to $S$. Hence,

\begin{equation}
\begin{aligned}
g_m^{(t)}(D,Q)
={}&
\max_{V\in\mathcal P_m^{t,Q}}
g(D,V)
\\
={}&
\max_{V\in\mathcal P_m^{t,Q}}
\sup
\Bigl\{
{\rm Tr}
\left[
D(
\varphi_1\otimes\varphi_2\otimes\cdots\otimes\varphi_t
)
\right]:
\varphi_j\in{\mathcal Pur}(H_{V_j}),
\
j=1,2,\ldots,t
\Bigr\}
\\
\leq{}&
\max_{S\in\mathcal P_m^{s,Q}}
\sup
\Bigl\{
{\rm Tr}
\left[
D(
\eta_1\otimes\eta_2\otimes\cdots\otimes\eta_s
)
\right]:
\eta_i\in{\mathcal Pur}(H_{S_i}),
\
i=1,2,\ldots,s
\Bigr\}
\\
={}&
g_m^{(s)}(D,Q).
\end{aligned}
\tag{H5}
\label{eqB5}
\end{equation}
%\end{widetext}

Eqs.~(\ref{eqB4}) and (\ref{eqB5}) give
$$
g_m^{(s)}(D,Q)
\geq
g_m^{(t)}(D\otimes I_{Q^c},P).
\eqno({\rm H6})
$$

By Eq.~(\ref{eq3.8}), we have
$$
\begin{array}{rl}
E_w^{(s,m)}(\rho_Q,Q)
=&
\displaystyle
\sup_{D\in\mathcal B_1^+(H_Q)}
\max
\left\{
0,
{\rm Tr}(D\rho_Q)-g_m^{(s)}(D,Q)
\right\}
\\
\leq&
\displaystyle
\sup_{D\otimes I_{Q^c}\in\mathcal B_1^+(H_P)}
\max
\left\{
0,
{\rm Tr}(D\rho_Q)
-
g_m^{(t)}(D\otimes I_{Q^c},P)
\right\}
\\
=&
\displaystyle
\sup_{D\otimes I_{Q^c}\in\mathcal B_1^+(H_P)}
\max
\left\{
0,
{\rm Tr}[(D\otimes I_{Q^c})\rho]
-
g_m^{(t)}(D\otimes I_{Q^c},P)
\right\}
\\
\leq&
\displaystyle
\sup_{L\in\mathcal B_1^+(H_P)}
\max
\left\{
0,
{\rm Tr}(L\rho)-g_m^{(t)}(L,P)
\right\}
\\
=&
E_w^{(t,m)}(\rho,P).
\end{array}
$$

Therefore, $E_w^{(s,m)}(\rho_Q,Q) \leq E_w^{(t,m)}(\rho,P), $ whenever
$(s,Q)\preccurlyeq^{c'}(t,P)$.
This completes the proof of Lemma~\ref{LB3.3}.
\hfill$\Box$

{\bf Proof of Theorem \ref{T3.5}}.
Now suppose that $(s,Q)\preccurlyeq'(t,P).$ By Definition~\ref{D3.1}, there exist sub-repartitions $R_1,R_2,\ldots,R_\alpha\in\mathcal{SP}^n$ and positive integers $k_1,k_2,\ldots,k_\alpha$ such that
\begin{equation}
\begin{aligned}
(s,Q)
\preccurlyeq^{x_1}
(k_1,R_1)
\preccurlyeq^{x_2}
(k_2,R_2)
\preccurlyeq^{x_3}
\cdots
\preccurlyeq^{x_\alpha}
(k_\alpha,R_\alpha)
\preccurlyeq^{x_{\alpha+1}}
(t,P),
\end{aligned}
\tag{H7}
\label{eqHchain}
\end{equation}
where $x_1,x_2,\ldots,x_{\alpha+1}\in\{a',b',c'\}.$
Let $m_j$ be the number of parties of $R_j$.
Applying Lemmas~\ref{LB3.1}--\ref{LB3.3} successively to the
consecutive pairs in Eq.~(\ref{eqHchain}), we obtain

$$
\begin{array}{rl}
E_w^{(s,r)}(\rho_Q,Q)
&\leq
E_w^{(k_1,m_1)}(\rho_{R_1},R_1)
\\
&\leq
E_w^{(k_2,m_2)}(\rho_{R_2},R_2)
\\
&\leq
\cdots
\\
&\leq
E_w^{(k_\alpha,m_\alpha)}(\rho_{R_\alpha},R_\alpha)
\\
&\leq
E_w^{(t,m)}(\rho_P,P).
\end{array}
$$

Hence $$E_w^{(s,r)}(\rho_Q,Q)\leq E_w^{(t,m)}(\rho_P,P).$$
This completes the proof of Theorem~\ref{T3.5}. \hfill$\Box$

%%\newpage

\section{ Algorithms}\label{SecAC}

The evaluation scheme developed in Appendix~\ref{SecComputationalDetails} relies on product-state optimizations of the form $g(L)$, which play a central role in the numerical evaluation of $k$-entanglement in multipartite quantum systems. In this appendix, we present a general iterative procedure for evaluating $g(L)$ in arbitrary finite-dimensional $n$-partite systems. For a prescribed $k$-partition $P=P_1|\cdots|P_k$, the same procedure evaluates $g(L,P)$ by treating $H_{P_1}\otimes\cdots\otimes H_{P_k}$ as a $k$-partite system. Since $g_n^{(k)}(L)=\max_{P\in\mathcal{P}_n^{vk}} g(L,P)$, this procedure provides the basic computational tool for evaluating
$E_w^{(k,n)}$ and constructing large-scale databases of EWs. The input is a positive operator $L\in\mathcal{B}_1^+(H_1\otimes\cdots\otimes H_n)$ with $\|L\|\leq1$, and the iteration is terminated according to a prescribed
convergence tolerance error.

\begin{breakablealgorithm}\label{alg1}
\caption{: General iterative algorithm for computing $g(L)$}
\begin{algorithmic}[1]

\REQUIRE
$L \in \mathcal{B}_1^+(H_1 \otimes \cdots \otimes H_n)$; $n \in \mathbb{N}_+$
\ENSURE $g(L)$

\STATE Randomly initialize product state $|b_1,\dots,b_n\rangle$
\STATE $|\xi_n\rangle \gets L|b_1,\dots,b_n\rangle$
\STATE $A^{\prime}\gets |\xi_n\rangle\langle\xi_n|$

\WHILE{not converged}
  \STATE \textbf{Phase 1: Update $|a_1^{\prime}\rangle$ via forward iteration}
  \STATE $m \gets n-1$
  \WHILE{not converged}
    \FOR{$m=n-1$ \TO $1$}
      \STATE $L' \gets {\rm Tr}_{m+1}(A^{\prime}) + I$
      \STATE $|\xi_m\rangle \gets L^{\prime} |b_1,\dots,b_m\rangle$
      \STATE $A^{\prime} \gets |\xi_m\rangle\langle\xi_m|$
    \ENDFOR
    \STATE Normalize: $|a_1^{\prime}\rangle \gets \frac{|\xi_1\rangle}{\|\xi_1\|} . $
  \ENDWHILE

  \STATE \textbf{Phase 2: Backward update of $|a_2^{\prime}\rangle,\dots,|a_n^{\prime}\rangle$}
  \FOR{$m=2$ \TO $n$}
    \STATE Schmidt decomposition  $|\xi_m\rangle = \sum_j \sqrt{\mu_j} |\phi_j\rangle |z_j\rangle$
    \STATE $|c_m\rangle \gets (\otimes_{i=1}^{m-1} \langle a_i^{\prime}|)\,|\xi_m\rangle$
    \STATE Normalize: $|a_m^{\prime}\rangle \gets \frac{|c_m\rangle}{\|c_m\|}$
  \ENDFOR

  \STATE $|a_1,\dots,a_n\rangle \gets |a_1^{\prime},\dots,a_n^{\prime}\rangle$
\ENDWHILE

\STATE $g(L) \gets \langle a_1,\dots,a_n | L | a_1,\dots,a_n\rangle$
\RETURN $g(L)$

\end{algorithmic}
\end{breakablealgorithm}

We now consider $n$-qubit systems for $n=2,3,4$. For each case, we construct a finite dataset $\widetilde{\mathcal{EW}}^{(n)}$ as in Eqs.(\ref{eq5.1})-(\ref{eq5.3}).
The dataset $\widetilde{\mathcal{EW}}^{(n)}$ approximates the full set $\mathcal{EW}^{(n)}$.
To do this,  it suffices to consider respectively
$\widetilde{\mathcal{EW}}^{k} _n= \{ ( L, { g_n(L, P) : P \in \mathcal{P}_n^{v_k} } ) : L \in \tilde{\mathcal{B}}_1^+ \}
$
for $k=2,\ldots,n$, where $\tilde{\mathcal{B}}_1^+ \subset \mathcal{B}_1^+(\mathbb{C}^2 \otimes \cdots \otimes \mathbb{C}^2)$ is a randomly sampled finite subset. Each $g_n(L,P)$ corresponds to a valid partition $P$.
Fully separable cases such as $g_2(L)$, $g_3(L, 1|2|3)$, and $g_4(L, 1|2|3|4)$ are computed via Algorithm~\ref{alg1} and omitted here. We focus instead on evaluating $\{g_n(L,P):P\in\mathcal P_n^{vk}\}$ for relevant partitions  with $k<n$.

%{\it Establishing the database $\mathcal{EW}_3^2$ for 3-qubits}

As introduced in Section~\ref{Sec5}, the finite three-qubit bipartition database is $$\widetilde{\mathcal{EW}}_3^{(2)}=\left\{\left(L,\{g_3(L,P):P\in\mathcal P_3^{v2}\}\right):L\in\widetilde{\mathcal B}_1^+(\mathbb C^2\otimes\mathbb C^2\otimes\mathbb C^2)\right\},$$ where $\mathcal P_3^{v2}=\{1|23,\,12|3,\,13|2\}.$ For each $P\in\mathcal P_3^{v2}$, the quantity $g_3(L,P)$ is evaluated by the corresponding partition-adapted iterative procedure. {To avoid excessive algorithmic details, we present only the representative case $P=12|3$ below. The other cases are implemented analogously in the numerical program, and further implementation details are available from the authors upon request.}

\begin{breakablealgorithm}\label{alg:SPI2}
\caption{Representative iteration algorithm for the two-partition $12|3$ in a three-qubit system}
\begin{algorithmic}[1]

\REQUIRE
$L \in \mathcal{B}_1^+(\mathbb{C}^2 \otimes \mathbb{C}^2 \otimes \mathbb{C}^2 )$; $n \in \mathbb{N}_+$
\ENSURE $g_{3}(L, 12|3)$

\STATE Randomly initialize product state  $|b_{12}\rangle \in \mathbb{C}^2 \otimes \mathbb{C}^2 $ and the state $|b_3\rangle \in \mathbb{C}^2$.

\STATE $|\xi_3\rangle \gets L |b_{12},  b_3\rangle$,
\STATE $A^{\prime} \gets |\xi_3\rangle \langle \xi_3|$.

\WHILE{not converged}
  \STATE \textbf{Phase 1: Update $|a_{12}^{\prime}\rangle$ via forward iteration}
  \WHILE{not converged}

      \STATE $L^{\prime}_2 \gets  {\rm Tr}_{3}(A^{\prime}) + I$;
      \STATE $|\xi_{12}\rangle \gets  L^{\prime}_2 \left| b_{12} \right\rangle$;

    \STATE $\left| a_{12}^{\prime} \right\rangle \gets  \frac{|\xi_{12}\rangle}{\sqrt{\langle \xi_{12} | \xi_{12} \rangle}}$.
  \ENDWHILE

  \STATE \textbf{Phase 2: Backward update of $|a_3^{\prime}\rangle$}

    \STATE Schmidt decomposition of $|\xi_3\rangle$ with respect to the bipartite decomposition $\{1, 2\}$ and $\{3\}$:
     $ |\xi_3\rangle \gets  \sum_{j=1}^{r} \sqrt{\mu_j} |\phi_j\rangle |z_j\rangle.$
    \STATE $|c_3\rangle\gets \langle a_{12}^{\prime} | \cdot |\xi_3\rangle$;
    \STATE Normalize:  $|a_3^{\prime}\rangle \gets  \frac{|c_3\rangle}{\sqrt{\langle c_3 | c_3 \rangle}}$.

  \STATE \( | a_{12} a_3\rangle \gets | a_{12}^{\prime},  a_3^{\prime}\rangle \).
\ENDWHILE

\STATE $g_{3}(L, 12|3)\gets \langle a_{12} a_{3} | L | a_{12} a_{3}\rangle.$

\RETURN  $g_{3}(L, 12|3)$.

\end{algorithmic}
\end{breakablealgorithm}

Using Algorithm~\ref{alg1} and the corresponding iterative procedures for the valid bipartitions, we construct the three-qubit dataset
\begin{equation*}
\begin{aligned}
\widetilde{\mathcal{EW}}^{(3)} = \big\{
&(L,\ g_3 (L, 1|2|3),\ g_3(L, 12|3),  g_3(L, 1|23),\ g_3(L,13|2)) : L\in \tilde{\mathcal{B}}_1^+(\mathbb{C}^2\otimes\mathbb{C}^2\otimes \mathbb{C}^2)
\big\}.
\end{aligned}
\end{equation*}

For the four-qubit tripartition database introduced in
Section~\ref{Sec5},
$\widetilde{\mathcal{EW}}_4^{(3)}
=
\left\{
\left(
L,\{g_4(L,P):P\in\mathcal P_4^{v3}\}
\right):
L\in\widetilde{\mathcal B}_1^+((\mathbb C^2)^{\otimes4})
\right\},
$
where $\mathcal P_4^{v3}=\{12|3|4,\,1|23|4,\,1|2|34,\,13|2|4,\,14|2|3,\,1|3|24\}.$ For each $P\in\mathcal P_4^{v3}$, the quantity $g_4(L,P)$ is evaluated by the corresponding partition-adapted iterative procedure. {To avoid excessive algorithmic details, we present
below only three representative cases, $12|3|4$, $1|23|4$, and $1|2|34$. The remaining cases are computed by their corresponding
iterative procedures and are not displayed here.}

\begin{breakablealgorithm}\label{alg:SPI5}
\caption{Iteration algorithm for the representative 3-partition $12|3|4$ in a 4-qubit system}
\begin{algorithmic}[1]

\REQUIRE
$L \in \mathcal{B}_1^+(\mathbb{C}^2 \otimes \mathbb{C}^2 \otimes \mathbb{C}^2 \otimes \mathbb{C}^2)$; $n=4$
\ENSURE $g_{4}(L, 12|3|4)$

\STATE Randomly initialize product state $|b_{12}\rangle \in \mathbb{C}^2 \otimes \mathbb{C}^2$, $|b_3\rangle \in \mathbb{C}^2$, and $|b_4\rangle \in \mathbb{C}^2$.

\STATE $|\xi_4\rangle \gets L |b_{12}, b_3, b_4\rangle$
\STATE $A^{\prime} \gets |\xi_4\rangle \langle \xi_4|$

\WHILE{not converged}
  \STATE \textbf{Phase 1: Update $|a_{12}^{\prime}\rangle$ via forward iteration}
  \WHILE{not converged}
    \STATE $L^{\prime}_2 \gets {\rm Tr}_{4}(A^{\prime}) + I$
    \STATE $|\xi_3\rangle \gets L^{\prime}_2 |b_{12}, b_3\rangle$
    \STATE $A^{\prime} \gets |\xi_3\rangle \langle \xi_3|$

    \STATE $L^{\prime}_3 \gets {\rm Tr}_{3}(A^{\prime}) + I$
    \STATE $|\xi_{12}\rangle \gets L^{\prime}_3 |b_{12}\rangle$
    \STATE $\left|a_{12}^{\prime}\right\rangle \gets \frac{|\xi_{12}\rangle}{\sqrt{\langle \xi_{12} | \xi_{12} \rangle}}$
  \ENDWHILE

  \STATE \textbf{Phase 2: Backward update of $|a_3^{\prime}\rangle$ and $|a_4^{\prime}\rangle$}

    \STATE Schmidt decomposition of $|\xi_3\rangle$ with respect to $\{1,2\}$ and $\{3\}$:
    $|\xi_3\rangle \gets \sum_{j=1}^{r} \sqrt{\mu_j} |\phi_j\rangle |z_j\rangle$
    \STATE $|c_3\rangle \gets \langle a_{12}^{\prime} | \cdot |\xi_3\rangle$
    \STATE Normalize: $|a_3^{\prime}\rangle \gets \frac{|c_3\rangle}{\sqrt{\langle c_3 | c_3 \rangle}}$

    \STATE Schmidt decomposition of $|\xi_4\rangle$ with respect to $\{1,2,3\}$ and $\{4\}$:
    $|\xi_4\rangle \gets \sum_{j=1}^{r'} \sqrt{\mu_j'} |\phi_j'\rangle |z_j'\rangle$
    \STATE $|c_4\rangle \gets \langle a_{12}^{\prime} | \otimes \langle a_3^{\prime} | \cdot |\xi_4\rangle$
    \STATE Normalize: $|a_4^{\prime}\rangle \gets \frac{|c_4\rangle}{\sqrt{\langle c_4 | c_4 \rangle}}$

  \STATE $|a_{12} a_3 a_4\rangle \gets |a_{12}^{\prime}, a_3^{\prime}, a_4^{\prime}\rangle$
\ENDWHILE

\STATE $g_{4}(L, 12|3|4) \gets \langle a_{12} a_3 a_4 | L | a_{12} a_3 a_4\rangle$

\RETURN $g_{4}(L, 12|3|4)$

\end{algorithmic}
\end{breakablealgorithm}

\begin{breakablealgorithm}\label{alg:SPI6}
\caption{: Iteration algorithm for the representative 3-partition $1|23|4$ in a 4-qubit system}
\begin{algorithmic}[1]

\REQUIRE
$L \in \mathcal{B}_1^+(\mathbb{C}^2 \otimes \mathbb{C}^2 \otimes \mathbb{C}^2 \otimes \mathbb{C}^2)$; $n=4$
\ENSURE
$g_{4}(L, 1|23|4)$

\STATE Randomly initialize product state $|b_1\rangle \in \mathbb{C}^2$, $|b_{23}\rangle \in \mathbb{C}^2 \otimes \mathbb{C}^2$, and $|b_4\rangle \in \mathbb{C}^2$.

\STATE $|\xi_4\rangle \gets L |b_1, b_{23}, b_4\rangle$
\STATE $A^{\prime} \gets |\xi_4\rangle \langle \xi_4|$

\WHILE{not converged}
  \STATE \textbf{Phase 1: Update $|a_1^{\prime}\rangle$ via forward iteration}
  \WHILE{not converged}
    \STATE $L^{\prime}_2 \gets {\rm Tr}_{4}(A^{\prime}) + I$
    \STATE $|\xi_{23}\rangle \gets L^{\prime}_2 |b_1, b_{23}\rangle$
    \STATE $A^{\prime} \gets |\xi_{23}\rangle \langle \xi_{23}|$

    \STATE $L^{\prime}_3 \gets {\rm Tr}_{23}(A^{\prime}) + I$
    \STATE $|\xi_1\rangle \gets L^{\prime}_3 |b_1\rangle$
    \STATE $|a_1^{\prime}\rangle \gets \frac{|\xi_1\rangle}{\sqrt{\langle \xi_1 | \xi_1 \rangle}}$
  \ENDWHILE

  \STATE \textbf{Phase 2: Backward update of $|a_{23}^{\prime}\rangle$ and $|a_4^{\prime}\rangle$}

    \STATE Schmidt decomposition of $|\xi_{23}\rangle$ with respect to $\{1\}$ and $\{2,3\}$:
    $|\xi_{23}\rangle \gets \sum_{j=1}^{r} \sqrt{\mu_j} |\phi_j\rangle |z_j\rangle$
    \STATE $|c_{23}\rangle \gets \langle a_1^{\prime} | \cdot |\xi_{23}\rangle$
    \STATE $|a_{23}^{\prime}\rangle \gets \frac{|c_{23}\rangle}{\sqrt{\langle c_{23} | c_{23} \rangle}}$

    \STATE Schmidt decomposition of $|\xi_4\rangle$ with respect to $\{1,2,3\}$ and $\{4\}$:
    $|\xi_4\rangle \gets \sum_{j=1}^{r'} \sqrt{\mu_j'} |\phi_j'\rangle |z_j'\rangle$
    \STATE $|c_4\rangle \gets \langle a_1^{\prime}| \otimes \langle a_{23}^{\prime}| \cdot |\xi_4\rangle$
    \STATE $|a_4^{\prime}\rangle \gets \frac{|c_4\rangle}{\sqrt{\langle c_4 | c_4 \rangle}}$

  \STATE $|a_1, a_{23}, a_4\rangle \gets |a_1^{\prime}, a_{23}^{\prime}, a_4^{\prime}\rangle$
\ENDWHILE

\STATE $g_{4}(L, 1|23|4) \gets \langle a_1, a_{23}, a_4 | L | a_1, a_{23}, a_4\rangle$

\RETURN $g_{4}(L, 1|23|4)$

\end{algorithmic}
\end{breakablealgorithm}
\begin{breakablealgorithm}\label{alg:SPI7}
\caption{:  Iteration algorithm for the representative 3-partition $1|2|34$ in a 4-qubit system}
\begin{algorithmic}[1]

\REQUIRE
$L \in \mathcal{B}_1^+(\mathbb{C}^2 \otimes \mathbb{C}^2 \otimes \mathbb{C}^2 \otimes \mathbb{C}^2)$; $n=4$
\ENSURE
$g_{4}(L, 1|2|34)$

\STATE Randomly initialize product state $|b_1\rangle \in \mathbb{C}^2$, $|b_2\rangle \in \mathbb{C}^2$, $|b_{34}\rangle \in \mathbb{C}^2 \otimes \mathbb{C}^2$.

\STATE $|\xi_{34}\rangle \gets L |b_1, b_2, b_{34}\rangle$
\STATE $A^{\prime} \gets |\xi_{34}\rangle \langle \xi_{34}|$

\WHILE{not converged}
  \STATE \textbf{Phase 1: Update $|a_1^{\prime}\rangle$ via forward iteration}
  \WHILE{not converged}
    \STATE $L^{\prime}_2 \gets {\rm Tr}_{34}(A^{\prime}) + I$
    \STATE $|\xi_2\rangle \gets L^{\prime}_2 |b_1, b_2\rangle$
    \STATE $A^{\prime} \gets |\xi_2\rangle \langle \xi_2|$

    \STATE $L^{\prime}_3 \gets {\rm Tr}_{2}(A^{\prime}) + I$
    \STATE $|\xi_1\rangle \gets L^{\prime}_3 |b_1\rangle$
    \STATE $|a_1^{\prime}\rangle \gets \frac{|\xi_1\rangle}{\sqrt{\langle \xi_1 | \xi_1 \rangle}}$
  \ENDWHILE

  \STATE \textbf{Phase 2: Backward update of $|a_2^{\prime}\rangle$ and $|a_{34}^{\prime}\rangle$}

    \STATE Schmidt decomposition of $|\xi_2\rangle$ with respect to $\{1\}$ and $\{2\}$:
    $|\xi_2\rangle \gets \sum_{j=1}^{r} \sqrt{\mu_j} |\phi_j\rangle |z_j\rangle$
    \STATE $|c_2\rangle \gets \langle a_1^{\prime} | \cdot |\xi_2\rangle$
    \STATE $|a_2^{\prime}\rangle \gets \frac{|c_2\rangle}{\sqrt{\langle c_2 | c_2 \rangle}}$

    \STATE Schmidt decomposition of $|\xi_{34}\rangle$ with respect to $\{1,2\}$ and $\{3,4\}$:
    $|\xi_{34}\rangle \gets \sum_{j=1}^{r'} \sqrt{\mu_j'} |\phi_j'\rangle |z_j'\rangle$
    \STATE $|c_{34}\rangle \gets \langle a_1^{\prime}| \otimes \langle a_2^{\prime}| \cdot |\xi_{34}\rangle$
    \STATE $|a_{34}^{\prime}\rangle \gets \frac{|c_{34}\rangle}{\sqrt{\langle c_{34} | c_{34} \rangle}}$

  \STATE $|a_1, a_2, a_{34}\rangle \gets |a_1^{\prime}, a_2^{\prime}, a_{34}^{\prime}\rangle$
\ENDWHILE

\STATE $g_{4}(L, 1|2|34) \gets \langle a_1, a_2, a_{34} | L | a_1, a_2, a_{34}\rangle$

\RETURN $g_{4}(L, 1|2|34)$

\end{algorithmic}
\end{breakablealgorithm}

%\subsubsection{ Establishing the database $\mathcal{EW}_4^2$ for 4-qubits}
The finite four-qubit bipartition database is $\widetilde{\mathcal{EW}}_4^{(2)} =\bigl\{\bigl(L,\{g_4(L,P):P\in\mathcal P_4^{v2}\}\bigr):
L\in\widetilde{\mathcal B}_1^+((\mathbb C^2)^{\otimes4})\bigr\}$, where $\mathcal P_4^{v2}=\{1|234,2|134,3|124,123|4,12|34,13|24,14|23\}$. For each $P\in\mathcal P_4^{v2}$, $g_4(L,P)$ is evaluated by
the corresponding partition-adapted iterative procedure. {To avoid excessive algorithmic details, we present below two representative cases, $1|234$ and $12|34$. The remaining cases are computed by their corresponding iterative procedures and are not displayed here.}

\begin{breakablealgorithm}\label{alg:SPI11}
\caption{Iteration algorithm for the representative 2-partition $1|234$ in a 4-qubit system}
\begin{algorithmic}[1]

\REQUIRE
$L \in \mathcal{B}_1^+(\mathbb{C}^2 \otimes \mathbb{C}^2 \otimes \mathbb{C}^2 \otimes \mathbb{C}^2)$; $n=4$
\ENSURE
$g_{4}(L, 1|234)$

\STATE Randomly initialize product state $|b_1\rangle \in \mathbb{C}^2$, $|b_{234}\rangle \in \mathbb{C}^2 \otimes \mathbb{C}^2 \otimes \mathbb{C}^2$ (entangled state on subsystems 2, 3, 4).

\STATE $|\xi_{234}\rangle \gets L |b_1, b_{234}\rangle$
\STATE $A^{\prime} \gets |\xi_{234}\rangle \langle \xi_{234}|$

\WHILE{not converged}
  \STATE \textbf{Phase 1: Update $|a_1^{\prime}\rangle$ via forward iteration}
  \WHILE{not converged}
    \STATE $L^{\prime}_1 \gets {\rm Tr}_{234}(A^{\prime}) + I$
    \STATE $|\xi_1\rangle \gets L^{\prime}_1 |b_1\rangle$
    \STATE $|a_1^{\prime}\rangle \gets \frac{|\xi_1\rangle}{\sqrt{\langle \xi_1 | \xi_1 \rangle}}$
  \ENDWHILE

  \STATE \textbf{Phase 2: Backward update of $|a_{234}^{\prime}\rangle$}

  \STATE Schmidt decomposition of $|\xi_{234}\rangle$ with respect to $\{1\}$ and $\{2,3,4\}$: $  |\xi_{234}\rangle \gets \sum_{j=1}^r \sqrt{\mu_j} |\phi_j\rangle |z_j\rangle$
  \STATE $|c_{234}\rangle \gets \langle a_1^{\prime} | \cdot |\xi_{234}\rangle$
  \STATE $|a_{234}^{\prime}\rangle \gets \frac{|c_{234}\rangle}{\sqrt{\langle c_{234} | c_{234} \rangle}}$

  \STATE $|a_1, a_{234}\rangle \gets |a_1^{\prime}, a_{234}^{\prime}\rangle$
\ENDWHILE

\STATE $g_4(L, 1|234) \gets \langle a_1, a_{234} | L | a_1, a_{234} \rangle$

\RETURN $g_4(L, 1|234)$

\end{algorithmic}
\end{breakablealgorithm}

\begin{breakablealgorithm}\label{alg:SPI15}
\caption{Iteration algorithm for the representative 2-partition $12|34$ in a 4-qubit system}
\begin{algorithmic}[1]

\REQUIRE
$L \in \mathcal{B}_1^+(\mathbb{C}^2 \otimes \mathbb{C}^2 \otimes \mathbb{C}^2 \otimes \mathbb{C}^2)$; $n=4$
\ENSURE
$g_{4}(L, 12|34)$

\STATE Randomly initialize product state $|b_{12}\rangle \in \mathbb{C}^2 \otimes \mathbb{C}^2$, $|b_{34}\rangle \in \mathbb{C}^2 \otimes \mathbb{C}^2$, where both are entangled in their respective subgroups.

\STATE $|\xi_{34}\rangle \gets L |b_{12}, b_{34}\rangle$
\STATE $A^{\prime} \gets |\xi_{34}\rangle \langle \xi_{34}|$

\WHILE{not converged}
  \STATE \textbf{Phase 1: Update $|a_{12}^{\prime}\rangle$ via forward iteration}
  \WHILE{not converged}
    \STATE $L^{\prime}_{12} \gets {\rm Tr}_{34}(A^{\prime}) + I$
    \STATE $|\xi_{12}\rangle \gets L^{\prime}_{12} |b_{12}\rangle$
    \STATE $|a_{12}^{\prime}\rangle \gets \frac{|\xi_{12}\rangle}{\sqrt{\langle \xi_{12} | \xi_{12} \rangle}}$
  \ENDWHILE

  \STATE \textbf{Phase 2: Backward update of $|a_{34}^{\prime}\rangle$}

  \STATE Schmidt decomposition of $|\xi_{34}\rangle$ with respect to  $\{1,2\}$ and $\{3,4\}$:
$|\xi_{34}\rangle \gets \sum_{j=1}^r \sqrt{\mu_j} |\phi_j\rangle |z_j\rangle$
  \STATE $|c_{34}\rangle \gets \langle a_{12}^{\prime} | \cdot |\xi_{34}\rangle$
  \STATE $|a_{34}^{\prime}\rangle \gets \frac{|c_{34}\rangle}{\sqrt{\langle c_{34} | c_{34} \rangle}}$

  \STATE $|a_{12}, a_{34}\rangle \gets |a_{12}^{\prime}, a_{34}^{\prime}\rangle$
\ENDWHILE

\STATE $g_4(L, 12|34) \gets \langle a_{12}, a_{34} | L | a_{12}, a_{34} \rangle$

\RETURN $g_4(L, 12|34)$

\end{algorithmic}
\end{breakablealgorithm}
This completes the construction of the four-qubit dataset, which contains the values $g_4(L,P)$ associated with all valid 2-, 3-, and 4-partitions.
\end{widetext}
\bibliography{12}

\end{document}